%% file: main.tex
\documentclass[fleqn,usenatbib]{mnras}
 
\usepackage{newtxtext,newtxmath}
\usepackage[T1]{fontenc}
\usepackage{graphicx}
\usepackage{amsmath}

\DeclareRobustCommand{\VAN}[3]{#2}
\let\VANthebibliography\thebibliography
\def\thebibliography{\DeclareRobustCommand{\VAN}[3]{##3}\VANthebibliography}

\newcommand{\Mpc}{\mathrm{Mpc}}
\newcommand{\Gpc}{\mathrm{Gpc}}
\newcommand{\Msun}{M_{\odot}}
\newcommand{\Hz}{\mathrm{Hz}}
\newcommand{\yr}{\mathrm{yr}}
\newcommand{\Gyr}{\mathrm{Gyr}}
\newcommand{\Msmbh}{M_{\rm SMBH}}
\newcommand{\kms}{{\rm km/s}}
 
\renewcommand{\theenumi}{(\arabic{enumi}}

 \input{Title_Abstract/Title.tex}

\begin{document}
 
 \label{firstpage}
 
 \pagerange{\pageref{firstpage}--\pageref{lastpage}}
 
 \maketitle

\input{Title_Abstract/Abstract.tex}

 \input{Sections/Introduction.tex}

\input{Sections/Properties_GNs.tex}

\input{Sections/BBH_form_evol.tex}

\input{Sections/DistribParams_SingleGN.tex}

\input{Sections/Binaries_within_aLIGOhorizon.tex}

\input{Sections/Results.tex}

\input{Sections/Summary_Conclusions.tex}

\input{Sections/Data_Availability.tex}

\input{Sections/Acknowledgments.tex}

 \bibliographystyle{mnras}
 \bibliography{refs}

 \appendix

\input{Appendices/Appendix_Properties_StellarPops.tex}

\input{Appendices/Appendix_Rates_1PN.tex}

\input{Appendices/Appendix_EvolBBHparams.tex}

\input{Appendices/Appendix_RespGWdetector}

\input{Appendices/Appendix_DetDistance}

\input{Appendices/Appendix_MCvolume}

 \bsp
 \label{lastpage}
 \end{document}

%% file: Title_Abstract/Title.tex
\title[Characterizing GW Captures in Galactic Nuclei]{High Eccentricities and High Masses Characterize Gravitational-wave Captures in Galactic Nuclei as Seen by Earth-based Detectors}

\author[L. Gond\'an \& B. Kocsis]{
L\'aszl\'o Gond\'an,$^{1}$\thanks{E-mail: lgondan@caesar.elte.hu}
Bence Kocsis$^{2}$
\\
$^{1}$ Institute of Physics, E\"otv\"os University, P\'azm\'any P. s. 1/A, Budapest 1117, HU\\
$^{2}$ Rudolf Peierls Centre for Theoretical Physics, Clarendon Laboratory, Parks Road, Oxford OX1 3PU, UK
}

\date{Accepted XXX. Received YYY; in original form ZZZ}

\pubyear{2021}

%% file: Title_Abstract/Abstract.tex
\begin{abstract}
 {
 The emission of gravitational waves (GWs) during single--single close encounters in galactic nuclei (GNs) leads to the formation and rapid merger of highly eccentric stellar--mass black hole (BH) binaries. The distinct distribution of physical parameters makes it possible to statistically distinguish this source population from others. Previous studies determined the expected binary parameter distribution for this source population in single GNs. Here we take into account the effects of dynamical friction, post-Newtonian corrections, and observational bias to determine the detected sources' parameter-distributions from all GNs in the Universe. We find that the total binary mass distribution of detected mergers is strongly tilted towards higher masses. The distribution of initial peak GW frequency is remarkably high between $1-70$ Hz, $\sim50\%$ of GW capture sources form above $10$ Hz with $e\gtrsim0.95$. The eccentricity when first entering the LIGO/Virgo/KAGRA band satisfies $e_{\rm10Hz}>0.1$ for over $92\%$ of sources and $e_{\rm10Hz}>0.8$ for more than half of the sources. At the point when the pericenter reaches $10GM/c^2$ the eccentricity satisfies $e_{\rm10M}>0.1$ for over $\sim70\%$ of the sources, making single--single GW capture events in GNs the most eccentric source population among the currently known stellar--mass binary BH merger channels in the Universe. We identify correlations between total mass, mass ratio, source detection distance, and eccentricities $e_{\rm10Hz}$ and $e_{\rm10M}$. The recently measured source parameters of GW190521 lie close to the peak of the theoretical distributions and the estimated escape speed of the host environment is \mbox{$\sim7.5\times10^3\,\rm km/s-1.2\times10^4\,\rm km/s$}, making this source a candidate for this astrophysical merger channel.
 }
\end{abstract}


%% file: Sections/Introduction.tex
\section{Introduction} 
\label{sec:Intro}
 
 The Advanced Laser Interferometer Gravitational--Wave Observatory\footnote{\url{http://www.ligo.caltech.edu/}} (aLIGO; \citealt{Aasietal2015}) and Advanced Virgo\footnote{\url{http://www.ego-gw.it/}} (AdV; \citealt{Acerneseetal2015}) have announced the detection of 46 binary black hole (BBH) mergers during their first two observing runs and the first half of the third observing run (O1, O2, and O3a;  \citealt{Abbottetal2016b,Abbottetal2016c,Abbottetal2017a,Abbottetal2017c,Abbottetal2017b,Abbottetal2019,Abottetal2020e,Abbottetal2020a,Abbottetal2020c}), and more than eighty public alerts\footnote{\url{https://gracedb.ligo.org/}} have been reported during the third observing run. Several additional BBH mergers were identified in the publicly available data from the first and second observing runs \citep{Venumadhavetal2019a,Venumadhavetal2019b,Zackayetal2019a,Zackayetal2019b,Nitzetal2020}. Based on the detected BBHs, the BBH merger rate density was observationally constrained by the LIGO--Virgo Collaboration to the range $15.3 - 38.2 \, {\rm Gpc}^{-3} {\rm yr}^{-1}$ \citep{Abottetal2020_O3population}.
 Upgrades on aLIGO and AdV, and by expanding the system to five stations with the involvement of the Japanese KAGRA\footnote{\url{https://gwcenter.icrr.u-tokyo.ac.jp/en/}} \citep{KagraColletal2018} and LIGO India\footnote{\url{http://www.gw.iucaa.in/ligo-india/}} \citep{Unnikrishnan2013} in the upcoming years \citep{Abbottetal2018}, will broaden the detector network's access \citep{Abbottetal2018}. The number of GW detections is expected to grow at an unprecedented rate in the upcoming years (e.g., \citealt{Abbottetal2018,Baibhavetal2019}).
 
 A large number of astrophysical processes have been proposed to explain the observed GW events \citep{Abadieetal2010,Baracketal2019}. One possibility is the isolated evolution of massive stellar binaries in galactic fields, either via common envelope evolution (e.g., \citealt{Giacobboetal2018,Kruckow2018,MapelliGiacobbo2018,Mapellietal2019,Neijsseletal2019,Speraetal2019}, and references therein) or via chemically homogeneous evolution \citep{deMinkMandel2016,MandeldeMink2016,Marchantetal2016}. In both cases, the resulting merging BH binaries are circular within measurement errors within the GW frequency band of both  LIGO/Virgo and the Laser Interferometer Space Antenna (LISA; \citealt{AmaroSeoaneetal2017}). Eccentricity has not been detected in the O1 and O2 observing runs of LIGO/Virgo \citep{Abbottetal2019d,RomeroShaw2019,Wuetal2020}. However, one BBH merger event in O3a, GW190521, has been shown to be possibly highly eccentric \citep{RomeroShaw2020,Gayathrietal2020}.
 
 Eccentricity may be significant in the LIGO/Virgo frequency band in several astrophysical merger pathways. In the late-inspiral phase, the highest eccentricities are expected following a single--single GW capture in GNs \citep{OLearyetal2009,KocsisLevin2012,HongLee2015,Gondanetal2018b}. Single--single \citep{Rodriguezetal2018b,Rodriguezetal2018,Samsingetal2020}, binary--single \citep{Samsingetal2014,SamsingRamirezRuiz2017,Rodriguezetal2018b,Rodriguezetal2018,Samsing2018,Samsingetal2018,Samsingetal2018_2} and binary--binary \citep{Zevinetal2018} interactions may lead to systems with significantly nonzero eccentricity in globular clusters (GCs) and the gas disks of active galactic nuclei \citep{Samsingetal2020b,Tagawaetal2020c} as well. In hierarchical triples, the Kozai--Lidov (KL) oscillation \citep{Kozai1962,Lidov1962,LidovZiglin1976} may lead to eccentric BBH mergers in the LIGO/Virgo band too, where the tertiary companion may be a star or another stellar-mass compact object, such as a BH, a neutron star (NS), or a white dwarf (WD) in the galactic field \citep{Antoninietal2017,LiuLai2017,SilsbeeTremaine2017,LiuLai2018,RodriguezAntonini2018,FragioneKocsis2020,Liuetal2020,MichaelyPerets2020}, in GCs \citep{Wen2003,Aarseth2012,Antoninietal2014,Antoninietal2016,Breiviketal2016,Martinezetal2020}, or in young massive and open star clusters \citep{Kimpsonetal2016}. Alternatively, the tertiary can be a supermassive black hole (SMBH) in GNs  \citep{AntoniniPerets2012,Hamersetal2018,Hoangetal2018,RandallXianyu2018a,RandallXianyu2018b,Fragioneetal2019b,Yuetal2020} or an intermediate-mass BH in GCs \citep{FragioneBromberg2019}. Furthermore, eccentric BBH mergers are also produced in the LIGO/Virgo band in hierarchical \citep{Fragioneetal2019} and non-hierarchical \citep{ArcaSeddaetal2018b} triples in the vicinity of a SMBH, and in $2 + 2$ \citep{FragioneKocsis2019,LiuLai2019} and hierarchical $3 + 1$ \citep{HamersSafarzadeh2020} quadruples. Eccentric mergers may also involve NSs in several astrophysical pathways \citep{Leeetal2010,Thompson2011,AntoniniPerets2012,Tsang2013,Samsingetal2014,PetrovichAntonini2017,Fragioneetal2019b,Fragioneetal2019,FragioneLoeb2019,ArcaSedda2020,Hoangetal2020}. 
 
 The predicted merger rate density differs for the various eccentric BBH merger channels, generally in the range \mbox{$0.001 - 10 \, {\rm Gpc}^{-3} {\rm yr}^{-1}$}. In these channels, BBHs typically form with eccentricities beyond $0.9$. Since GW emission circularizes the orbits \citep{Peters1964,Hinderetal2008}, depending on the binary separation at formation, they may either only be eccentric in the frequency bands of future space-based detectors (LISA; \mbox{(B-)DECIGO}, \citealt{Setoetal2001,Kawamuraetal2011,Kawamuraetal2020}; TianQin, \citealt{Luoetal2016,Meietal2020}), or if they form at sufficiently high GW frequencies they may remain eccentric all the way to the LIGO/Virgo band. Single--single GW capture BBH mergers in GNs are unique among other formation channels as the majority of them form directly in the LIGO/Virgo frequency band or very close to it, such that they retain a high eccentricity, $e > 0.1$, when their peak GW frequency \citep{Wen2003} enters the LIGO/Virgo band \citep{Gondanetal2018b}. This fraction is typically $\lesssim 20 \%$ for other channels producing stellar-mass eccentric compact binaries.
 
 The measurement of eccentricity offers a unique way to distinguish different astrophysical merger pathways. It can be measured for stellar-mass compact binaries at $10 \, \Hz$ band for $e \gtrsim 0.02 - 0.1$  \citep{BrownZimmerman2010,HuertaBrown2013,Huertaetal2017,Huertaetal2018,Loweretal2018,GondanKocsis2019,RomeroShaw2019,Lenonetal2020,Wuetal2020}. Furthermore, both LISA and TianQin will have the capability to measure orbital eccentricity for such sources as well \citep{Nishizawaetal2016,Liuetal2020b}. Together with other suggested parameters such as spins and masses, their distributions and correlations among them may be used as an indicator to statistically disentangle among the contributions of the different BBH formation channels. This includes the measurement of eccentricity in the LISA  \citep{Nishizawaetal2017,DOrazioSamsing2018,Samsingetal2018,SamsingDorazio2018,Hoangetal2019,Kremeretal2019,RandallXianyu2019,Demeetal2020,EmamiLoeb2020b}, DECIGO/TianQin \citep{ChenAmaroSeoane2017,Samsingetal2020}, and LIGO bands \citep{ArcaSeddaetal2018b,DOrazioSamsing2018,Gondanetal2018b,Samsing2018,Samsingetal2018,SamsingDorazio2018,Zevinetal2018}; the spin distribution \citep{FishbachHolz2017}, spin orientation \citep{Rodriguezetal2016,Farretal2017,LiuLai2017,Stevensonetal2017,TalbotThrane2017,Vitaleetal2017,Farretal2018,Gerosaetal2018,LiuLai2018,RodriguezAntonini2018,Lopezetal2018,Yangetal2019}, the projected effective spin parameter \citep{Antoninietal2018,Ngetal2018,Schoderetal2018,Zaldarriagaetal2018}, and its distribution \citep{FragioneKocsis2020}; the mass distribution \citep{Mandeletal2015,Stevensonetal2015,OLearyetal2016,Fishbachetal2017,Mandeletal2017,Zevinetal2017,Kocsisetal2018,Pernaetal2019}; the identification of correlations between mass and eccentricity \citep{Wen2003,Breiviketal2016,Gondanetal2018b,RandallXianyu2018b}, total mass and spins \citep{ArcaSedaBenacquista2018}, and total mass and effective spin \citep{Safarzadehetal2020}; together with Doppler effects related to a possible movement of the BBH's center of mass (e.g., \citealt{Inayoshietal2017,Meironetal2017}). Initial orbital parameters and the characteristic velocity before a hardening encounter may also be measured with sufficient accuracy with the aLIGO--AdV--KAGRA detector network at design sensitivity for BBHs and other stellar-mass compact binaries as well, and their reconstruction may help in constraining astrophysical formation scenarios \citep{GondanKocsis2019}. Finally, \citet{Samsingetal2020} has shown that the distribution of peak GW frequency at the time of binary formation peaks at different frequencies for dynamical BBH merger channels in GCs, thereby this quantity may also be used to distinguish among these formation channels.
 
 So far, studies on stellar-mass eccentric compact mergers listed above have predicted merger rate densities (and in some cases detection rates) and distributions of binary parameters in different host environments both in the local Universe ($z \simeq 0$) and at higher redshifts. Merger rate densities and binary parameter distributions have also been estimated for a (weighted) grid of host models in the local Universe, covering a wide range of initial conditions for the evolution of hosts across cosmic time. Moreover, these studies have suggested several binary parameters and correlations among them to distinguish among GW source populations.
 
 We focus on the single--single GW capture process, in which two single compact objects undergo a close encounter and lose a sufficient amount of energy due to GW emission to become bound. This was first introduced in \citet{OLearyetal2009} for BHs in GN hosts. They predicted merger rate densities together with detection rate estimates for aLIGO at design sensitivity and investigated the radial distributions and distributions of initial binary separation for merging BBHs for single GNs and binaries within aLIGO's effective horizon. Subsequent studies have refined the merger rate density estimates \citep{KocsisLevin2012,Tsang2013,RasskazovKocsis2019} and determined the distributions of orbital parameters in single GNs and the distributions of several mass-dependent parameters in single GNs and within aLIGO's effective horizon, by taking into account variations in the underlying BH mass function and SMBH mass \citep{Gondanetal2018b}.
 
 In this paper, we extend our previous analysis in \citet{Gondanetal2018b} to predict the distributions of binary parameters as observed by aLIGO at design sensitivity for the single-single GW captures in GNs accounting for both the intrinsic variations of these quantities among merging systems and observational bias. We determine the observable distribution of eccentricity at $10 \, \Hz$, or when binaries form at a higher frequency. Furthermore, we investigate the distributions of mass-dependent parameters (e.g., total mass, chirp mass, and mass ratio) for aLIGO detections. We identify possible correlations among binary parameters using their observable distributions, which may help to statistically disentangle this channel among other eccentric formation channels. We also explore the characteristics of the peak GW frequency distribution at the time of binary formation that may distinguish the GW capture BBH merger channel in GNs from dynamical BBH merger channels in GCs.
 
 To examine the distributions of binary parameters, we generate mock Monte Carlo (MC) catalogs of GW capture BBHs as observed by aLIGO. To do so, we first generate BBH populations in single GNs following \citet{Gondanetal2018b} using analytical fits to the Fokker--Planck models of multi-mass BH populations in GNs in equilibrium \citep{OLearyetal2009}. We improve upon \citet{OLearyetal2009} and \citet{Gondanetal2018b} with respect to the modeling of GW capture BBH populations in GNs by (i) adding the first post-Newtonian corrections beyond leading order to the description of GW capture process and binary evolution, (ii) accounting for the possibility of strong mass segregation \citep{AlexanderHopman2009,Keshetetal2009}, and (iii) the accumulation of massive objects from the neighborhood of the GN through dynamical friction (DF) as in \citet{RasskazovKocsis2019}. We use a semi-analytical inspiral-only waveform model of eccentric BBHs in the frequency domain, which includes the leading order GW dissipation and neglects 1PN precession  \citep{MorenoGarridoetal1994,MorenoGarridoetal1995,Mikoczietal2012}, to calculate the signal--to--noise ratio (${\rm S/N}$) values. Furthermore, we define the maximum distance of detection and the detection volume in the MC sample separately for the given inclination, polarization, and sky position, and other physical parameters, as a function of redshifted binary mass, initial orbital separation, and initial orbital eccentricity using the ${\rm S/N}$. This improves upon previous works which adopted an inclination, polarization, and sky-position averaged detection range. We generate mock samples of GNs with the observed distribution of SMBHs in the detection volume, sample GW capture BBHs for each GN host from the corresponding redshift-dependent merger rate distribution, and finally discard the sources that do not reach the ${\rm S/N}$ limit of detection.
 
 The paper is organized as follows. In Sections \ref{sec:Prop_GNs}, we introduce the adopted models for the GN host environment properties. In Section \ref{sec:BinayFormEvol}, we review the phases of GW capture-induced binary evolution. In Section \ref{sec:GWcaptureBBHsSingleGN}, we describe the numerical methods we applied in our MC simulations to generate mock samples of BBHs forming through the GW capture process in single GNs. In Section \ref{sec:BBHs_in_LIGOsHorizon}, we describe the setup of MC simulations resulting in mock catalogs of GW capture BBHs detectable by an advanced GW detector. We present our main results in Section \ref{sec:Results}. Finally, we summarize the results of the paper and draw conclusions in Section \ref{sec:SummAndConc}. Several details of our methodology are included in the Appendix. We summarize the characteristics of relaxed stellar populations in GNs and determine the impact of DF on these populations in Appendices \ref{sec:StellPopsProp} and \ref{sec:ImpactDF_GWcBBHs}, respectively. In Appendix \ref{sec:CoordSyst_RespGWdet}, we introduce the geometric conventions we use to describe how the GWs interact with an L-shaped Earth-based GW detector. Then, in Appendix \ref{sec:HorizonDistance}, we determine the horizon distance as a function of binary parameters and the maximum luminosity distance of detection for initially highly eccentric BBHs for the advanced GW detectors. Ultimately, in Appendix \ref{sec:MCsampling}, we introduce the MC routine with which we generate mock catalogs of GW capture BBHs detectable by an advanced GW detector and mock catalogs in the local Universe.
 
 We use geometric units ($G = 1 = c$) when presenting equations related to the description of BH and GW capture BBH populations in GNs and the formation and evolution of these binaries, where mass $M$ and distance $r$ have units of time $GM/c^3$.

%% file: Sections/Properties_GNs.tex
\section{Galactic Nuclei} 
\label{sec:Prop_GNs}
 
 We start by briefly summarizing the characteristics of GN hosts (Sections \ref{subsec:GN}-\ref{subsec:GNs_Relax}) together with the quantities of their stellar populations relevant in generating mock samples of GNs and merging GW capture BBHs in single GN hosts (Sections \ref{subsec:GNs_NumbDist} and \ref{subsec:GNs_VelDist}).

\subsection{Galactic nucleus}
\label{subsec:GN}
 
 GNs are dense and massive assemblies of stars and compact objects gravitationally bound to a central SMBH and are found at the centers of most galaxies (see \citealt{Neumayeretal2020} for a review). In these environments, the population of stars is significantly dominated by old main-sequence stars of mass $\lesssim 1 \, \Msun$ (MSs) \citep{Pfuhletal2011}, while compact object populations consist of WDs, NSs, and BHs (e.g., \citealt{Genzeletal2010,Generozovetal2018,Haileyetal2018}). The orbital evolution of the components of stellar populations is dominated by the SMBH within the GN's radius of influence:
\begin{equation}  \label{eq:rmax}
  r_{\rm max} = \frac{ \Msmbh }{ \sigma_*^2 } 
\end{equation}
 \citep{Peebles1972}. Here, $\Msmbh$ is the mass of the SMBH, and $\sigma_*$ is the velocity dispersion of the underlying stellar populations in the nucleus near the SMBH. To estimate $r_{\rm max}$, we use the $\Msmbh-\sigma_*$ fit of \citet{KormendyHo2013},
\begin{equation}  \label{eq:Msigma}
  \Msmbh \simeq 3.097 \times 10^8 \,\Msun \left( \frac{ \sigma_* }{ 200\ \mathrm{km}\ \mathrm{s}^{-1}} \right)^{4.384} \, .
\end{equation}
 
 We neglect the BH population outside $r_{\rm max}$ as they marginally contribute to the total merger rate of GW capture BBHs because their number density distribution falls down quickly beyond $r_{\rm max}$ \citep{OLearyetal2009}.

\subsection{Mass distribution of supermassive black holes} 
\label{subsec:SMBH_MassFunc} 
 
 To sample SMBH masses in the MC simulations, we take the mass distribution of SMBHs from \citet{Shankaretal2004}, 
\begin{equation}  \label{eq:P_Msmbh}
  P(\Msmbh) = \frac{ C_{\rm N} }{ \Msmbh } \left( \frac{ \Msmbh }{ M_* }\right)^{ \gamma + 1} \exp{ \left( -\frac{ \Msmbh^{ \kappa } }{ M_*^{ \kappa }}  \right) } \, ,
\end{equation} 
 where $\gamma = -1.11$, $\kappa = 0.49$, $M_* = 6.4 \times 10^7 \Msun$, and hereafter $C_{\rm N}$ is a normalization constant of the introduced distribution function.\footnote{The term $\Msmbh^{-1}$ in Equation (\ref{eq:P_Msmbh}) appears due to the fact that the fitting formula in \citet{Shankaretal2004} is originally given in ${\rm log}_{10}( \Msmbh)$ space.} This function is consistent within uncertainties with the other SMBH mass distribution estimates in the literature inferred either from observations (e.g., \citealt{Grahametal2007,Hopkinsetal2007,Shankaretal2009,Vikaetal2009,Shankar2013,Uedaetal2014}), numerical simulations (e.g., \citealt{Sijackietal2015}), or the galaxy bulge mass distribution \citep{Thanjavuretal2016}. Observations based on hard X-ray survey compilation (e.g., \citealt{Uedaetal2014}) and numerical simulations (e.g., \citealt{Sijackietal2015}) pointed out that the SMBH mass distribution weakly evolves out to $z \simeq 1$ over the SMBH mass range $ \Msmbh \in \left[ 10^5 \, \Msun,  10^{10} \, \Msun \right]$. Since we find that these binaries may be detected by single advanced GW detectors at design sensitivity to a maximum redshift of $z \simeq 0.99$, (Section \ref{subsec:DLdistribGNs}), we neglect the redshift dependence of $P(\Msmbh)$ in this study.

\subsection{Mass range of supermassive black holes} 
\label{subsec:SMBH_MassRange} 
 
 Observations demonstrated that Equation (\ref{eq:Msigma}) extends to SMBHs with masses as low as of order $10^5 \Msun$ \citep{Barthetal2005,GreeneHo2006}, and the smallest SMBH mass ever reported is $\sim 5 \times 10^4 \Msun$ \citep{Baldassareetal2015}. Considering these two arguments, we set the lower limit of the SMBH mass range of interest to be $10^5 \, \Msun$.
 
 Recent observations showed evidence for SMBHs with masses up to $\sim 6.6 \times 10^{10} \, \Msun$ \citep{Shemmeretal2004}.\footnote{Such high masses are expected from theoretical models on the merger history of galaxies (e.g., \citealt{MilosavljevicMerrit2001,Merritt2006})} By sampling SMBH masses from $P(\Msmbh)$ between $10^5 \, \Msun$ and $6.6 \times 10^{10} \, \Msun$, we find a negligible fraction of SMBHs \mbox{($\lesssim 0.5 \%$)} above $10^7 \, \Msun$. Since the merger rate of GW capture BBHs in GNs is weakly sensitive to $\Msmbh$ (\mbox{Table 1} in \citealt{RasskazovKocsis2019}),\footnote{At least for dynamical systems in isotropic thermal equilibrium.} we conclude that the absence of SMBHs with masses $\Msmbh \gtrsim 10^7 \, \Msun$ in MC samples of GN hosts only marginally biases our results.  Accordingly, we cut off the SMBH mass range at $10^7 \, \Msun$.

\subsection{Spatial distribution of galactic nuclei} 
\label{subsec:SMBH_SpatDist} 
 
 As GW capture BBHs may be detected from comoving distances out to $\sim 2.47 - 3.37 \, \Gpc$ for the inspiral phase with single advanced GW detectors at design sensitivity (Section \ref{subsec:DLdistribGNs}), their detection range significantly exceeds the largest scale of inhomogeneity in the mass distribution of the Universe. Thus, we assume that anisotropies and inhomogeneities average out over the considered volume. Since most galaxies are expected to harbor a SMBH in their center (e.g., \citealt{KormendyRichstone1995,KormendyHo2013}), we consider a homogeneous and isotropic spatial distribution of GN hosts in comoving coordinates. Indeed, decades of observations (e.g., \citealt{GiovanelliHaynes1985,GellerHuchra1989,Fisheretal1995,Collessetal2001,Abazajian2009,Aguado2019}) and recent N-body simulations (e.g., \citealt{Springeletal2005,BoylanKolchinetal2009,Anguloetal2012,Geneletal2014,Vogelsbergeretal2014a,Vogelsbergeretal2014b,Sijackietal2015}) pointed out that the large-scale inhomogeneity is limited to smaller scales: galaxies form clusters \citep{HubbleHumason1931,Zwicky1933} on megaparsec scales that are organized in superclusters \citep{Abell1958} on a few tens of megaparsec scales, superclusters form filaments \citep{Tully1986} and walls \citep{GellerHuchra1989} that surround cosmic voids \citep{GregoryThompson1978}, of maximum radii in the range of $20 - 100 \, h^{-1} \, \Mpc$ \citep{Mao2017} with Hubble parameter $h \simeq 0.674$ \citep{Planck2018}.

\subsection{Relaxation of multi-mass stellar populations around supermassive black holes} 
\label{subsec:GNs_Relax} 
 
 The relaxation of spherically symmetric, multi-mass, stellar populations around an SMBH has been investigated in detail (e.g., \citealt{BahcallWolf1977,AmaroSeoaneetal2004,Baumgardtetal2004,Freitagetal2006,HopmanAlexander2006,AlexanderHopman2009,Keshetetal2009,OLearyetal2009,PretoAmaroSeoane2010,AharonPerets2016,Alexander2017,Vasiliev2017,Baumgardtetal2018,FragioneSari2018,Generozovetal2018,Panamarevetal2019,EmamiLoeb2020}). These studies showed that, within the SMBH's radius of influence ($r \leqslant r_{\rm max}$), stellar objects undergo dynamical mass segregation and form an approximately power-law number density profile, $n(r,m) \propto r^{-\alpha(m)}$. Here, $r$  is the radius measured from the SMBH at the center of the GN, and the $\alpha(m)$ exponent is higher for more massive objects. Observations of the stellar distribution in the Galactic center of old main-sequence stars justified the existence of a cusp with a slope consistent within uncertainties with theoretical expectations (e.g., \citealt{Schodeletal2007,Trippeetal2008,Gillessenetal2009,YusefZadehetal2012,Feldmeieretal2014,Schodeletal2018,GallegoCanoetal2018}).
 
 Recently, \citet{SzolgyeneKocsis2018} has shown that the heavy objects may form a flattened distribution (i.e. a disk-like structure) within GNs due to vector resonant relaxation, and the mass-segregated radial profile of these objects in equilibrium may be different in these configurations from that of spherically symmetric populations \citep{Fouvryetal2018}. Such distribution was observed for young massive stars within the center of the Milky Way (e.g., \citealt{Paumardetal2006,Bartkoetal2009,Luetal2009,Bartkoetal2010,Doetal2013,Yeldaetal2014}) and other galaxies (e.g. \citealt{Sethetal2008,Lockhartetal2018}). So far, the number density profiles of BHs in BH disks have been uncertain, we, therefore, restrict our investigations to spherically symmetric BH populations around SMBHs in this study. Populations of light stars (MSs, NSs, and WDs) are modeled with spherically symmetric number density profiles according to observations for old main-sequence stars in the Milky Way.
 
 Nuclear star clusters around SMBHs of mass \mbox{$\Msmbh \lesssim 10^7 \Msun$} reach an equilibrium distribution within a Hubble time (e.g., \citealt{Merrit2013,BarOr2013,Gondanetal2018b}). The equilibrium state is reached within a few $\Gyr$ in stellar populations with either massive components in the nucleus or deep inside the nucleus \citep{Mastrobuono-Battisti2014,Panamarevetal2019,EmamiLoeb2020}. Specifically for the Galactic Center, a quasi-steady state is established for the innermost regions after $\sim 3 \, \Gyr$, and the spatial distribution of BHs can be characterized with the predicted power-law density profile after $\sim 5 \, \Gyr$ \citep{Panamarevetal2019}. Since we find that inspiraling GW capture BBHs are detectable with single advanced GW detectors at design sensitivity only within a redshift of $z \simeq 1$ (Section \ref{subsec:DLdistribGNs}), we assume that the number density distribution has reached equilibrium.

\subsection{Mass and number density distribution of objects and related quantities} 
\label{subsec:GNs_NumbDist} 
 
 Assuming spherically symmetric and relaxed stellar systems, we parameterize the 3D number density distribution of mass $m$ objects as
\begin{equation}  \label{eq:n(r)}
  n(r,m) =  C_{\rm frac} n_{\rm inf} \mathcal{F}(m) \left( \frac{r}{r_{\rm max}} \right)^{-\alpha(m)} \, . 
\end{equation}
 Here, $m$ ranges between $m_{\min} \leqslant m \leqslant m_{\max}$, $\mathcal{F}(m)$ is the normalized mass distribution, and $C_{\rm frac}$ is related to the number fraction ratio of the considered stellar population. These parameters together with $\alpha(m)$ are specified separately for the different stellar populations in Appendix \ref{sec:StellPopsProp}. In particular, we assume $m_{\rm BH,min} = 5 \Msun$ and $m_{\rm BH,max} = 50 \Msun$, and consider several BH mass functions including $\mathcal{F}(m_{\rm BH}) \propto m_{\rm BH}^{-\beta}$ with $\beta\in\{ 1,2,3\}$, and that obtained by population synthesis in \citet{Belczynskietal2016}. $n_{\rm inf}$ is the number density of stars at $r_{\rm max}$, which can be estimated by assuming that the enclosed mass of MSs within $r_{\rm max}$ is $2 \Msmbh$ (e.g. \citealt{Merritt2004,OLearyetal2009}). Specifically, if the radial density distribution of MSs follows the profile $n_{\rm MS} \propto r^{-3/2}$, $ n_{\inf}$ results in
\begin{equation} \label{eq:n_MS}
  n_{\inf} \simeq  1.38 \times 10^5 \mathrm{pc}^{-3} \sqrt{ \frac{ 10^6 \Msun }{ \Msmbh } } 
\end{equation}
 \citep{OLearyetal2009}. Note that these relations hold only in regions where the heavy BHs are outnumbered by low-mass BHs. In the innermost region, this may be violated by the above model, and the radial distribution of the heavy BHs will become shallower there, eventually approaching a Bahcall-Wolf cusp (e.g., \citealt{AlexanderHopman2009,Vasiliev2017}). Furthermore, mutual scattering of BHs into the loss cone may also flatten the BH number density profile at small radii (e.g., \citealt{AlexanderPfuhl2014}). We neglect these complications for simplicity.
 
 We calculate total number density distribution of a stellar population at radius $r$ as
\begin{equation}
  n(r) = \int n(r,m) \, dm \, .
\end{equation}
 The combined number density of all stellar population at radius $r$, $n_{\rm tot}(r,m)$, is computed as
\begin{equation}  \label{eq:ntot_rm}
  n_{\rm tot}(r,m) = n_{\rm MS}(r,m) + n_{\rm WD}(r,m) + n_{\rm NS}(r,m) + n_{\rm BH}(r,m) \, ,
\end{equation}
 the corresponding total number density is
\begin{equation}  \label{eq:ntot_r}
  n_{\rm tot}(r) = \int n_{\rm tot}(r,m) \, dm \, ,
\end{equation}
 and the second moment of the mass distribution at each $r$ is
\begin{equation}  \label{eq:secMomentM}
  \langle M^2 \rangle (r) = n_{\rm tot}^{-1}(r,m) \int n_{\rm tot}(r,m) \, m^2 \, dm \, .
\end{equation}

\subsection{Distribution of relative velocity} 
\label{subsec:GNs_VelDist} 
 
 The phase space distribution of mass $m$ objects in a spherically symmetric and relaxed stellar population around an SMBH can be given using the \citet{BahcallWolf1976} one-body phase space distribution function generalized for a multi-mass system by \citet{OLearyetal2009} as
 \begin{equation}  \label{eq:frv}
   f_m(\mathbf{r},\mathbf{v}) = C_{\rm N} E(r,v)^{p(m)} \, .
\end{equation}
 Here, $p(m)$ is a mass-dependent parameter due to mass segregation  \citep{OLearyetal2009} (Equation \ref{eq:pmbh}), and $E(r,v)$ is the Keplerian binding energy per unit mass of an object in the field of the SMBH,
\begin{equation}  \label{eq:E_bind}
  E(r,v) = \frac{ \Msmbh}{ r } - \frac{ v^2 }{ 2 } \, .
\end{equation}
 The adopted phase space distribution function is applicable outside the loss cone \citep{ShapiroLightman1976,SyerUlmer1999}, where objects are not removed by the SMBH, and interior to $r_{\rm max}$. That is, $ r_{\rm min} \leqslant r \leqslant r_{\rm max}$, where $r_{\rm min}$ is an inner radius at which $n(r,m)$ exhibits a cutoff. Furthermore, $v$ must satisfy \mbox{$0 \leqslant v \leqslant v_{\rm esc}$} over the radius range of interest, where $v_{\rm esc}$ denotes the local escape velocity at radius $r$,
\begin{equation}
  v_{\rm esc} = \sqrt{\frac{2 \, \Msmbh }{r}} \, .
\end{equation}
 The $ C_{\rm N}$ normalization in Equation \eqref{eq:frv} can be obtained by calculating the total number of mass $m$ objects in the GN as
\begin{equation}  \label{eq:Cnorm_f}
 \frac{dN}{dm} = \int_{ r_{\rm min} }^{ r_{\rm max} } dr 4 \pi r^2 \int_0 ^{ v_{\rm esc} } dv 4 \pi v^2 f_m(\mathbf{r},\mathbf{v}) 
\end{equation} 
 \citep{Gondanetal2018b}. 
 Equations \eqref{eq:n(r)}, \eqref{eq:frv}, and \eqref{eq:Cnorm_f} lead to the following relation between $\alpha(m)$ and $p(m)$:
\begin{equation}  \label{eq:alpha_p}
  \alpha(m) = \frac{3}{2} + p(m) \, .
\end{equation}
 We specify $\alpha$, $p(m)$, $r_{\rm min}$ in Appendix \ref{subsec:GNs_BHPops}.
 
 Using $f_m(\mathbf{r},\mathbf{v})$, the distribution of the magnitude of relative velocity $w$ between two encountering objects A and B around an SMBH at radius $r$ can be given for $0 \leqslant w \leqslant v_{\rm max}$ as
\begin{align}  \label{eq:P_ABw}
  P_{AB,r}(w) &= \frac{ 4 \pi^2 w \, C_{\rm N} }{\left( p_B + 1 \right) v_{\rm max} }
 \nonumber \\
 & \times \int_{w - v_{\rm max}}^{v_{\rm max} } \frac{dv \, v }{ v_{\rm max}^2 } \left( 1 - \frac{ v^2 }{ v_{\rm max}^2 } \right)^{p_A} \left[ 1 - \frac{ \left(v - w \right)^2 }{ v_{\rm max}^2 }\right]^{p_B + 1}
 \end{align}
 \citep{Gondanetal2018b}, where $p_A = p(m_A)$ and $p_B = p(m_B)$, $v_{\rm max} = 2 v_{\rm esc}$, and $P_{AB,r}(w)=0$ otherwise.

%% file: Sections/BBH_form_evol.tex
\section{Formation and Evolution of Gravitational-wave Capture Binary Black Holes} 
\label{sec:BinayFormEvol}
 
 The formation and evolution of GW capture BBHs are described in Sections \ref{subsec:Formation} and \ref{subsec:Evolution}, respectively, and we account for the interaction with a third object in Section \ref{subsec:BBHevaporation}.
 
 We use the following notations. We denote the binary component masses by $m_A$ and $m_B$, where $m_B \leqslant m_A$, the total mass of the binary by $M_{\rm tot} = m_A + m_B$, and the mass ratio by $q = m_B / m_A$. The reduced mass and symmetric mass ratio satisfy $\mu = \eta M_{\rm tot}$ and $\eta = q / (1 + q)^2$, respectively. The dimensionless pericenter distance is defined as $\rho_{\rm p} = r_{\rm p}/M_{\rm tot}$, where $r_{\rm p} = a ( 1- e)$ is the pericenter, $e$ is the orbital eccentricity, $a$ is the semi-major axis. A $0$ in the lower index denotes the initial value at formation, i.e., after the first close approach.

\subsection{Formation of binaries}
\label{subsec:Formation}
 
 We follow \citet{OLearyetal2009} and \citet{Gondanetal2018b} to determine initial orbital parameters of GW capture BBHs, but here we account for the 1PN order corrections as well.
 
 Encounters between two BHs in GNs are almost always nearly parabolic due to the relativistic nature of these events and the low-velocity dispersion in GNs compared to the speed of light \citep{QuinlanShapiro1987,LeeBradley1993}.\footnote{Note that the parabolic approximation for non-spinning encountering objects with a mass ratio $1/16 \lesssim q \leqslant 1$ may break down if the magnitude of the relative velocity is higher than $\sim 10 - 20 \%$ of the speed of light \citep{Baeetal2017}. However, since the SMBH removes objects efficiently from inside $r \lesssim 10^{3} \Msmbh$, the velocities are typically smaller than $v \sim (\Msmbh / r)^{1/2} \sim 0.04$ and so the assumption of parabolic encounters is justified. Our MC simulations confirm this expectation.} The distance of closest approach between encountering BHs can be given as
\begin{equation}  \label{eq:rp0}
  r_{\rm p0} = \left( \sqrt{ \frac{1}{ b^2 } + \frac{ M_{\rm tot}^2 }{ b^4 w^4 } } + \frac{ M_{\rm tot} }{ b^2 w^2 } \right)^{-1}
\end{equation}
 \citep{OLearyetal2009}, where $b$ is the impact parameter, and the relative velocity satisfies $0 \leqslant w \leqslant v_{\rm max}(r)$ at a certain radius $r$ (Section \ref{subsec:GNs_VelDist}). The energy and angular momentum loss due to GW emission during an encounter on a parabolic orbit (i.e., $e \sim 1$) can be given to 1PN order as
\begin{equation}  \label{eq:dEGW}
  \delta E_{\rm GW} = - \frac{ 85 \pi \eta^2 M_{\rm tot}^{9/2} }{ 12 \sqrt{2} r_{\rm p0}^{7/2} } \times \left[ 1 + \frac{ M_{\rm tot}}{ r_{\rm p0} } \frac{ 282387 - 157780 \eta }{ 47600 } \right] 
\end{equation}
 \citep{BlanchetSchaefer1989} and as
\begin{equation}  \label{eq:dLGW}
  \delta L_{\rm GW} =  - \frac{ 6 \pi \eta^2 M_{\rm tot}^4 }{ r_{\rm p0}^2 } \times \left[ 1 - \frac{ M_{\rm tot}}{ r_{\rm p0} } \frac{ 68355 - 68460 \eta }{ 30240 } \right]
\end{equation}
 \citep{JunkerSchaefer1992}, where the terms in the brackets $[\,]$ are PN correction factors relative to the leading-order expressions \citep{PetersMathews1963,Peters1964,Hansen1972,Turner1977}.\footnote{The correction factor adopted to $\delta E_{\rm GW}$ $\left( \delta L_{\rm GW} \right)$ yields deviation from the leading-order term by more than $5 \%$ for $r_{\rm p0} / M_{\rm tot} \lesssim 100$ ($55$) up to $\sim 70 \%$ ($\sim 30 \%$) at $r_{\rm p0} / M_{\rm tot} \simeq 8$, where most of the GW capture BBHs are expected to form in single GNs \citep{Gondanetal2018b}.} The final energy ($E_{\rm fin}$) and final angular momentum ($L_{\rm fin}$) of the system after the encounter can be given as
\begin{equation}  \label{eq:E_fin}
 E_{\rm fin} = \frac{ \mu w^2 }{2} +  \delta E_{\rm GW} \, ,
\end{equation}
\begin{equation}\label{eq:L_fin}
 L_{\rm fin} = \mu b w + \delta L_{\rm GW}  \, ,
\end{equation}
 and the BHs form a bound system if the final energy is negative after the encounter, i.e., $E_{\rm fin} < 0$. This criterion sets an upper limit on the impact parameter, which can be given in the corresponding 1PN order as
\begin{align}  \label{eq:bmax}
  b_{\rm max} & = \frac{ M_{\rm tot} }{ w^{9/7} } \left( \frac{340 \pi \eta}{ 3 }\right)^{1/7} 
  \nonumber \\
  & \times \left[ 1 + \left( \frac{ 5763 - 3220 \eta }{ 3400 } \right) 
    \left( \frac{ 3 }{ 340 \pi \eta }\right)^{2/7} w^{4/7} \right]
\end{align}
 \citep{JunkerSchaefer1992}, where the first row describes the leading-order result \citep{PetersMathews1963,Hansen1972}, and the second row accounts for the 1PN order correction factor. A lower limit on $b$ is set by the fact that we require the two BHs to avoid direct coalescence during the first close approach,
\begin{equation}  \label{eq:bmin}
 b_{\rm min} = \frac{4 M_{\rm tot}}{ w } 
\end{equation}
 \citep{Kocsisetal2006,OLearyetal2009}. Therefore, if $b$ satisfies the criterion\footnote{These bounds are in agreement within $10\%$ with that obtained by integrating the instantaneous 2.5PN and 3.5PN order equations of motion if $w \lesssim 0.01$ \citep{KocsisLevin2012}.}
\begin{equation}  \label{eq:Cond_b}
 b_{\rm min} < b < b_{\rm max} \, ,
\end{equation}
 an inspiraling binary forms with an initial semi-major axis of
\begin{equation}  \label{eq:a0}
 a_0 = \frac{\eta M_\mathrm{tot}^2 }{2 |E_{\rm fin}|} \times \left[ 1 + \frac{\eta - 7}{2} \frac{ |E_{\rm fin}| }{ \eta M_{\rm tot} } \right] \, ,
\end{equation} 
 and with an initial orbital eccentricity of
\begin{equation}  \label{eq:e0}
 e_0  = \sqrt{1 - \frac{ 2 |E_{\rm fin}| L_{\rm fin}^2 }{ \eta^3 M_{\rm tot}^2 } - \left[ \frac{ 2 (\eta - 6) |E_{\rm fin}| }{ \eta } - \frac{ 5 (\eta - 3) E_{\rm fin}^2 L_{\rm fin}^2 }{ \eta^4 M_{\rm tot}^2 } \right] } \, ,
\end{equation}
 where the terms in the brackets $[ \, ]$ also account for the 1PN order correction factors \citep{JunkerSchaefer1992}. The initial dimensionless pericenter distance $\rho_{\rm p0}$ can be expressed in terms of $a_0$, $M_{\rm tot}$, and $e_0$ as\footnote{Note that the adopted 1PN order corrections have a negligible ($\lesssim 1 \%$) impact on both $a_0$ and $e_0$, while their impact on $\rho_{\rm p0}$ is significant (up to $\sim 37 \%$) for a non-negligible fraction of GW capture BBHs in MC samples because \mbox{$\rho_{\rm p0} \propto (1-e_0)$} and $e_0 \sim 1$.}
\begin{equation}   \label{eq:rhop0}
  \rho_{\rm p0} = \frac{r_{\rm p0}}{M_{\rm tot}} = \frac{ a_0 ( 1 - e_0) }{ M_{\rm tot} } \, .
\end{equation}

\subsection{Evolution of binaries}
\label{subsec:Evolution}
 
 After a binary is formed with initial orbital parameters $e_0$ and $\rho_{p0}$, it evolves due to the GW radiation reaction. We adopt the orbit averaged evolution equations in 1PN order with radiation damping up to 3.5PN order derived in \citet{JunkerSchaefer1992} for the semi-major axis
\begin{align}  \label{eq:dadt}
  \frac{da}{dt} & = - \frac{64}{5} \frac{\eta M_{\rm tot}^3}{a^3 (1-e^2)^{7/2}} \left( 1 + \frac{73}{24} e^2 + \frac{37}{96} e^4 \right) 
  \nonumber \\ 
 & + \frac{ \eta \, M_{\rm tot}^4 }{ 210 } \frac{(14008 + 4704 \, \eta) + (80124 + 21560 \, \eta) \, e^2 }{ a^4 \, (1-e^2)^{9/2} } 
 \nonumber \\ 
 & + \frac{ \eta \, M_{\rm tot}^4 }{ 210 } \frac{ (17325 + 10458 \, \eta) \, e^4 - 0.5 \, (5501 - 1036 \, \eta) \, e^6 }{ a^4 \, (1-e^2)^{9/2} } 
\end{align} 
 and orbital eccentricity
\begin{align}  \label{eq:dedt}
   \frac{de}{dt} & = -\frac{304}{15} \frac{e \eta M_{\rm tot}^3}{a^4 (1-e^2)^{5/2}} \left( 1 + \frac{121}{304} e^2 \right) 
   \nonumber \\
   & + \frac{ \eta \, M_{\rm tot}^4 }{ 840 } \frac{ (133640 + 37408 \eta) \, e + (108984 + 33684 \eta) \, e^3 }{ a^5 \, (1-e^2)^{7/2} }
   \nonumber \\
   & - \frac{\eta \, M_{\rm tot}^4 }{ 840 } \frac{ (25211 - 3388 \eta) \, e^5 }{ a^5 \, (1-e^2)^{7/2} }
\end{align}
 to follow the evolution of orbital parameters. Here the first rows account for the evolution in leading order \citep{Peters1964}.
 
 Eccentric binaries emit GWs with a broad frequency spectrum, we therefore use the peak GW frequency defined by \citet{Wen2003} as
\begin{equation}   \label{eq:f_GW}
  f_{\rm GW} = \frac{ 1 }{\pi \rho_{\rm p}^{3/2} M_{\rm tot} (1 + e)^{0.3046} } 
\end{equation} 
 to determine the residual eccentricity $e_{\rm 10 Hz}$ with which GW capture BBHs enter the $10 \, \Hz$ frequency band of aLIGO/AdV/KAGRA \citep{Abbottetal2018} or when they form at a higher frequency. We assume that a binary is formed above $f_{\rm GW,10Hz} = 10 \, \Hz$ if \mbox{$f_{\rm GW,0} \geqslant 10 \, \Hz$}, otherwise it forms below the frequency range of interest. Specifically for GW capture BBHs, the condition \mbox{$f_{\rm GW,0} \geqslant 10 \, \Hz $} is equivalent to $\rho_{\rm p0} \leqslant \rho_{\rm p0, 10 \, Hz}$ with 
\begin{equation}   \label{eq:rhoaLIGO}
  \rho_{\rm p0,10 Hz} \equiv  \left( 2^{0.3046} \pi M_{\rm tot}  f_{\rm GW,10Hz} \right)^{-2/3} 
\end{equation}
 \citep{Gondanetal2018b}, due to the assumed high initial orbital eccentricity ($e_0 \sim 1$). Note that we account for the redshifted total mass $M_{{\rm tot},z} = M_{\rm tot} (1 + z)$ instead of $M_{\rm tot}$ in Equations \eqref{eq:f_GW} and \eqref{eq:rhoaLIGO} when considering binaries at redshift $z$.
 
 We define, $e_{\rm 10 Hz} \equiv e_0$ if $\rho_{\rm p0} < \rho_{\rm p0,10 Hz}$. For binaries with \mbox{$\rho_{\rm p0} \geqslant \rho_{\rm p0,10 Hz}$}, we define $e_{\rm 10 Hz}$ as the eccentricity at which \mbox{$f_{\rm GW}=10 \,\rm Hz$}. In this regime $f_{\rm GW}$ increases strictly monotonically in time due to the shrinking of the orbit \citep{Peters1964,Hinderetal2008}. Accordingly, we assume that a binary enters the $10 \, \Hz$ band if $f_{\rm GW}$ reaches $\rm 10 \, Hz$. To obtain $e_{\rm 10 \, Hz}$ corresponding to this point of binary evolution, we determine $\rho_{\rm p} (e)$ from $f_{\rm GW} \propto \rho_{\rm p}^{-3/2} (1 + e)^{-0.3046}$. To do so, we express $d\rho_{\rm p} / dt$ in terms of $da / dt$ and $de / dt$ as
\begin{equation}  \label{eq:drhopdt}
  \frac{d \rho_{\rm p} }{ dt } = \frac{1 - e}{ M_{\rm tot} } \frac{ da }{ dt } - \frac{ a }{ M_{\rm tot} } \frac{ de }{ dt } \, ,
\end{equation}  
 then change variable from $t$ to $e$ using the chain rule as
\begin{equation}  \label{eq:drhopde_3p5PN}
  \frac{d \rho_{\rm p} }{ de } = \frac{ d \rho_{\rm p} }{ dt } \left( \frac{ de }{ dt } \right)^{-1} \, .
\end{equation} 
 Finally, we numerically solve Equation \eqref{eq:drhopde_3p5PN} with the initial condition $\rho_{\rm p} (e_0) = \rho_{\rm p 0}$ utilizing a fourth-order Runge--Kutta--Fehlberg method with adaptive step size until \mbox{$f_{\rm GW} = 10 \, \Hz$}.\footnote{We adopt the analytical result of \citet{Mikoczietal2015} for the orbital frequency as a function of orbital eccentricity up to 1PN order to validate the developed numerical integrator.} Note that the adopted 1PN order equations of motion lead to somewhat higher $e_{\rm 10 Hz}$ than corresponding equations in leading order \citep{JunkerSchaefer1992}.
 
 We also determine the eccentricity $e_{\rm 10 M}$ with which binaries reach $\rho_{\rm p} = 10$ or when they form at a lower $\rho_{\rm p0}$ value. In the latter case for binaries with $\rho_{\rm p0} \leqslant 10$, we define \mbox{$e_{\rm 10 M} \equiv e_0$} , and for the former if $\rho_{\rm p0} > 10$ we evolve binaries from their initial orbit to $\rho_{\rm p} = 10$ using Equation \eqref{eq:drhopde_3p5PN}. Note that there is a one-to-one correspondence between $e_{\rm 10 M}$ and $\rho_{\rm p0}$ due to $e_0 \simeq 1$ and the weak dependence of $\rho_{\rm p}(e)$ on $M_{\rm tot}$, and a lower $\rho_{\rm p0}$ yields a higher $e_{\rm 10 M}$.

\subsection{Possible interaction with a third object}
\label{subsec:BBHevaporation}

 GNs are among the densest environments in the Universe, therefore a third object may interact with the newly formed binary during its eccentric inspiral phase, i.e., the binary undergoes binary--single interaction \citep{Samsingetal2014}. However, \citet{OLearyetal2009} and \citet{Gondanetal2018b} have shown that such events are very rare in high-velocity dispersion isotropic systems because GW capture BBHs at formation are so tight that they typically merge before encountering a third object.
 
 We conservatively discard all binaries in the MC sample if the typical timescale for a close encounter between the binary and a single object ($t_{\rm enc}$) is shorter than the binary's merger timescale ($t_{\rm GW,0} = \arrowvert a \, (dt / da) \arrowvert _{a_0, e_0} $), i.e.,
\begin{equation}  \label{eq:Cond_disrupt}
  t_{\rm GW,0} > t_{\rm enc} \, .
\end{equation} 
 Here, $t_{\rm GW,0}$ is given to 1PN beyond leading order as
\begin{equation}  \label{eq:t_GW_ecc_0}
 t_{\rm GW,0} = \frac{5 \, a_0^4 }{64  \mu  M_{\rm tot}^2} \frac{\left(1 - e_0^2 \right)^{7/2}}{\left(1 +  \frac{73}{24} e_0^2 + \frac{37}{96} e_0^4 \right)} 
 \left[ 1 + \frac{7 \, (6 \, q^2 + 11 \, q + 6) \, r_{\rm S}}{ 8 \, a_0  (1 - e_0) \left( q + 1 \right) ^2} \right] 
\end{equation} 
 \citep{Zwicketal2019}, where $r_{\rm S} = 2 M_{\rm tot}$ is the two-body Schwarzschild radius. Finally, we follow \citet{OLearyetal2009} and adopt their estimate on $t_{\rm enc}$ for soft binaries,
\begin{equation}  \label{eq:t_enc}
  t_{\rm enc} \approx \frac{ 1 }{ 12 \pi w a_0^2 n_{\rm tot} (r) } \, .
\end{equation}
 
 The remaining binaries in the MC sample undergo isolated dynamical evolution driven by GW radiation reaction.

%% file: Sections/DistribParams_SingleGN.tex
\section{Distributions of Binary Parameters for a Single Galactic Nucleus} 
\label{sec:GWcaptureBBHsSingleGN}
 
 We introduce the components of developed MC framework in Section \ref{subsec:MergRates}, describe the framework itself in Section \ref{subsec:Setup_MC}, and summarize the conclusions of MC experiments carried out for single GN hosts in Section \ref{subsec:ResMC_SingleGN}.

\subsection{Merger rate distributions}
\label{subsec:MergRates}

 We generate MC samples of GW capture events in single GNs as follows. We consider non-spinning binaries, for which the GW capture process can be parameterized with four parameters in each radial shell of radius $r$ around the SMBH: the component masses $m_A$ and $m_B$, the impact parameter $b$, and the magnitude of the relative velocity $w$. The differential rate of encounters can be given in the fourteen-dimensional phase space for a single GN as
\begin{equation}  \label{eq:diff_formrate}
 d^{14} \Gamma_{\rm 1GN}  = \sigma w f_A(\mathbf{r}_A, \mathbf{v}_A) f_B(\mathbf{r}_B, \mathbf{v}_B) d^3 r_A d^3 r_B  d^3 v_A d^3 v_B  d m_A  d m_B
\end{equation}
 \citep{OLearyetal2009}, where $\sigma = \pi b_{\rm max}^2 - \pi b_{\rm min}^2$ is the cross section for two BHs to form a GW capture BBH in a close encounter. The differential cross section for fixed $r$, $w$, and impact parameter between $b$ and $b + db$ is \mbox{$d \sigma = 2 \pi b \, db$}. To express the differential rate in terms of the parameter set $\{ m_A, m_B, w, b, r \}$, i.e., to obtain $\partial^{5} \Gamma_{\rm 1GN} / \partial r \partial b \partial w \partial m_A \partial m_B $ analytically, one can perform the following calculations: (i) differentiate $d^{14} \Gamma_{\rm 1GN}$ with respect to $\sigma$ and substitute \mbox{$d \sigma = 2 \pi b \, db$}, (ii) assume short-range encounters (i.e., $ \mathbf{r}_A \approx \mathbf{r}_B $ and $\left\arrowvert \mathbf{r}_A \right\arrowvert \approx \left\arrowvert \mathbf{r}_B \right\arrowvert = r$), (iii) expand $d^3 r = 4 \pi r^2 dr$, and (iv) change from variables $\mathbf{v}_A$ and $\mathbf{v}_B$ to $w$; details of calculations can be found in Appendices B.3 and B.4 in \citet{Gondanetal2018b}. The obtained five-dimensional merger rate distribution accounts for encountering BHs.
 
 We use this together with the conditions introduced in Equations \eqref{eq:P_ABw}, \eqref{eq:Cond_r}, \eqref{eq:Cond_b}, and \eqref{eq:Cond_disrupt} to generate MC samples of GW capture BBHs in single GNs similar to \citet{Gondanetal2018b} as follows. First we marginalize $\partial^{5} \Gamma_{\rm 1GN} / \partial r \partial b \partial w \partial m_A \partial m_B $ over the subspaces $(w,b)$ and $(w,b,r)$, respectively, to obtain  $\partial^{3} \Gamma_{\rm 1GN} / \partial m_A \partial m_B \partial r$ and  $\partial^{2} \Gamma_{\rm 1GN} / \partial m_A \partial m_B$ (see Appendix \ref{sec:Rates_1PN}). After correcting for the effects of dynamical friction (Equation \ref{eq:eq:rate_mAmB_DF}), we generate $(m_A, m_B)$ BH mass pairs by randomly sampling the corresponding distribution function (Equation \ref{eq:PDF_mAmB} below). Second, for each $(m_A, m_B)$ mass pair, we assign a radial shell within $r \in [r_{\rm min}^{A,B}, r_{\rm max}]$ using the radius dependent merger rate distribution for fixed component masses, \mbox{$d \Gamma_{{\rm 1GN},AB} / dr = \left( \partial^{3} \Gamma_{\rm 1GN} / \partial m_A \partial m_B \partial r \right) \arrowvert _{m_A, m_B}$}. Third, for each radial shell, we restrict the $w - b$ plane to the subspace in which inspiraling binaries can form to eliminate encounters leading to direct plunge or escape to infinity using Equation \eqref{eq:Cond_b}. Then, we sample pairs of $w$ and $b$ using the merger rate distribution in the $w - b$ plane for fixed component masses and radius defined as $\partial^2 \Gamma_{{\rm 1GN},AB,r} / \partial w \partial b = \left( \partial^{5} \Gamma_{\rm 1GN} / \partial r \partial b \partial w \partial m_A \partial m_B \right) \arrowvert _{m_A, m_B, r}$, using Equation \eqref{eq:rate_wb} below. Finally, we discard the binaries possibly undergoing binary--single interactions.
 
 The distribution functions used in this analysis are summarized next.
 \citet{Gondanetal2018b} showed that $\partial^2 \Gamma_{{\rm 1GN},AB,r} / \partial w \partial b $ for a fixed radius and component masses is
\begin{equation}  \label{eq:rate_wb}
 \frac{\partial^2 \Gamma_{\rm 1GN,AB,r} }{ \partial w \partial b} \propto b w P_{AB,r}(w) \, ,
\end{equation} 
 hence the two-dimensional PDF in the $w - b$ plane is
\begin{equation}  \label{eq:PDF_wb}
 P_{AB,r}(w,b) = C_{\rm N} b w  P_{AB,r}(w) 
\end{equation}
 if $0 \leqslant w \leqslant v_{\rm max}$ and $b_{\rm min} \leqslant b \leqslant b_{\rm max}$, and zero otherwise. Furthermore, we use the radial distribution at fixed component masses $P_{AB}(r)$ (Equation \ref{eq:PDF_r}) to sample radial distances.
 
 For a BH population with a certain BH mass distribution $\mathcal{F}_{\rm BH}(m)$, the component-mass-dependent merger rate of GW capture BBH events in a single GN $\partial^{2} \Gamma_{\rm 1GN} / \partial m_A \partial m_B $ is given in Equation \eqref{eq:rate_mAmB}. In GNs, BHs can migrate within $r_{\rm max}$ from outside of the GN due to DF \citep{Chandrasekhar1943a,Chandrasekhar1943b,Chandrasekhar1943c}, which increases the total number of BHs $N_{\rm BH}$ in the GN leading to the enhancement of merger rates. The impact of DF on the GW capture BBH merger rate can be taken into account by adding multiplying factors to $\partial^{2} \Gamma_{\rm 1GN} / \partial m_A \partial m_B $ as
\begin{equation}  \label{eq:eq:rate_mAmB_DF}
 \frac{\partial^2\Gamma_{\rm 1GN,DF}}{\partial m_A \partial m_B} = \zeta_A \zeta_B \frac{\partial^2\Gamma_{\rm 1GN}}{\partial m_A \partial m_B} 
\end{equation}
 where the enhancement factor $\zeta \equiv \zeta (m_{\rm BH}, \Msmbh, t_{\rm DF})$ denotes the relative increase in the number of mass $m_{\rm BH}$ BHs in the BH population around an SMBH of mass $\Msmbh$ within the DF timescale $t_{\rm DF}$ \citep[see][for details]{RasskazovKocsis2019}. Here $\zeta$ is determined in two limiting cases with respect to the formation rate of BHs in the centers of galaxies, when BHs formed $12 \, \Gyr$ ago, or they formed with a constant rate during the last $12 \, \Gyr$. We refer to the former and later model as "instantaneous" and "continuous", respectively. Accordingly, we set $t_{\rm DF} = 12 \, \Gyr$ for galaxies at $z = 0$, i.e. for galaxies in the local Universe, and we calculate $t_{\rm DF}$ for galaxies at $0 < z \lesssim 0.99$ as discussed in Section \ref{subsec:NumbGenBBHsperGN}. The age of the Universe at redshift $0$ and $1$ are $\sim 13.8 \, \Gyr$ and $\sim 5.9 \, \Gyr$ (Section \ref{subsec:NumbGenBBHsperGN}), respectively, implying that $4.1 \, \Gyr \lesssim t_{\rm DF} \leqslant 12 \, \Gyr$. Sampling $(m_A, m_B)$ pairs from the normalized $\partial^2\Gamma_{\rm 1GN,DF} / \partial m_A \partial m_B$ for both models, we find almost the same mass distributions over the possible values of $\{ t_{\rm DF}, p_0, \mathcal{F}_{\rm BH}, \Msmbh \}$. This can be explained by the fact that the fractional increase in the number of BHs relative to one another is similar in both cases; see Figure 2 in \citet{RasskazovKocsis2019} for an example. Therefore, without loss of generality we simply adopt the "continuous" model when computing $\zeta$. Note that we find the same trend for $t_{\rm DF}$ down to $t_{\rm DF} = 0$ under the assumption of relaxed GNs. Hereafter, we draw $(m_A, m_B)$ mass pairs in MC experiments from the PDF
\begin{equation}  \label{eq:PDF_mAmB}
  P(m_A, m_B) = C_{\rm N} \frac{\partial^2 \Gamma_{\rm 1GN,DF}}{\partial m_A \partial m_B} \, .
\end{equation}
 Note that $P(m_A, m_B)$ does not depend on $N_{\rm BH}$ explicitly because of the normalization of $\partial^2\Gamma_{\rm 1GN,DF} / \partial m_A \partial m_B$. Moreover, $\zeta$ systematically increases with $m_{\rm BH}$ thereby facilitating the merger of more massive binaries. As a consequence, $P(M_{\rm tot})$ and $P(\mathcal{M})$ shift toward higher masses, while $P(q)$ and thereby $P(\eta)$ are shifted toward equal-mass systems. However, the degree of these shifts is mild because $\zeta$ is only $\sim 20 - 60 \%$ higher for $50 \, \Msun$ BHs than for $5 \, \Msun$ BHs after $t_{\rm DF} = 12 \, \Gyr$, where higher ratios correspond to more massive SMBHs.
 
 The quantities that have redshift dependence in the distributions $\{ P_{AB,r}(w,b), P_{AB}(r), P(m_A, m_B) \}$ due to DF or the relaxation process are $\{p_0, \mathcal{F}_{\rm BH}, r_{\rm min}^{A,B}, \zeta \}$.\footnote{All three PDFs depend on $N_{\rm BH}$ only through $r_{\rm min}^{A,B}$.} As this effect is negligible for $\{p_0, \mathcal{F}_{\rm BH}, r_{\rm min}^{A,B} \}$ (Appendices \ref{subsec:GNs_BHPops} and \ref{sec:ImpactDF_GWcBBHs}), the redshift evolution enters only in $P(m_A, m_B)$ through $\zeta$. Furthermore, under the assumption of a relaxed GN, $P(m_A, m_B)$ is not sensitive to the BH formation rate, and similarly for all mass-dependent parameters. Similarly, the impact of BH formation rate on the distributions of initial orbital parameters is negligible because they can be obtained by sampling the normalized $\partial^{5} \Gamma_{\rm 1GN} / \partial r \partial b \partial w \partial m_A \partial m_B $ (see Appendix B in \citealt{Gondanetal2018b} for details), where only the $m_A - m_B$ subspace is affected by the BH formation rate through $P(m_A, m_B)$.

\subsection{Monte Carlo sampling for single galactic nuclei}
\label{subsec:Setup_MC}

 In this section, we describe how we obtain the MC distributions of initial orbital parameters, component masses, and residual eccentricity for BBHs forming through the GW capture process in a single GN in the redshift range $0 \leqslant z \lesssim 1$.
 
 The MC framework is based on the methodology of sampling $\partial^{5} \Gamma_{\rm 1GN} / \partial r \partial b \partial w \partial m_A \partial m_B $ over the five-dimensional subspace in which GW capture BBHs may form as introduced in Section \ref{subsec:MergRates}.
 
 The MC routine implements the following steps.
 \begin{enumerate}
 
 \item First, we choose the free parameters related to the SMBH $\{\Msmbh,z\}$ and stellar populations $\{ p_0, \mathcal{F}_{\rm BH}, \alpha_{\rm MS}, \alpha_{\rm WD}, \alpha_{\rm NS} \}$ from their appropriate domain; see Sections \ref{subsec:SMBH_MassRange} and \ref{subsec:DLdistribGNs} and Appendix \ref{sec:StellPopsProp}.
 
 \item Next, we generate $N_{AB} = 10^3$ random component mass pairs $(m_A, m_B)$ using $P(m_A, m_B)$ (Equation \ref{eq:PDF_mAmB}). Here, $P(m_A, m_B)$ is calculated at a certain $z$ (Section \ref{subsec:MergRates}).
 
 \item For each $(m_A, m_B)$ pair, we randomly draw \mbox{$N_r = 250$} radius values between $ r_{\rm min}^{AB}$ and $r_{\rm max}$ from $P_{AB}(r)$ (Equation \ref{eq:PDF_r}). 
 
 \item Lastly, for each radius value we pick $N_{w,b} = 50$ pairs of $(w, b)$ from $P_{AB,r}(w,b)$ (Equation \ref{eq:PDF_wb}).
 
 \item We compute $a_0$ (Equation \ref{eq:a0}), $e_0$ (Equation \ref{eq:e0}), and $\rho_{\rm p0}$ (Equation \ref{eq:rhop0}) for each binary.
 
 \item We discard the binaries possibly undergoing binary--single interactions in the MC sample using the condition in Equation \eqref{eq:Cond_disrupt}. The remaining binaries merge after isolated binary evolution driven by GW radiation reaction. We randomly select $N_{\rm BBH} = 10^5$ binaries from the sample of remaining binaries.
 
 \item We record $e_{\rm 10 Hz} \equiv e_0$ for each binary in the sample that forms with $f_{\rm GW,0} \geqslant 10 \, \Hz$, i.e., forms with \mbox{$\rho_{\rm p0} \leqslant \rho_{\rm p0,10 Hz}$} (Equation \ref{eq:rhoaLIGO}). We evolve the binaries with $\rho_{\rm p0} > \rho_{\rm p0,10 Hz}$ from their initial orbit up until $f_{\rm GW} = 10 \, \Hz $ using Equation \eqref{eq:drhopde_3p5PN}, and record the resulting eccentricities, $e_{\rm  10 Hz}$.
 
 \item Similarly, we record $e_{\rm 10 M} \equiv e_0$ for binaries with $\rho_{\rm p0} \leqslant 10$, and evolve those with $\rho_{\rm p0} > 10$ until $\rho_{\rm p} = 10$ using Equation \eqref{eq:drhopde_3p5PN} and then record their eccentricity $e_{\rm 10 M}$.
 
 \item The output parameters for the generated binaries are $\{m_A, m_B, e_0, \rho_{\rm p0}, e_{\rm 10 Hz}, e_{\rm 10 M} \}$.
\end{enumerate}
 
 We verify the convergence of the resulting distributions as a function of sample sizes $\{ N_{AB}, N_r, N_{w,b}, N_{\rm BBH} \}$ by
 evaluating the Kolmogorov--Smirnov test with respect to the final distributions  for various choices of the parameter set $\{ \Msmbh, z, p_0, \mathcal{F}_{\rm BH}, \alpha_{\rm MS}, \alpha_{\rm WD}, \alpha_{\rm NS} \}$. We find that distributions converge for $\lesssim 3 \times 10^4$ samples.

\begin{figure*}
    \centering
    \includegraphics[width=75mm]{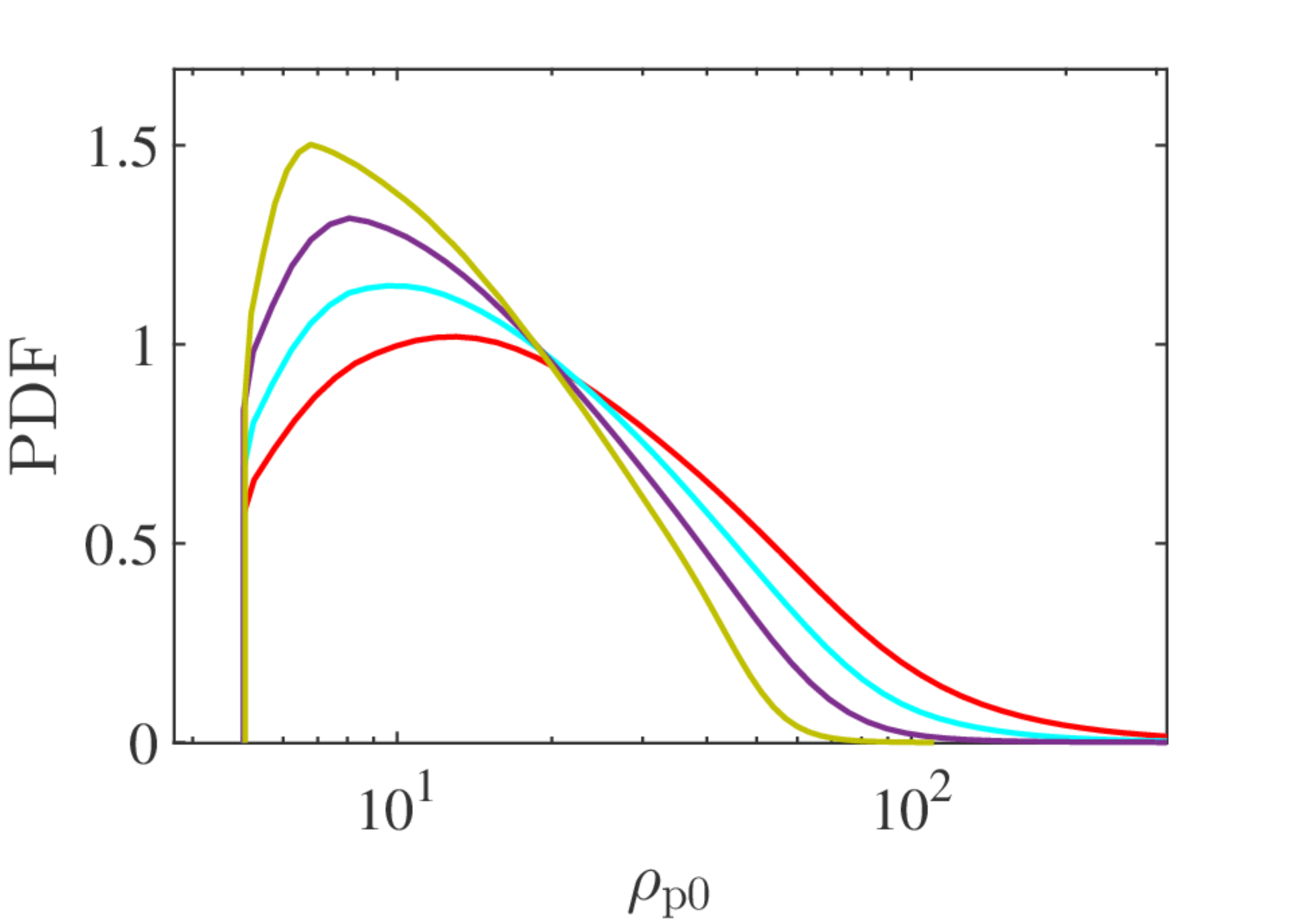}
    \includegraphics[width=75mm]{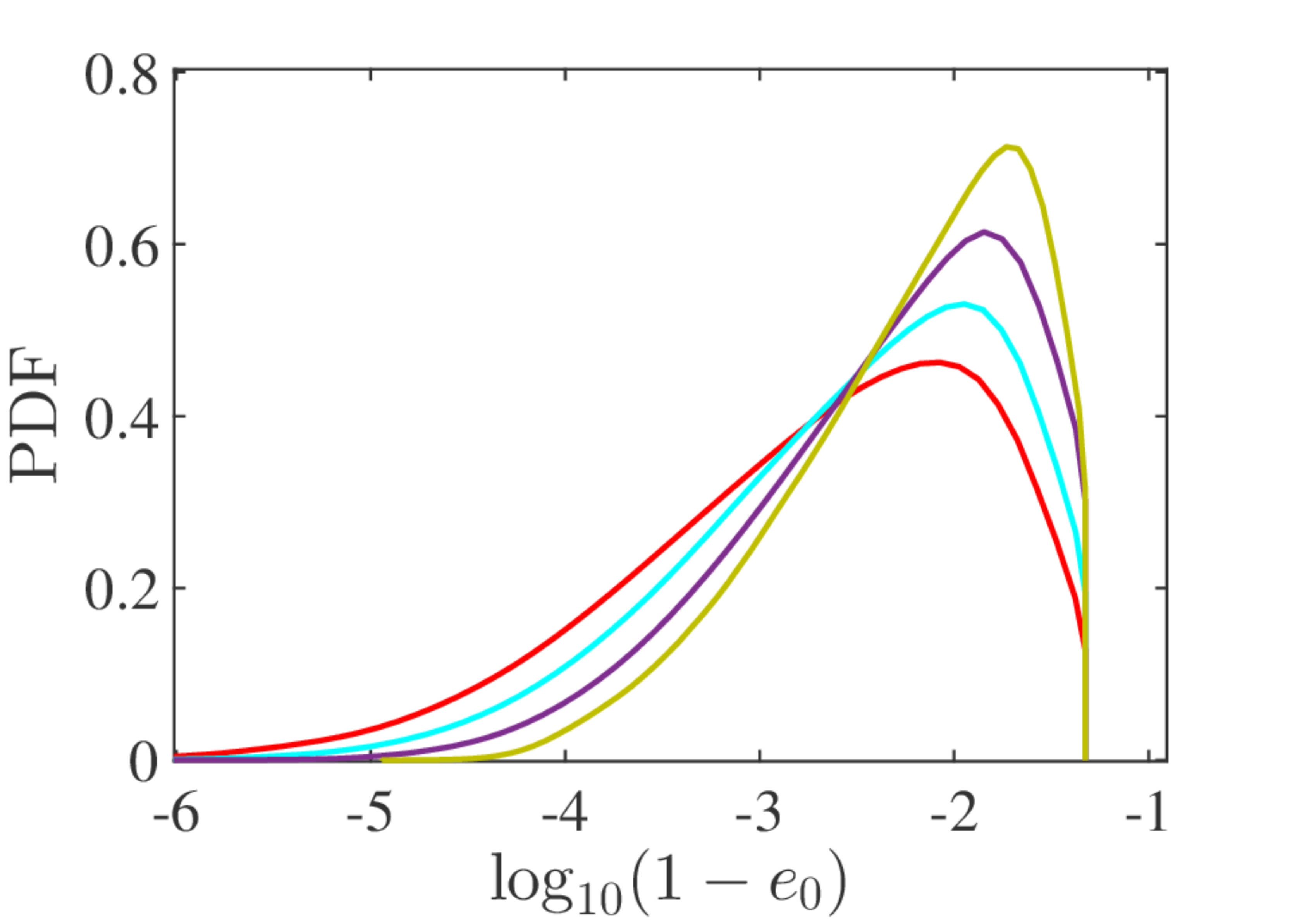}
    \\
    \includegraphics[width=75mm]{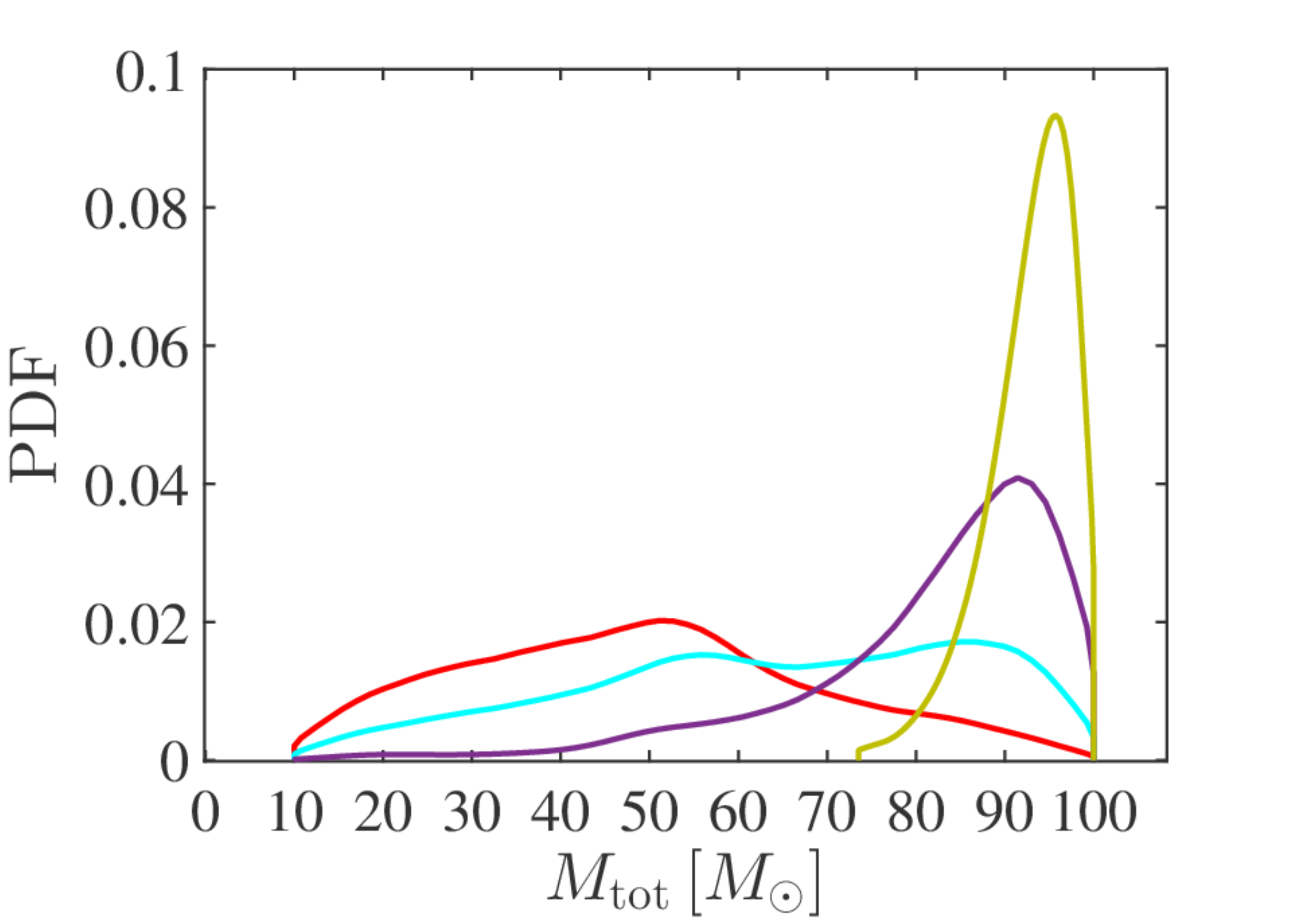}
    \includegraphics[width=75mm]{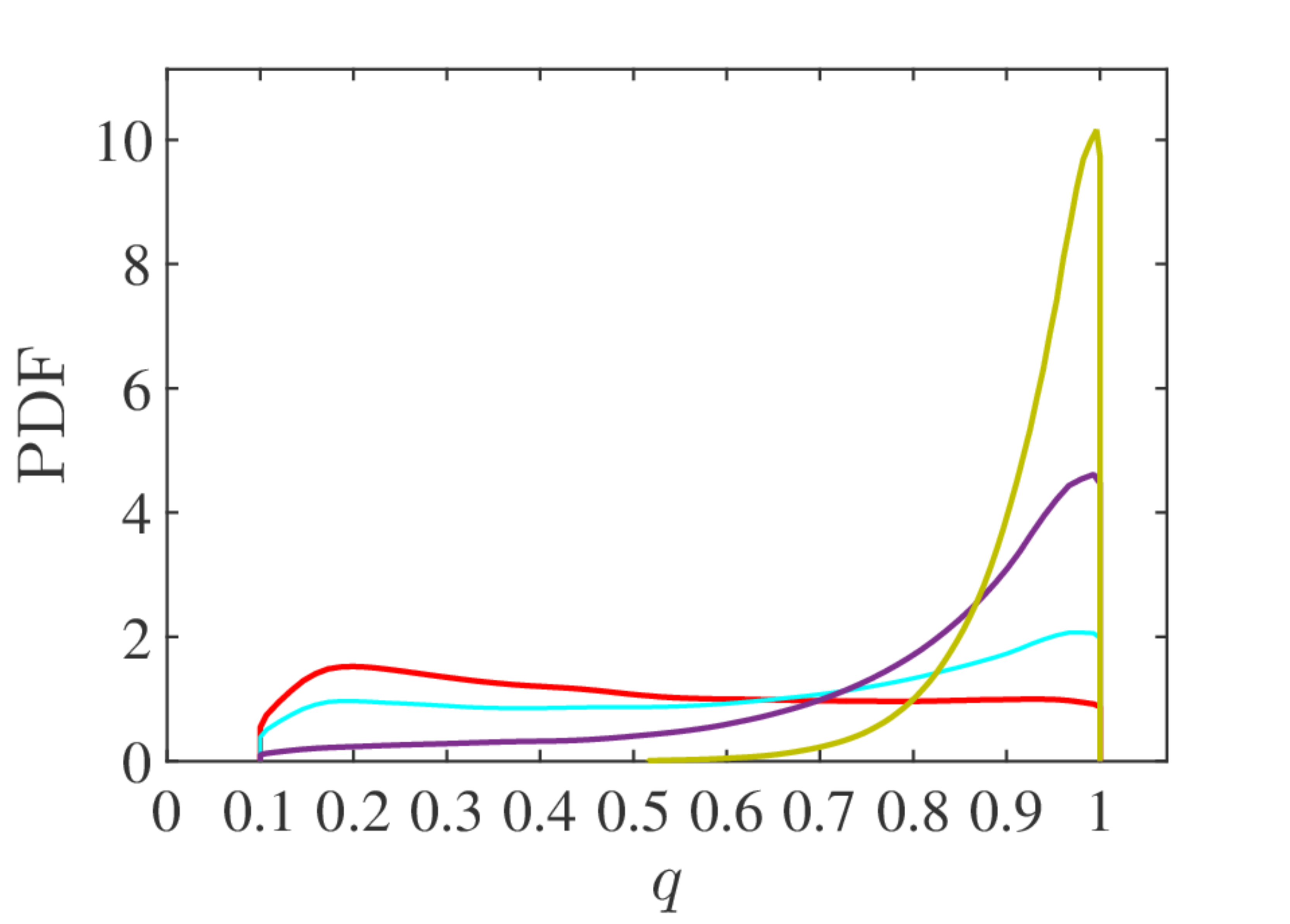}
    \\
    \includegraphics[width=75mm]{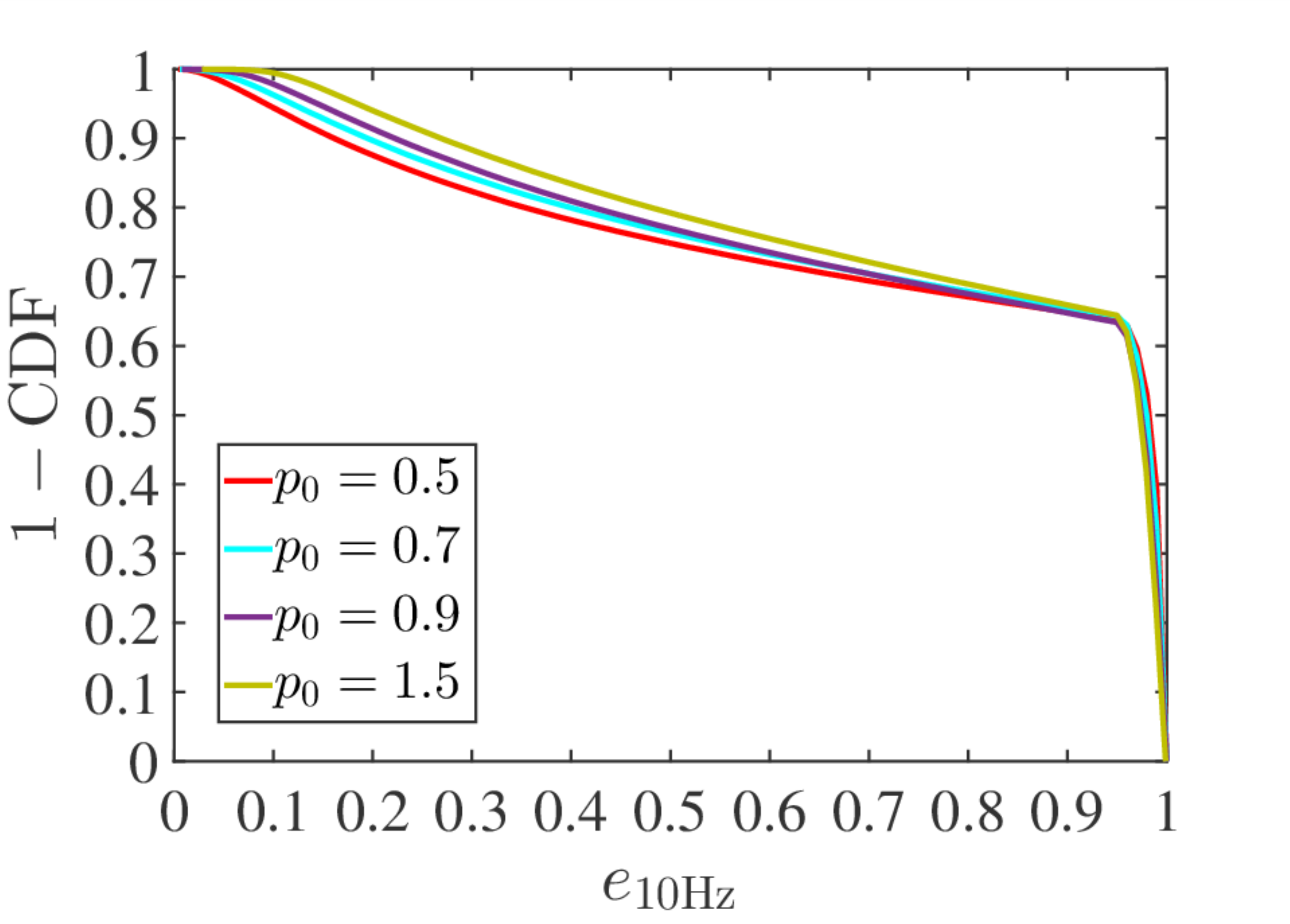}
    \includegraphics[width=75mm]{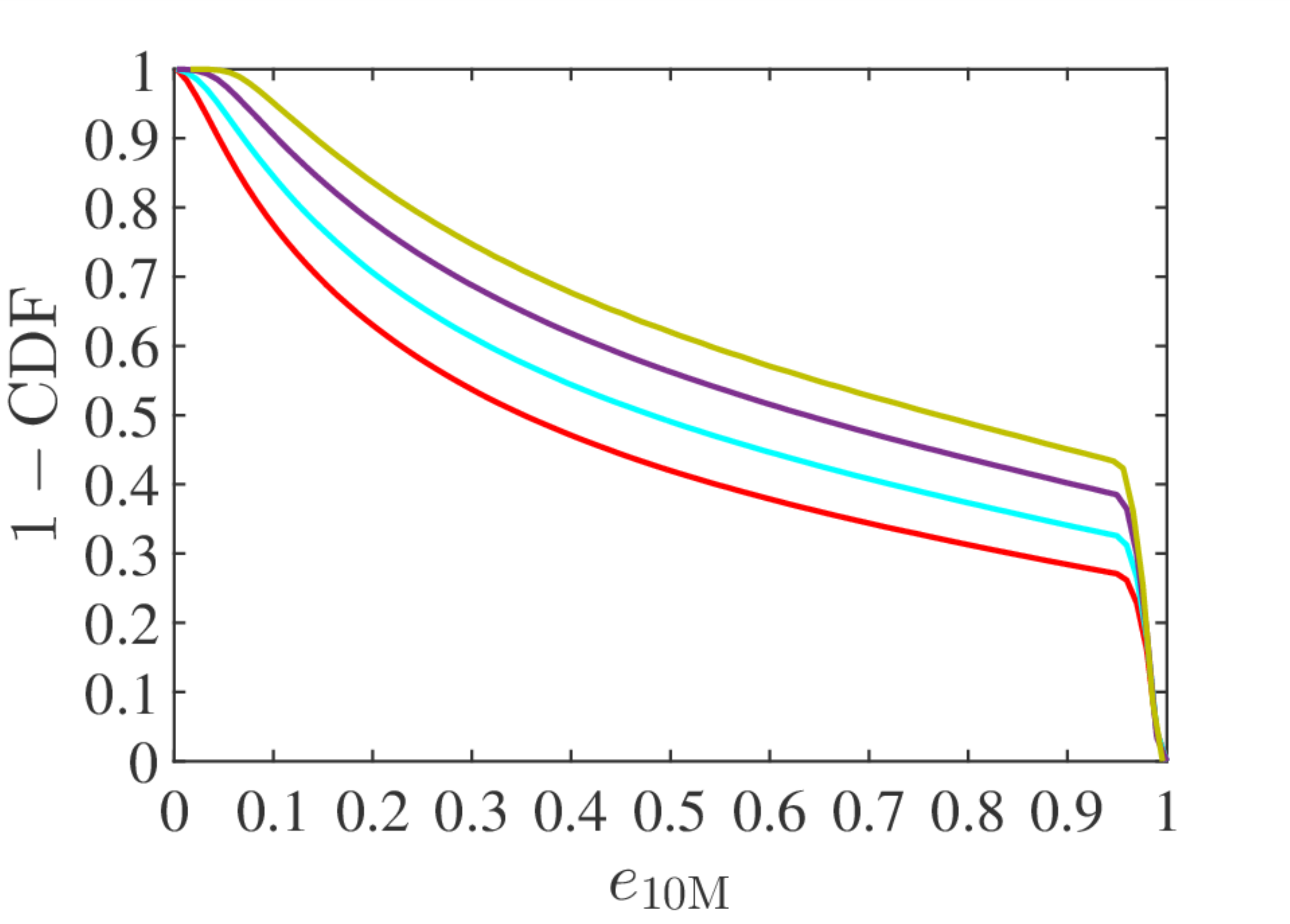}
  \caption{Impact of the $p_0$ mass--segregation parameter (Equation \ref{eq:pmbh}) on the probability density function (PDF) of initial dimensionless pericenter distance $\rho_{\rm p0}$ (row 1, left), initial orbital eccentricity $e_0$ (row 1, right), total mass $M_{\rm tot}$ (row 2, left), mass ratio $q$ (row 2, right), 1 minus the cumulative distribution function (CDF) of residual eccentricity $e_{\rm 10 Hz}$ (row 3, left), and $1 - {\rm CDF}$ of the residual eccentricity $e_{\rm 10 M}$ (row 3, right) for GW capture BBHs merging in a single Milky Way-size nucleus with a SMBH mass of $\Msmbh \approx 4.28 \times 10^6 \, \Msun$ \citep{Gillessenetal2017} in its center in the local Universe. The initial orbital eccentricity is typically very close to unity. $e_{\rm 10 Hz}$ is given when the binaries' emitted GW signals peak at the $10 \, \Hz$ frequency band of aLIGO/AdV/KAGRA or when they form at a higher frequency (binaries with $e_{\rm 10 Hz} \gtrsim 0.95$). Similarly, $e_{\rm 10 M}$ is the eccentricity with which binaries reach $\rho_{\rm p} = 10$ or when they form at a lower $\rho_{\rm p0}$ value (binaries with $e_{\rm 10 M } \gtrsim 0.95$). Systems that formed below $10 \, \Hz$ or with $\rho_{\rm p0} > 10$ were evolved from their initial orbit using the evolution equations of \citet{JunkerSchaefer1992}. Results are shown for binaries forming in a multi--mass BH population with a power-law mass distribution with exponent $\beta = 2$ and with BH masses between $5 \, \Msun$ and $50 \, \Msun$. We model the impact of dynamical friction on BH populations as introduced in \citet{RasskazovKocsis2019}.  \label{fig:ParamDist_GN_p0Dep} } 
\end{figure*}

\subsection{Monte Carlo results for single galactic nuclei}
\label{subsec:ResMC_SingleGN}
 
 Here we present the MC distributions of GW capture binary parameters in single GNs for different choices of the free parameters corresponding to the SMBH $\{ \Msmbh, z \}$ (Sections \ref{subsec:SMBH_MassRange} and \ref{subsec:DLdistribGNs}), and the BH and stellar populations $\{ p_0, \mathcal{F}_{\rm BH}, \alpha_{\rm MS}, \alpha_{\rm WD}, \alpha_{\rm NS} \}$ (Appendix \ref{sec:StellPopsProp}).\footnote{Similar investigations were carried out previously in \citet{Gondanetal2018b}, but only for GW capture BBH populations with fixed component masses and for the parameter space $\{ p_0, \beta, \Msmbh, m_{\rm BH, max} \}$, where $p_0$ was reduced to the range $0.5 - 0.6$ and the impact of DF on the BH population was neglected.}

\begin{figure*}
    \centering
    \includegraphics[width=75mm]{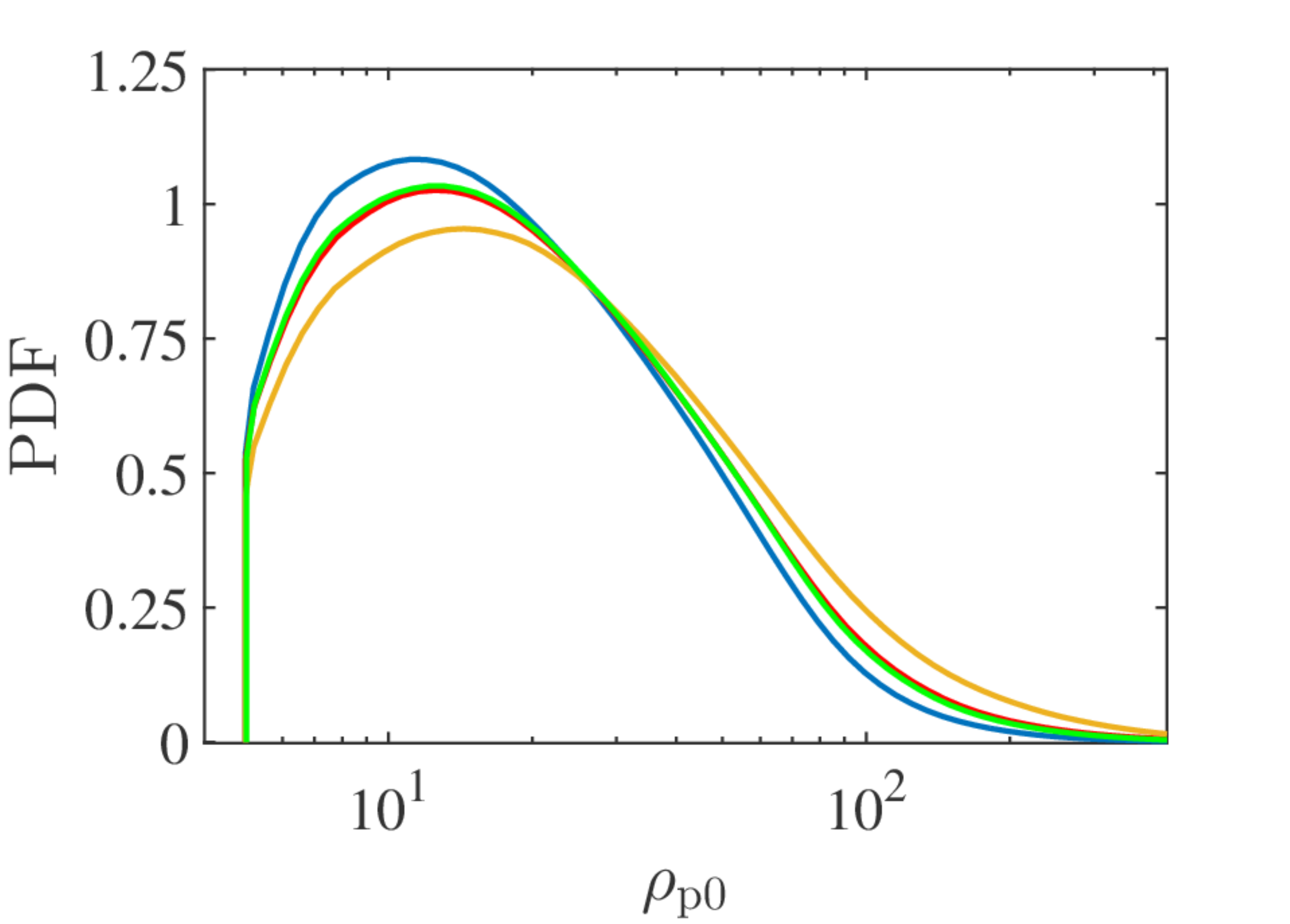}
    \includegraphics[width=75mm]{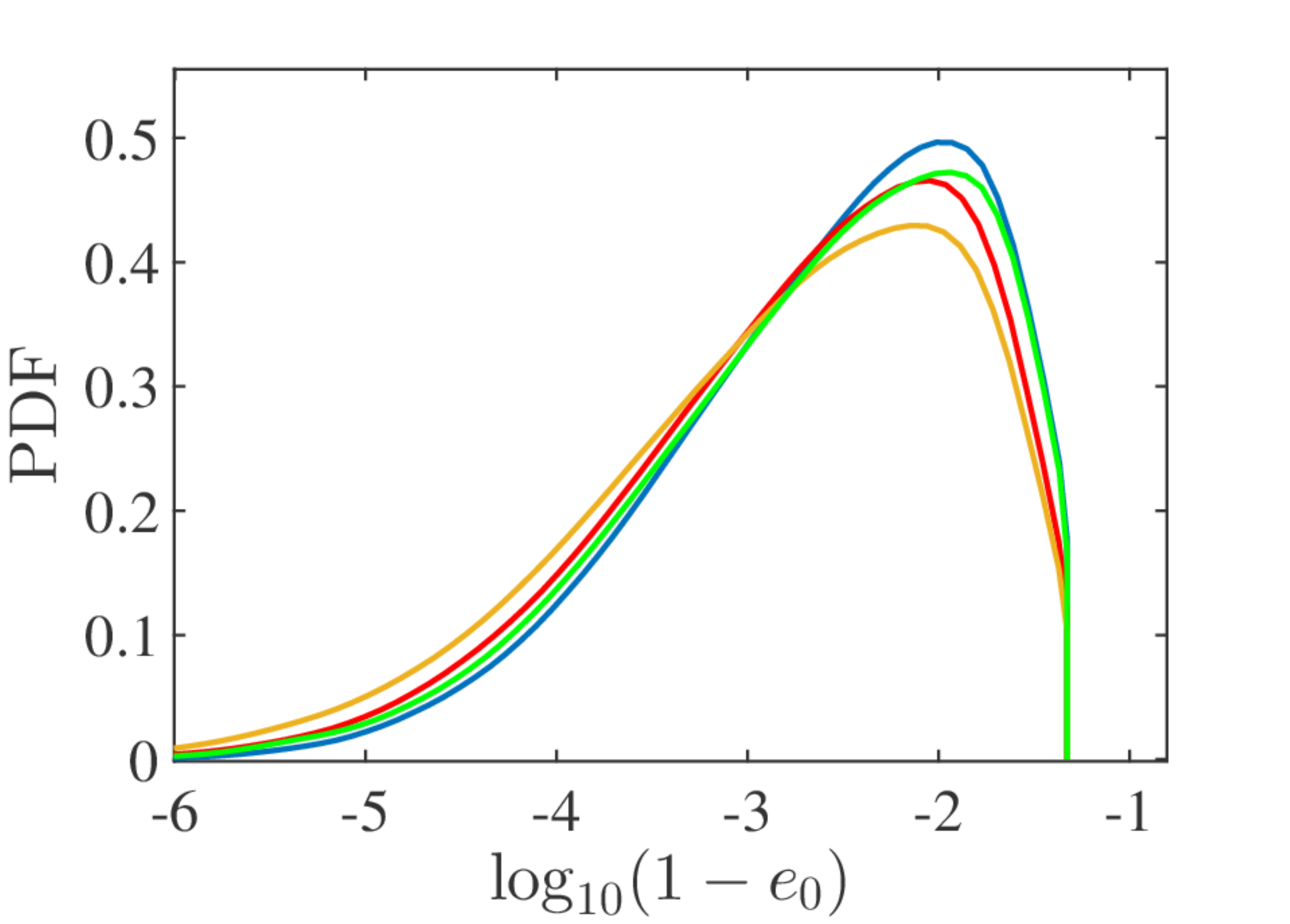}
    \\
    \includegraphics[width=75mm]{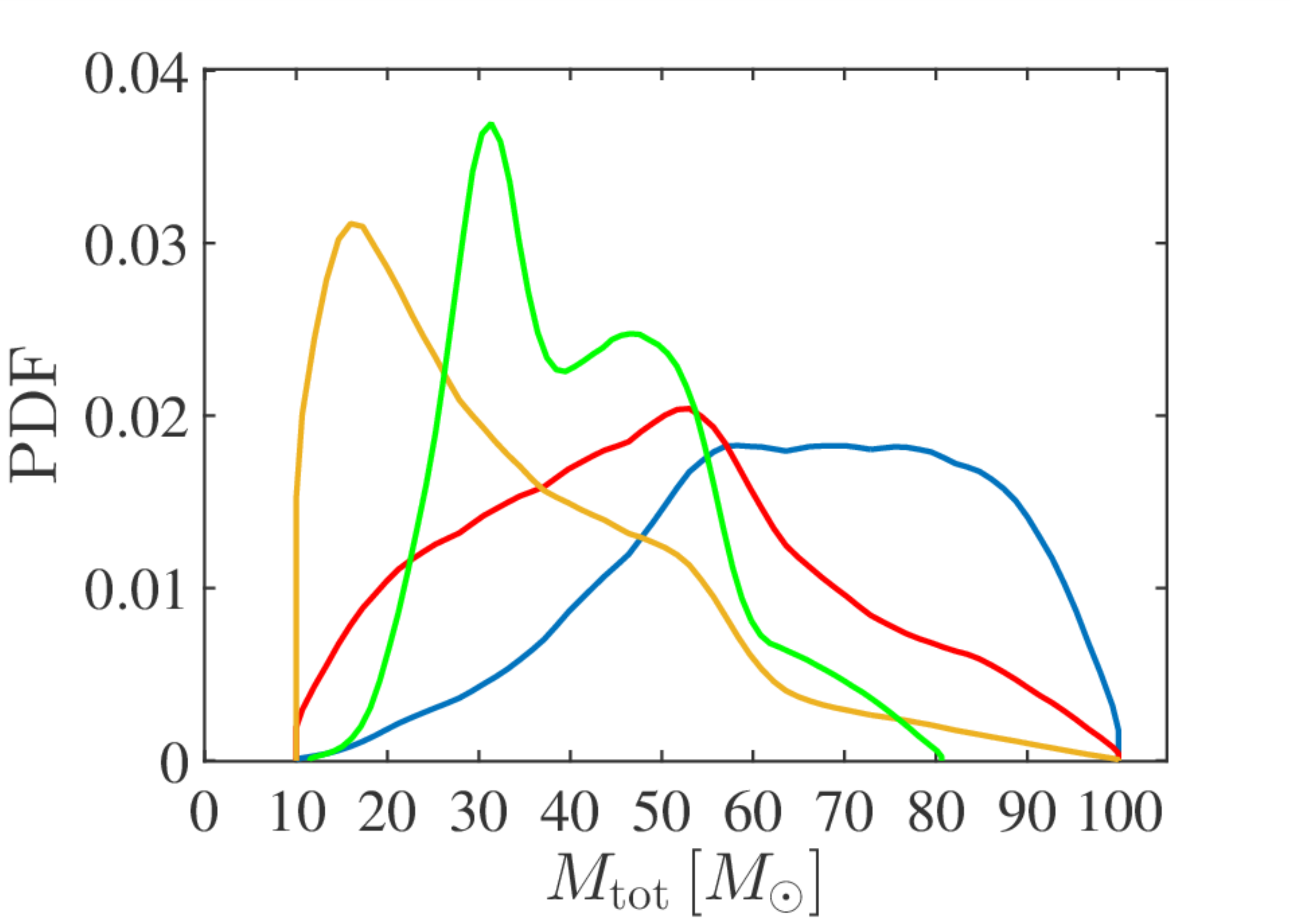}
    \includegraphics[width=75mm]{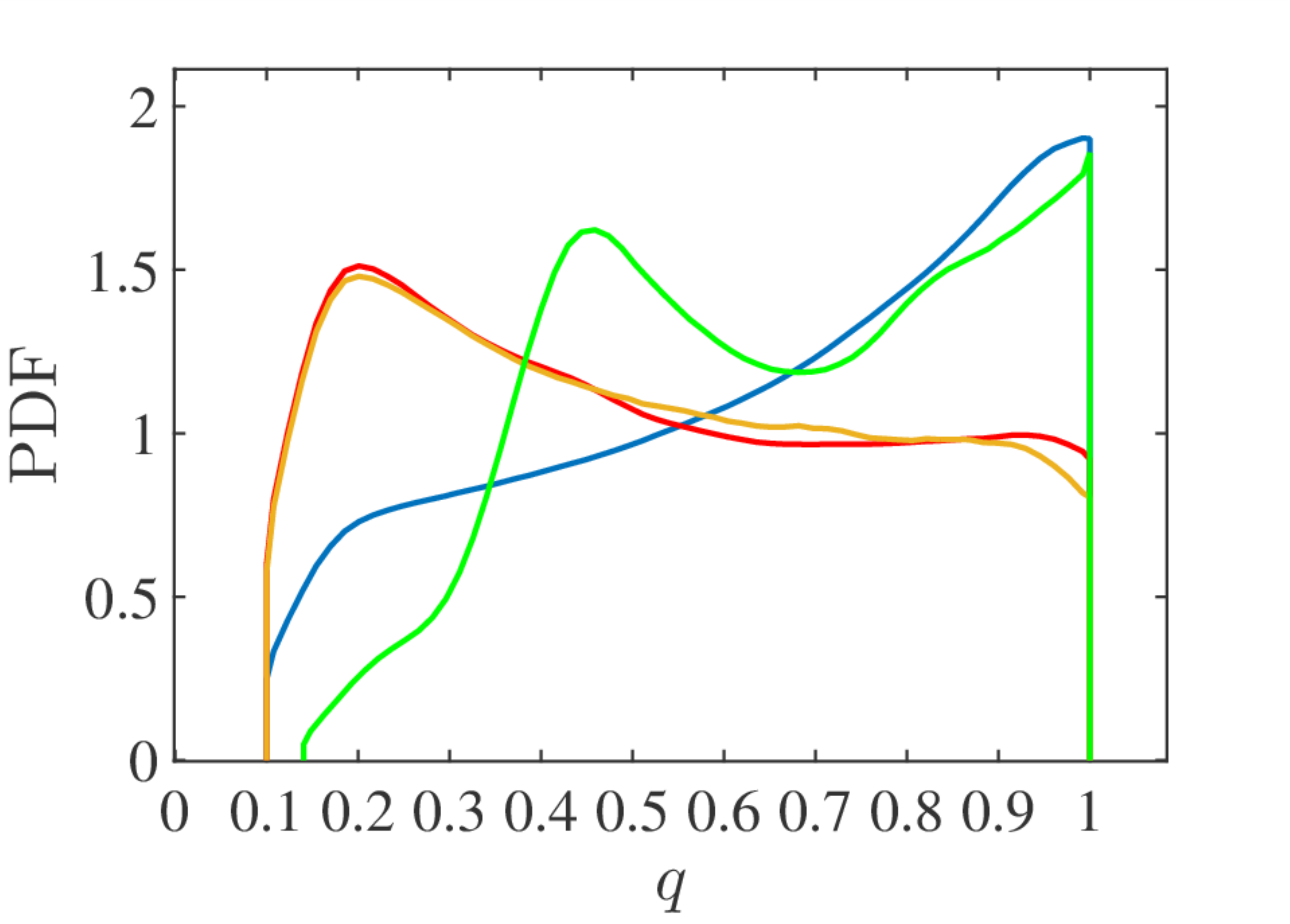}
    \\
    \includegraphics[width=75mm]{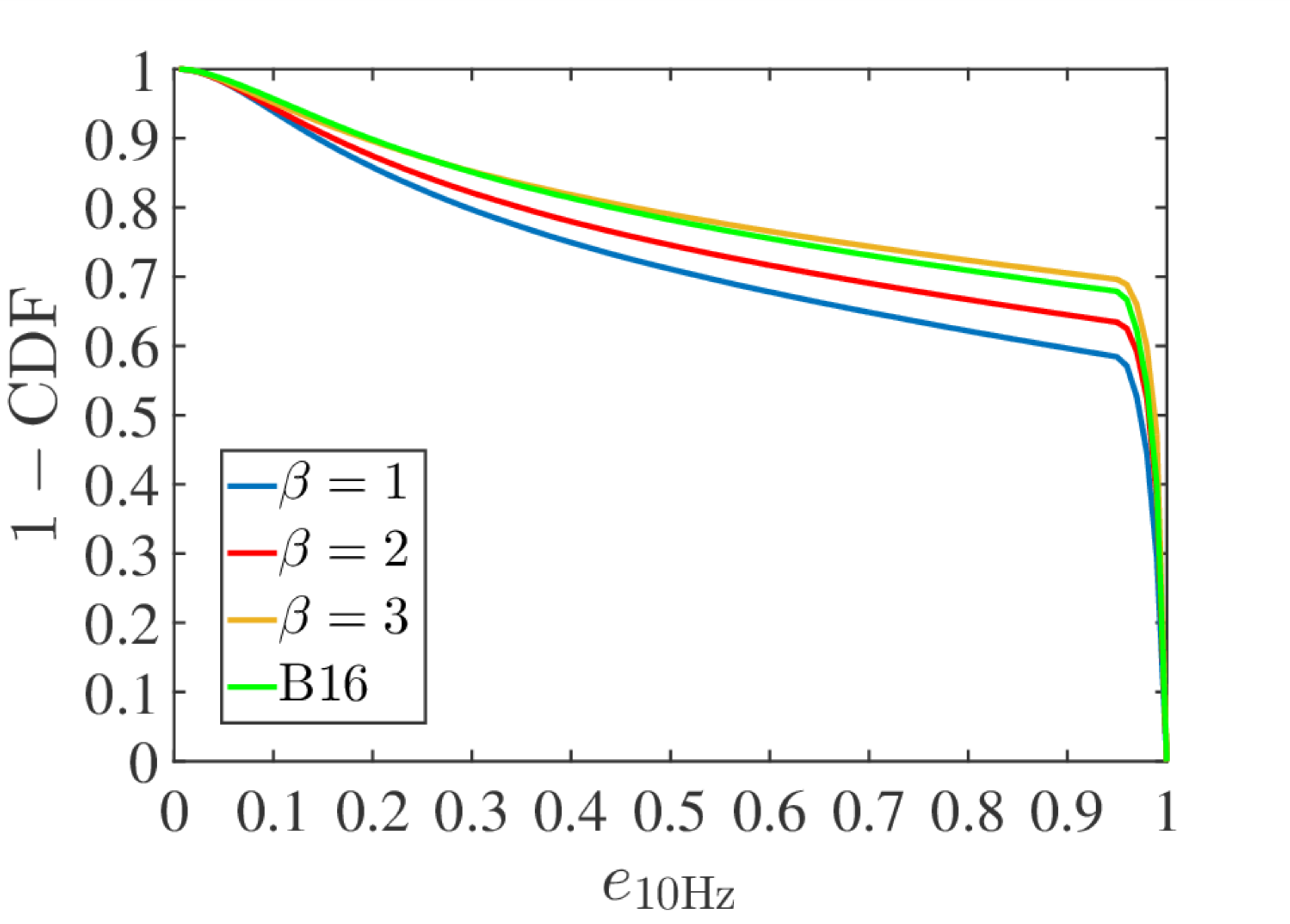}
    \includegraphics[width=75mm]{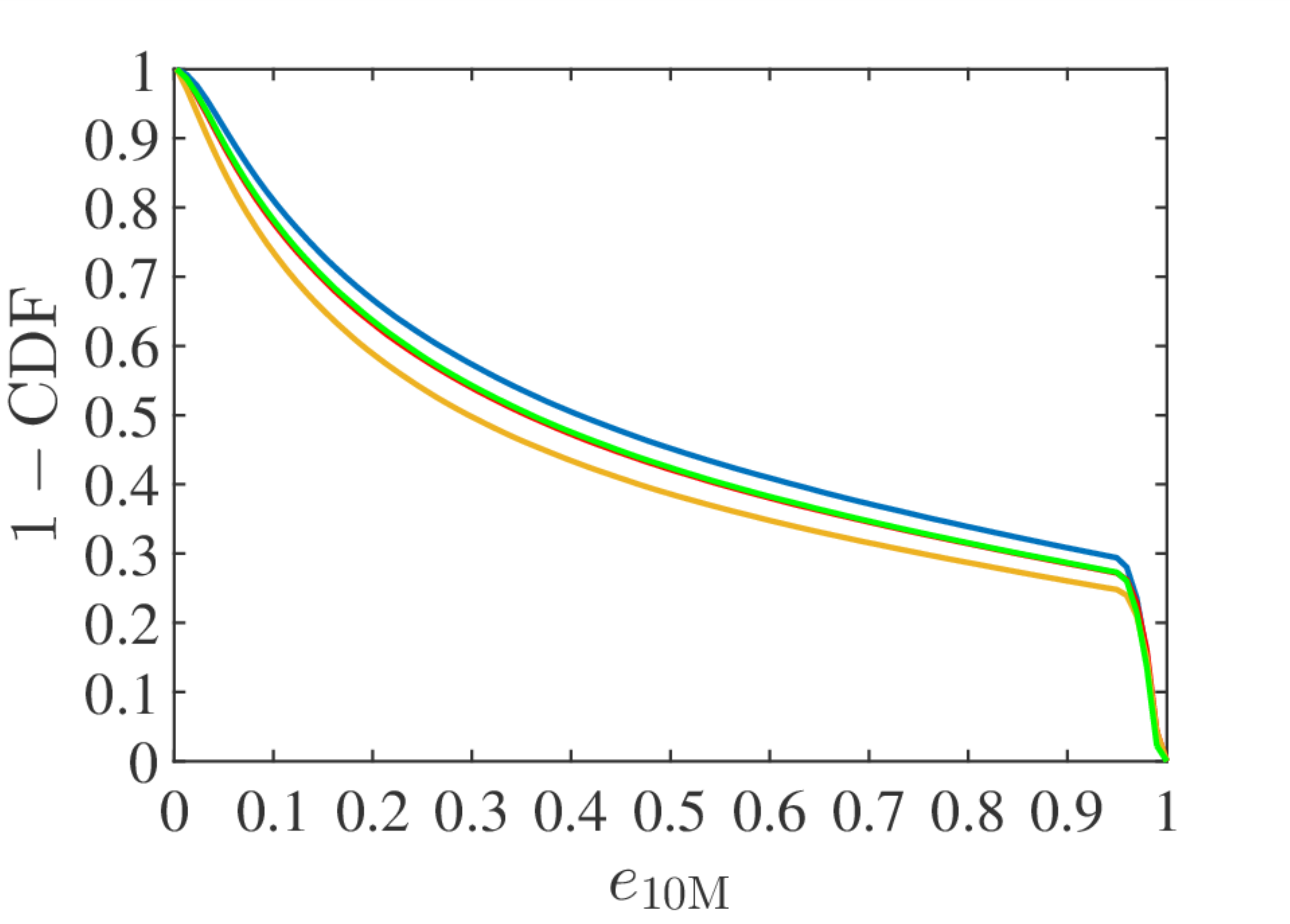}
  \caption{Same as Figure \ref{fig:ParamDist_GN_p0Dep}, but for various choices of the BH mass functions $m_{\rm BH}^{-\beta}$ between $5 \, \Msun \leqslant m_{\rm BH} \leqslant 50 \, \Msun$ as labeled in the legend and for the \citet{Belczynskietal2016} mass distribution (labeled as B16) at the fiducial mass--segregation parameter $p_0 = 0.5$.  \label{fig:ParamDist_GN_betaDep} } 
\end{figure*}

 In our numerical experiments, GW capture BBHs typically form in GNs with \mbox{$0.95 \lesssim e_0 \lesssim 0.9999$} and with $5 \lesssim \rho_{\rm p0} \lesssim 100$, where $P(\rho_{\rm p0})$ drops off quickly beyond $\simeq 10 - 20$ depending on $p_0$ and $\mathcal{F}_{\rm BH}$. Here the lower cutoffs in $P(\rho_{\rm p0})$ and $P(e_0)$ correspond to the conditions $b_{\rm min} < b$ and $E_{\rm fin} < 0$, respectively \citep{Gondanetal2018b}. Furthermore, our results show that more than $92 \%$ of the binaries form in or enter the aLIGO/AdV/KAGRA band with eccentricities larger than $0.1$, while $\sim 59 - 74 \%$ \mbox{($ \sim 55 - 72 \%$)} of them have \mbox{$e_{\rm 10 Hz} > 0.8$} \mbox{($e_{\rm 10 Hz} > 0.9$)} depending on $p_0 \in [0.5, 0.9]$ and the adopted $\mathcal{F}_{\rm BH}$ models. Specifically for an extreme mass-segregated cusp model referring to that of the young stars in the Galactic Center ($\{ p_0, \beta \} = \{ 1.5, 2 \}$; Appendix \ref{subsec:GNs_BHPops}), the corresponding range is $67 - 71\%$ ($63 - 66 \%$) and more than $97 \%$ of sources have $e_{\rm 10 Hz} > 0.1$. The quick drop off in the \mbox{$1-{\rm CDF}$} beyond $0.95 \simeq e_{\rm 10 Hz}$ corresponds to those binaries that formed above the $10 \, \Hz$ frequency band. Similarly, the majority ($\sim 71 - 94 \% $) of binaries reach $\rho_{\rm p} = 10$ with eccentricities above $0.1$ and $\sim 24 - 46 \%$ ($\sim 21 - 42 \%$) of them have $e_{\rm 10 M} > 0.8$ ($e_{\rm 10 M} > 0.9$), and systems with $e_{\rm 10 M} \gtrsim 0.95$ in the $1-{\rm CDF}$ formed with $\rho_{\rm p0} < 10$. For $\{p_0, \beta \} = \{1.5, 2 \}$, $91 - 95 \%$ of sources have $e_{\rm 10 M} > 0.1$ and $46 - 50 \%$ ($42 - 46\%$) of them have $e_{\rm 10 M} > 0.8$ ($e_{\rm 10 M} > 0.9$). Note that these findings are in a good agreement with that obtained in \citet{Gondanetal2018b} from results of MC simulations specified for binaries with fixed component masses. Moreover, the typical $M_{\rm tot}$ and $q$ significantly depend on both $p_0$ and $\mathcal{F}_{\rm BH}$. For instance, $P(q)$ is roughly uniform and the typical $M_{\rm tot}$ is roughly $\sim m_{\rm BH, max}$ if $p_0 \simeq 0.5 - 0.7$. However, binaries form with a strong preference for equal--mass systems with the heaviest components in the BH population if $p_0 \gtrsim 0.8$. Examples for these findings are displayed in Figures \ref{fig:ParamDist_GN_p0Dep} and \ref{fig:ParamDist_GN_betaDep}.
 
 We find that the distributions of mass-dependent parameters of a single GN are significantly affected by the BH mass function $\mathcal{F}_{\rm BH}$ and the mass--segregation parameter $p_0$. Furthermore, the distributions of initial dimensionless pericenter distance $P(\rho_{\rm p0})$ is highly sensitive to $p_0$ but mostly insensitive to $\mathcal{F}_{\rm BH}$. The initial eccentricity $e_0$ is well above $0.9$ in all cases. Examples are shown in Figures \ref{fig:ParamDist_GN_p0Dep} and \ref{fig:ParamDist_GN_betaDep}.
 
 Figure \ref{fig:ParamDist_GN_p0Dep} also shows that a higher $p_0$, i.e., a more mass segregated BH number density profile, leads to the merger of more massive binaries with systematically lower $\rho_{\rm p0}$ and with a preference for equal-mass systems. Indeed, more mass segregated clusters have an increased chance of mergers closer to the SMBH, where the velocity dispersion is high, but here GW capture is only possible during very close approaches where $\rho_{\rm p0}$ is low. A more top-heavy $\mathcal{F}_{\rm BH}$ has a similar effect as displayed in Figure \ref{fig:ParamDist_GN_betaDep}, since heavier objects are closer to the SMBH due to mass segregation.
 
 Nevertheless, the $e_{\rm 10 Hz}$ distribution $P(e_{\rm 10 Hz})$ in a single GN is weakly affected by $p_0$, e.g. $\sim 67 \%$ ($\sim 94 \%$) and $\sim 68 \%$ ($\sim 97 \%$) of sources have \mbox{$e_{\rm 10 Hz} > 0.8$} (\mbox{$e_{\rm 10 Hz} > 0.1$}), respectively, for standard ($p_0 = 0.5$) and strong ($p_0 = 0.9$) mass segregation, respectively, for $\{ \Msmbh, \beta \} = \{ 4.28 \times 10^6 \, \Msun, 2 \}$ (Figure \ref{fig:ParamDist_GN_p0Dep}). This trend holds for extreme high $p_0$ values, e.g. the corresponding source fraction is $\sim 69 \%$ ($\sim 98 \%$) for \mbox{$\{p_0, \beta \} = \{1.5, 2 \}$}. Indeed, the way $e$ and $\rho_{\rm p}$ decrease due to GW emission, $e(\rho_{\rm p})$, weakly depends on masses at 1PN order \citep{JunkerSchaefer1992}\footnote{$e(\rho_{\rm p})$ is mass independent in leading order \citep{Peters1964}.}, but $f_{\rm GW}\propto M_{\rm tot}^{-1}$ for fixed $\rho_{\rm p}$ and $e$. Hence, binaries with higher $M_{\rm tot}$ reach $10 \, \rm Hz$ with less evolved orbital parameters, i.e. higher $e$ and $\rho_{\rm p}$ if they form with $f_{\rm GW,0} < 10 \, \Hz$. The mergers in more mass-segregated clusters have higher $M_{\rm tot}$ and so somewhat higher $e_{\rm 10 Hz}$. $P(e_{\rm 10 Hz})$ is also not very sensitive to the BH mass function, e.g. \mbox{$e_{\rm 10 Hz} > 0.8$} (\mbox{$e_{\rm 10 Hz} > 0.1$}) has a probability of $\sim 62 \%$ ($\sim 93 \%$) and $\sim 72 \%$ \mbox{($\sim 96 \%$)} for very top - and bottom - heavy distributions, respectively, for $\Msmbh = 4.28 \times 10^6 \, \Msun$ and the fiducial $p_0 = 0.5$ value (Figure \ref{fig:ParamDist_GN_betaDep}).
 
 The properties of $P(\rho_{\rm p0})$ determines that of $P(e_{\rm 10 M})$ for different assumptions on mass segregation and on the BH mass function due to the one-to-one correspondence between $e_{\rm 10 M}$ and $\rho_{\rm p0}$ (Section \ref{subsec:Evolution}). In particular, a higher $p_0$ leads to the merger of binaries with systematically higher $e_{\rm 10 M}$, while $P(e_{\rm 10 M})$ is less sensitive to $\mathcal{F}_{\rm BH}$, as seen in Figures \ref{fig:ParamDist_GN_p0Dep} and \ref{fig:ParamDist_GN_betaDep}. We find that the distribution of $e_{\rm 10 M}$ in a single GN is sensitive to $p_0$, e.g. $\sim 77 \%$ ($\sim 31 \%$) and $\sim 90 \%$ ($\sim 43 \%$) of sources possess $e_{\rm 10 M} > 0.1$ \mbox{($e_{\rm 10 M} > 0.8$)} for standard and strong mass segregation, respectively, for $\Msmbh = 4.28 \times 10^6 \, \Msun$ and $\beta = 2$ (Figure \ref{fig:ParamDist_GN_p0Dep}). Note that $P(e_{\rm 10 M})$ weakly depends on $p_0$ even for extremely high values, e.g. $\sim 93 \%$ ($\sim 49 \%$) of sources have $e_{\rm 10 M} > 0.1$ ($e_{\rm 10 M} > 0.8$) for \mbox{$\{p_0, \beta \} = \{1.5, 2 \}$}. Nevertheless, $P(e_{\rm 10 M})$ is not affected significantly by the BH mass distribution, e.g. $e_{\rm 10 M} > 0.1$ ($e_{\rm 10 M} > 0.8$) has a probability of $\sim 73 \%$ ($\sim 29 \%$) and $\sim 81 \%$ ($ \sim 34 \%$) for very top - and bottom - heavy distributions, respectively, when $\{ \Msmbh, p_0 \} = \{ 4.28 \times 10^6 \, \Msun, 0.5 \}$ (Figure \ref{fig:ParamDist_GN_betaDep}).
 
 In order to investigate how the different radial-number density exponents of low mass objects, $\{ \alpha_{\rm MS}, \alpha_{\rm WD}, \alpha_{\rm NS} \}$, influence the distributions of binary parameters, we run MC simulations in the two limiting cases, when $\{ \alpha_{\rm MS}, \alpha_{\rm WD}, \alpha_{\rm NS} \} = 3/2$ and $ 7/4$. We find a negligible difference. The fraction of sources with binary--single interactions is always below $\lesssim 4 \%$.
 
 In simulations with different SMBH mass, we find that binaries form with somewhat lower $\rho_{\rm p0}$ around more massive SMBHs, which is consistent with the findings of \citet{Gondanetal2018b}. This originates from the fact that GNs with a more massive SMBH has a higher velocity dispersion according to the $M-\sigma$ relation (Equation \ref{eq:Msigma}), which leads to the formation of binaries with lower $\rho_{\rm p0}$ as discussed in the previous paragraph. Accordingly, $P(e_{\rm 10 M})$ shifts toward higher eccentricities by a small amount. The component-mass-dependent parameters' distributions do not depend significantly on $\Msmbh$ (Appendix B.7 in \citealt{Gondanetal2018b}), and $P(e_{\rm 10 Hz})$ increases by only a small amount toward higher eccentricities for larger $\Msmbh$ as binaries with lower $\rho_{\rm p0}$ enter the aLIGO/AdV/KAGRA band with higher $e_{\rm 10 Hz}$ (Equation \ref{eq:f_GW}). In summary, we conclude the assumptions on $\Msmbh$ do not influence significantly the distribution of source parameters in single GNs.
 
 The GW capture process in single GNs at lower $z$ produces heavier binary populations with lower $\rho_{\rm p0}$ and $e_0$ due to DF.\footnote{A certain redshift value $z$ is equivalent to a timescale $t_{\rm DF}$, where lower $z$ corresponds to a longer $t_{\rm DF}$ (Section \ref{subsec:SetupMC}). The number of BHs in a GN, predominantly the heavier ones, grows with $t_{\rm DF}$ \citep{RasskazovKocsis2019}, thereby shifting the merger rate toward higher masses, which will result in heavier binaries with lower $\rho_{\rm p0}$ as discussed above.} Note that this effect is not significant because $\zeta$ is proven to be only mildly steeper for smaller $z$ and thereby it does not enhance $P(m_A, m_B)$ for higher masses significantly. In consequence, the impact of $z$ on $P(e_{\rm 10 M})$ is weak. Comparing the impact of DF with that of Doppler--shift on $P(e_{\rm 10 Hz})$ we find that Doppler--shift has the larger effect on $P(e_{\rm 10 Hz})$ shifting it toward lower eccentricities for GNs at higher $z$.\footnote{As $f_{\rm GW} \propto M_{\rm tot,z}$ it leads to lower $e_{\rm 10 Hz}$ for higher $z$ after solving Equations \eqref{eq:f_GW}.}
 
 Finally, we note that the adopted PN corrections together with DF change the median of the $e_{\rm 10 M}$ distribution by $\lesssim 86 \%$ and of other binary parameter distributions also investigated in \citet{Gondanetal2018b} by less than $\simeq 26 \%$.

%% file: Sections/Binaries_within_aLIGOhorizon.tex
\section{Generation of Mock Catalogs for Binaries Detectable by an Advanced GW Detector}
\label{sec:BBHs_in_LIGOsHorizon}
 
 The radial distribution of GNs with respect to an observer is derived in Section \ref{subsec:DLdistribGNs}. We calculate the target number of binaries to be generated in the MC simulation per single GN in Section \ref{subsec:NumbGenBBHsperGN}. Finally, we outline the MC methods with which we generate mock catalogs of GW capture BBHs detectable by an advanced GW detector and mock catalogs in the local Universe in Section \ref{subsec:SetupMCaux}.

\begin{figure}
    \includegraphics[width=90mm]{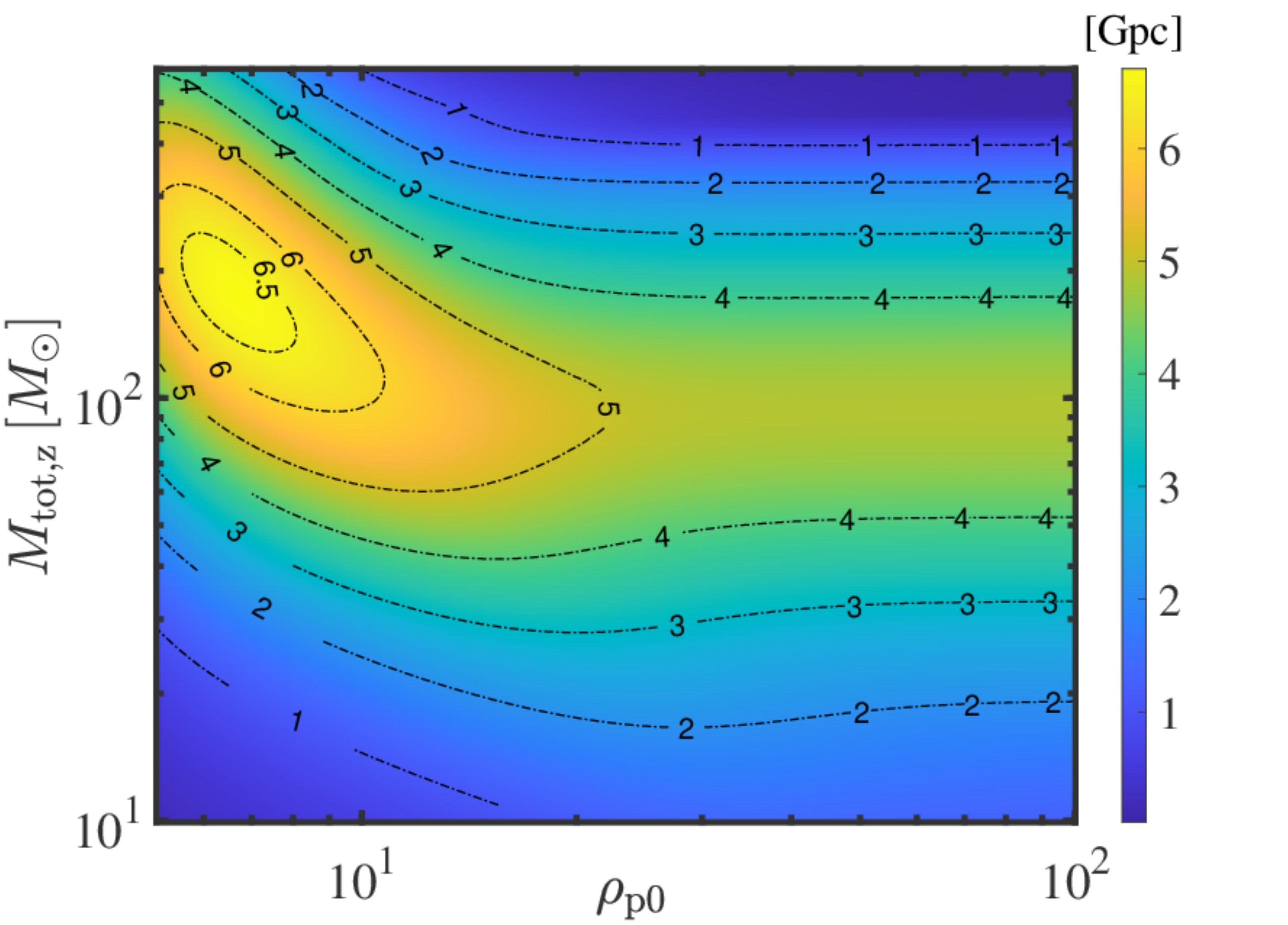}
  \caption{Horizon distance of a single aLIGO detector at design sensitivity for the inspiral phase of initially highly eccentric BBHs (i.e. inspiral phase with ${\rm S/N} = 8$, optimally oriented and overhead BBH) as a function of redshifted total mass $M_{\rm tot,z}$ and initial dimensionless pericenter distance $\rho_{\rm p0}$ for equal-mass ($\eta = 1/4$) and $e_0 = 0.99$. A semi-analytical inspiral-only waveform model is assumed for non-spinning eccentric BBH sources when calculating the $\rm S/N$ (Appendix \ref{sec:HorizonDistance}). Initially highly eccentric BBHs with $\rho_{\rm p0} \lesssim 20$ can be detected from cosmological distances out to $\sim 6.71 \, \Gpc$, while binaries in the circular limit are detectable from at most $\sim 4.86 \, \Gpc$. \label{fig:HorizonDistGWcBBHs} }
\end{figure}

 In Figure \ref{fig:HorizonDistGWcBBHs}, we show the horizon distance $D_{\rm hor}$ of a single aLIGO detector at design sensitivity for the inspiral phase of initially highly eccentric BBHs, where the matched filtering ${\rm S/N}$ equals to the detection threshold ${\rm S/N}_{\rm lim} = 8$ (e.g., \citealt{Abadieetal2010}), as a function of the parameters $\rho_{\rm p0}$ and $M_{\rm tot,z}$. Figure \ref{fig:HorizonDistGWcBBHs} assumes equal-mass ($\eta = 1/4$) binaries with initial orbital eccentricity $e_0 = 0.99$. $D_{\rm hor}$ is reduced by $(4 \eta)^{1/2}$ or increased by $8 / ({\rm S/N}_{\rm lim})$ for other choices of $\eta$ and ${\rm S/N}_{\rm lim}$, respectively, and it marginally increases with $e_0$ in the highly eccentric limit $e_0 \rightarrow 1$; see Appendix \ref{sec:HorizonDistance}. The result is very similar to that in \citet{OLearyetal2009}. The maximum detection distance converges to the circular result for high $\rho_{\rm p0}$. Eccentric sources with $\rho_{\rm p0} \lesssim 20$ are detectable to a much larger distance than circular inspirals sources. Similar trends are obtained for AdV and KAGRA as well; see Appendix \ref{sec:HorizonDistance} for further discussion on the $\rho_{\rm p0}$ and $M_{\rm tot,z}$ dependence of $D_{\rm hor}$.

\subsection{Radial distribution of galactic nuclei}
\label{subsec:DLdistribGNs}
 
 We utilize a spatially-flat $\Lambda$CDM cosmology model with model parameters \citep{Planck2018}: Hubble constant $H_0 = 67.4 \, \kms^{-1} \Mpc^{-1}$, and the density parameters for matter and dark energy \mbox{$\Omega_{\rm M} = 0.315$}, and $\Omega_{\Lambda} = 0.685$, respectively. Accordingly, the luminosity distance $D_{\rm L}$ for a given redshift $z$ is
\begin{equation}   \label{eq:CovDLz} 
  D_{\rm L} = \frac{ (1+z)c }{ H_0 } \int_0 ^z \frac{dz'}{ \left( \Omega_\mathrm{M} (1+z')^3 + \Omega_{\Lambda} \right)^{1/2} }
\end{equation}
 \citep{Hogg1999}. The comoving distance $D_{\rm C}$ is related to $D_{\rm L}$ by the relationship $D_{\rm C} = D_{\rm L} / (1 + z)$ \citep{Hogg1999}, which imply that for the largest detection distance of initially highly eccentric BBHs for aLIGO at design sensitivity $D_{\rm L} = 6.71 \, \Gpc$ (Appendix \ref{sec:HorizonDistance}) the corresponding comoving distance and redshift are \mbox{$D_{\rm C} \sim 3.37 \, \Gpc$} and $z \simeq 0.99$, respectively. Similarly, for AdV and KAGRA, the largest detection distances are $\sim 4.12 \, \Gpc$ and $\sim 4.48 \, \Gpc$ (Appendix \ref{sec:HorizonDistance}), respectively, and the corresponding $\{ D_{\rm C}, z\}$ pairs are approximately $\{ 2.47 \, \Gpc, 0.66 \}$ and $\{ 2.61 \, \Gpc, 0.71 \}$, respectively.
 
\begin{figure}
  \includegraphics[width=85mm]{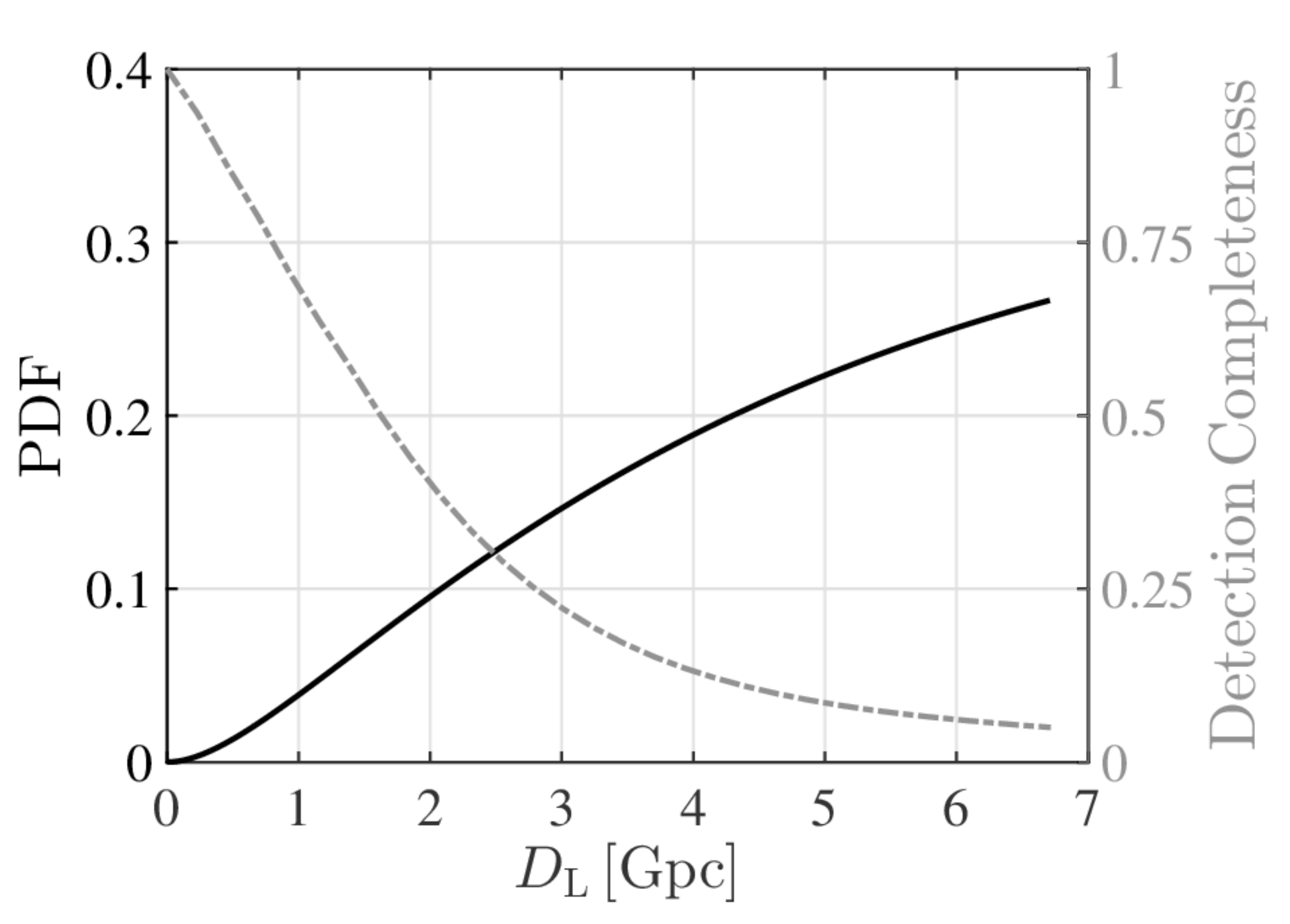}
  \caption{The intrinsic distribution of GNs with respect to the luminosity distance $D_{\rm L}$ (solid, left axis) (cf. Figure \ref{fig:HorizonDistGWcBBHs}), and the detection completeness (dashed-dotted, right axis) of binaries detected by aLIGO at design sensitivity with $\rm S/N > 8$ for the inspiral phase in the fiducial model ($\beta = 2$ and $p_0 = 0.5$, Appendix \ref{sec:StellPopsProp}). \label{fig:PDF_Dlum}   } 
\end{figure}

 GNs are expected to have a homogeneous and isotropic spatial distribution in comoving coordinates (Section \ref{subsec:SMBH_SpatDist}), thereby their radial distribution in terms of $D_{\rm C}$ is $P(D_{\rm C}) \propto D_{\rm C}^2$. This implies that
\begin{equation}  \label{eq:dens_DL}
 P(D_{\rm L}) = P(D_{\rm C}) \frac{d D_{\rm C}}{d D_{\rm L}} = C_{\rm N} \left( \frac{ D_{\rm L} }{ 1 + z} \right)^2 \frac{d D_{\rm C}}{d D_{\rm L}} \, ,
\end{equation}
 where $dD_{\rm C} / dD_{\rm L}$ and $z$  can be given as a function of $D_{\rm L}$ numerically using Equation \eqref{eq:CovDLz} and that $D_{\rm C} = D_{\rm L} / (1 + z)$. Here, $C_{\rm N}$ is given by normalizing $P(D_{\rm L})$ over the range $D_{\rm L} \in [0 \, \Gpc, \, 6.71 \, \Gpc]$. Figure \ref{fig:PDF_Dlum} shows $P(D_{\rm L})$ out to the maximum reach of an aLIGO detector at design sensitivity.
 
\subsection{Number of binaries generated in a single galactic nucleus in Monte Carlo simulations}
\label{subsec:NumbGenBBHsperGN}
 
 In case of the adopted spatially-flat $\Lambda$CDM cosmology model (Section \ref{subsec:DLdistribGNs}), the cosmic time--redshift relation is given as
\begin{equation}\label{eq:t_vs_z}
  t_z = \frac{1}{ H_0 } \int_{z}^{\infty} \frac{ (1 + z')^{-1} dz' }{ \left( \Omega_\mathrm{M} (1+z')^3 + \Omega_{\Lambda} \right)^{1/2} } 
  =\frac{2}{3 H_0 \Omega_{\Lambda}^{1/2} } {\rm ln} \left( \frac{1 + \cos{\gamma} }{ \sin{\gamma} } \right) 
\end{equation}
 \citep{Hogg1999}, where $\tan{\gamma} = ( \Omega_{\rm M} /\Omega_{\Lambda})^{1/2} ( 1 + z )^{3/2} $. The present age of the Universe $T_{\rm Age}$ follows by setting $z = 0$:
\begin{equation}  \label{eq:AgeUniv}
 T_{\rm Age} = \frac{2}{3 H_0 \Omega_{\Lambda}^{1/2} } {\rm ln} \left[ \frac{1 + \Omega_{\Lambda}^2}{ \left( 1 - \Omega_{\Lambda} \right)^2 } \right] \sim 13.8 \, \Gyr \, .
\end{equation}
 Using Equations \eqref{eq:t_vs_z} and \eqref{eq:AgeUniv}, the relationship between the DF time $t_{\rm DF}$ measured from $12 \, \Gyr$ ago (Section \ref{subsec:MergRates}) and redshift $z$ can be given by
\begin{equation}  \label{eq:tDF_vs_z}
  t_{\rm DF} = t_z - \left( T_{\rm Age} - 12 \, \Gyr \right) \, .
\end{equation}
 According to this equation, $\zeta$ in Equation \eqref{eq:eq:rate_mAmB_DF} can be given in terms of $z$ instead of $t_{\rm DF}$ by means of the derived relationship. Thus, we can parameterize $\partial^2\Gamma_{\rm 1GN,DF} / \partial m_A \partial m_B$ and thereby MC experiments for single GNs with redshift $z$ or equivalently with the corresponding $D_{\rm L}$ (Section \ref{subsec:Setup_MC}). Finally, using Equation \eqref{eq:tDF_vs_z} we find that the age of the Universe at $z \simeq 0.99$ is $\sim 5.9 \, \Gyr$.
 
 In order to accurately model the merging GW capture populations in the considered volume, we generate mock mergers for each GN host in a mock GN sample according to their cosmological merger rate per GN, $\Gamma_{\rm 1GN,Cosm}$. For each GN, we combine the intrinsic merger rate of GW capture BBHs $\Gamma_{\rm 1GN,DF}$ with the cosmological scale factor $1/(1+z)$ as
\begin{equation}  \label{eq:CosMergRate}
  \Gamma_{\rm 1GN,Cosm} = \frac{ \Gamma_{\rm 1GN,DF} }{ 1 + z } = \frac{ 1 }{ 1 + z } \iint_{ m_{\rm BH, min} }^{ m_{\rm BH, max} } \frac{\partial^2\Gamma_{\rm 1GN,DF}}{\partial m_A \partial m_B} \, dm_A \, dm_B  \, .
\end{equation}
 Here, $\partial^2\Gamma_{\rm 1GN,DF} / \partial m_A \partial m_B$ depends on $z$ implicitly through $\zeta$ and also depends on $\{ \Msmbh, \, p_0, \, \mathcal{F}_{\rm BH} \}$ explicitly (Section \ref{subsec:MergRates}).
 
 For a given $p_0$ and $\mathcal{F}_{\rm BH}$, we choose an MC sample size of a mock GW capture BBH population in a single GN host, $\mathcal{N}_{\rm 1GNMC}$, for different $\Msmbh$ and $z$ as follows. First, we set a fiducial MC sample size $\mathcal{N}_{\rm 1GNMC,fid}$ for a fiducial GN host with redshift and SMBH mass values $z_{\rm fid}$ and $M_{\rm SMBH,fid}$, respectively, as $\{ \mathcal{N}_{\rm 1GN,fid}, \, z_{\rm fid}, \, M_{\rm SMBH,fid} \}=\{ 100, \, 0, \, 10^5 \, \Msun \}$. Then, we assign $\mathcal{N}_{\rm 1GNMC}$ according to the merger rate as
\begin{equation}  \label{eq:SampleSize_GNhost}
 \mathcal{N}_{\rm 1GNMC} = \bigg\{ \mathcal{N}_{\rm 1GNMC,fid} \times \frac{ \Gamma_{\rm 1GN,Cosm}(\Msmbh, z) }{ \Gamma_{\rm 1GN,Cosm}(M_{\rm SMBH,fid}, z_{\rm fid}) } \bigg\} \, ,
\end{equation}
 where the bracket $\{ \, \}$ denotes the floor function.
 
 The MC simulations show that DF influences $\mathcal{N}_{\rm 1GNMC}$ significantly as shown in Figure \ref{fig:NumbGenBBHsperGN} as it helps to increase the number of BHs in GNs, so we account for DF in our calculations; see Appendix \ref{sec:ImpactDF_GWcBBHs} for details.

\subsection{Setup of Monte Carlo simulations for binaries detectable by an advanced GW detector}
\label{subsec:SetupMCaux}
 
 In this section, we give an overview of our methodology that we use in generating mock catalogs of GW capture BBHs detectable by a single aLIGO/AdV/KAGRA detector and mock catalogs in the local Universe. The free parameters of MC simulations are $p_0$ and $\mathcal{F}_{\rm BH}$ as other parameters are sampled from distributions or resulted from observations and/or numerical simulations. Accordingly, we assume identical GNs in terms of these parameters in MC runs.
 
 We generate the mock catalog for detectable binaries in the following  three main steps.
\begin{enumerate}
  \item First, we generate a mock sample of GNs with different SMBH masses and spatial coordinates in the considered volume (Appendix \ref{sec:HorizonDistance}) according to the observed SMBH mass distribution (Section \ref{subsec:SMBH_MassFunc} and \ref{subsec:SMBH_MassRange}) and radial distribution of GNs  (Sections \ref{subsec:DLdistribGNs}). We draw $(\theta_N, \, \phi_N)$, the sky position angles relative to the detector (Appendix \ref{sec:CoordSyst_RespGWdet}), from an isotropic distribution (Section \ref{subsec:SMBH_SpatDist}).
  
  \item Next, we generate a mock sample of GW capture BBH sources for each GN host. The number of binaries generated in a single GN depends on their cosmological merger rate per GN (Section \ref{subsec:NumbGenBBHsperGN}). We generate the initial orbital parameters $\{ \rho_{\rm p0}, \, e_0 \}$ and the masses as discussed in Section \ref{subsec:Setup_MC}. The angular momentum unit vector angles $(\theta_L, \, \phi_L)$ (Appendix \ref{sec:CoordSyst_RespGWdet}) in the GW detector frame are drawn from an isotropic distribution as we assume isotropic BH populations in single GNs (Sections \ref{subsec:GNs_Relax}). Finally, we draw the polarization angle $\psi$ (Appendix \ref{sec:CoordSyst_RespGWdet}) from a uniform distribution.
  
  \item We discard binaries which are undetectable, i.e those with ${\rm S/N} \leqslant 8$ (Appendix \ref{sec:HorizonDistance}). As the ${\rm S/N}$ depends on the following parameters $\{ M_{\rm tot, z}, \, \eta, \, \rho_{\rm p0}, \, e_0, \, D_{\rm L}, \, \theta_N, \, \phi_N, \, \theta_L, \, \phi_L, \, \psi \}$, the resulting mock catalog accounts for the observational bias for an aLIGO/AdV/KAGRA detector in terms of these parameters.
\end{enumerate}
 
 We restrict the above-introduced method to generate mock catalogs of binaries in the local Universe. The main steps are as follows.
\begin{enumerate}
  \item We generate a mock sample of GNs with different SMBH masses according to the observed SMBH mass distribution (Section \ref{subsec:SMBH_MassFunc} and \ref{subsec:SMBH_MassRange}).
  
  \item Finally, a mock sample of GW capture BBH sources is generated for each GN host. The number of binaries generated in a single GN depends on their merger rate per GN in the local Universe (Section \ref{subsec:NumbGenBBHsperGN}). The corresponding initial orbital parameters $\{ \rho_{\rm p0}, \, e_0 \}$ and the masses are generated as discussed in Section \ref{subsec:Setup_MC}.
\end{enumerate}
 We list the steps of simulations in full detail in Appendix \ref{sec:MCsampling}.

%% file: Sections/Results.tex
\section{Results} 
\label{sec:Results}
 
 Here we present our main results: the distribution of GW frequency at binary formation in Section \ref{subsec:BBHsInDetBandsAtForm}, the distributions of orbital parameters and mass-dependent parameters for binaries detectable by aLIGO in Section \ref{subsec:DistaLIGOdet}, possible correlations among source parameters for aLIGO detections in Section \ref{subsec:CorrBBHparams}, and constraints on the escape speed of the host environment based on eccentricity in Section \ref{subsec:VelDist_Host}.

\begin{figure}
    \centering
    \includegraphics[width=84mm]{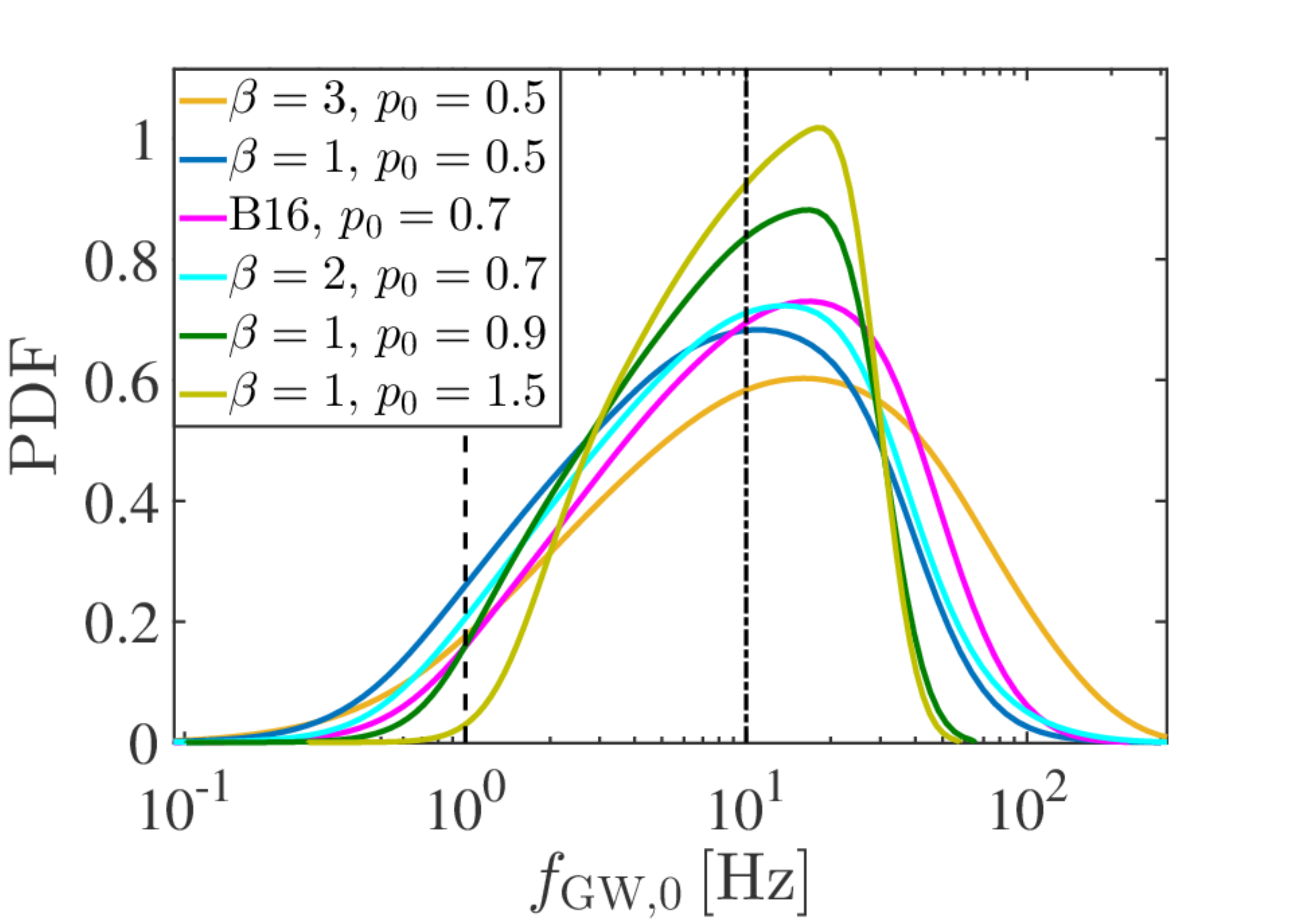}
    \\
    \includegraphics[width=84mm]{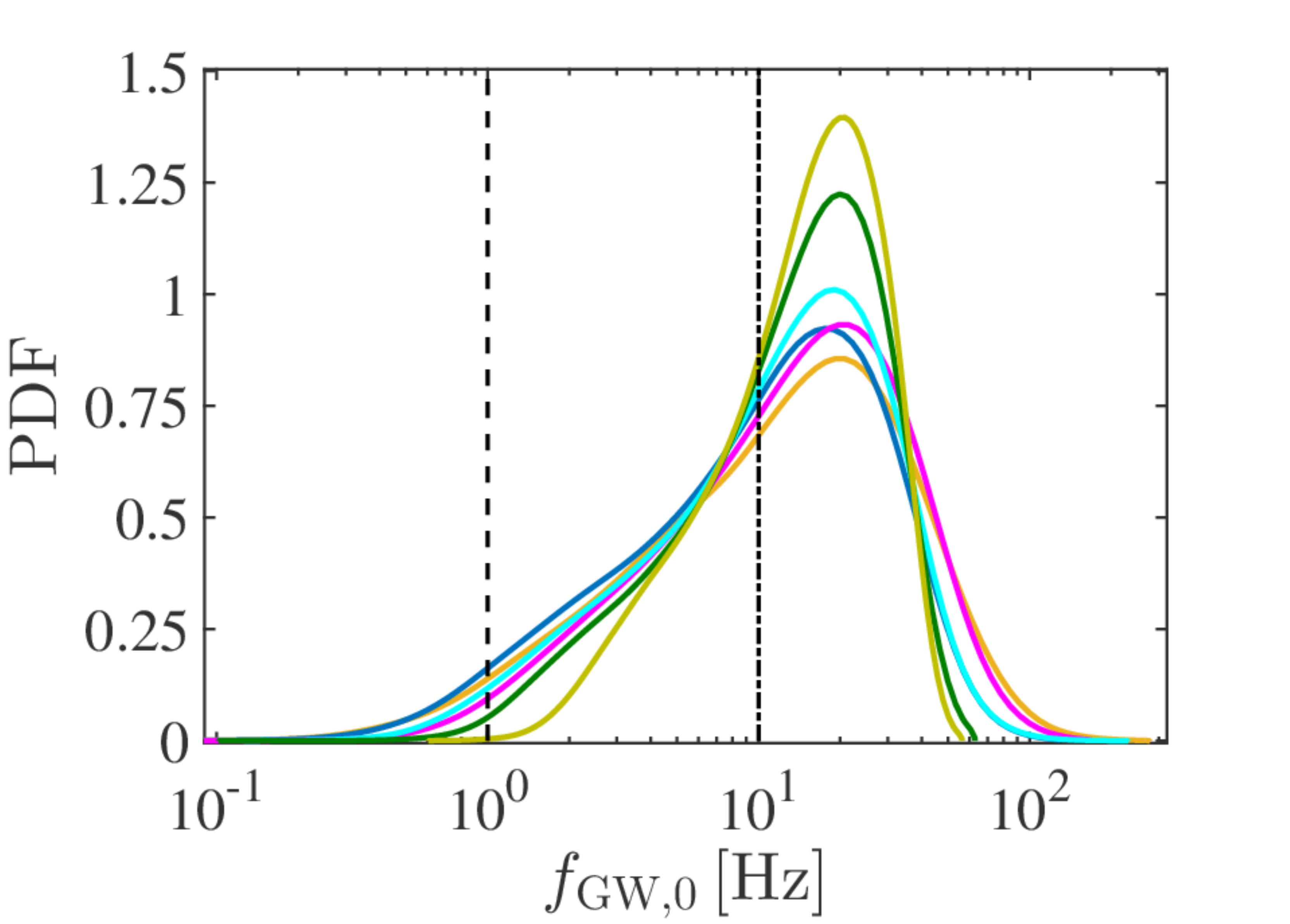}
    \\
    \includegraphics[width=84mm]{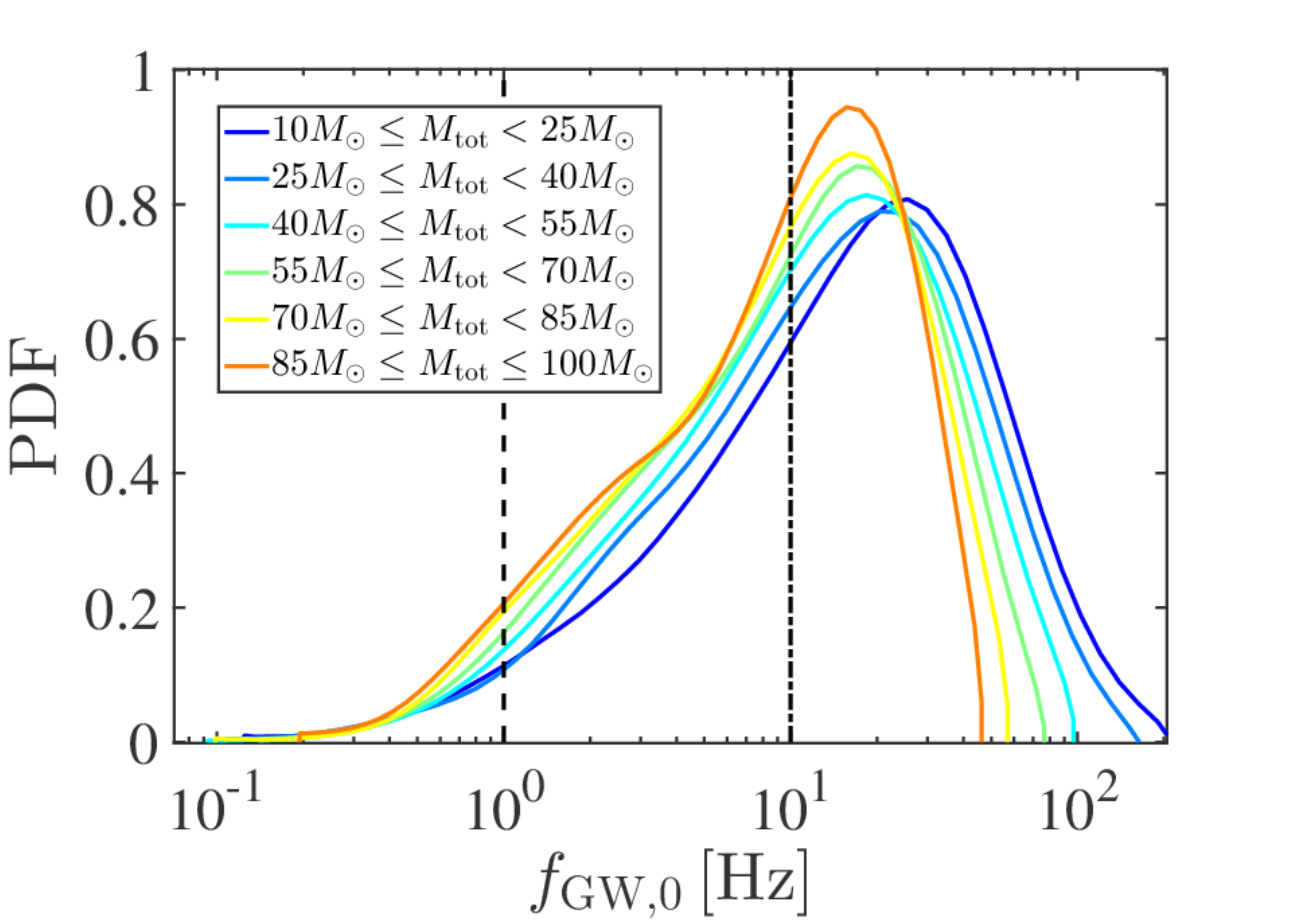}
  \caption{Distribution of peak GW frequency $f_{\rm GW,0}$ (Equation \ref{eq:f_GW}) at the time of GW capture BBH formation for all binaries in GNs out to $z \simeq 0.99$  (top panel), the binaries detected by aLIGO with $\rm S/N > 8$ for the inspiral phase (middle panel), and binaries detected by aLIGO in various mass ranges in the fiducial model ($\beta = 2$ and $p_0 = 0.5$, bottom panel). The top and middle panels show the results for different BH mass functions and radial distributions (cf. Figures \ref{fig:ParamDist_GN_p0Dep}--\ref{fig:ParamDist_GN_betaDep} and \ref{fig:ParamDistGWcBBHs_Volume_Limits}). In all cases, the distributions peak between $\sim 10 - 20 \, \Hz$, where the detected distribution shows a systematic increase of $f_{\rm GW,0}$ with decreasing binary mass. Thus, a large fraction of single--single GW capture BBHs form in the aLIGO/AdV/KAGRA band with $e_0 \gtrsim 0.95$; see Section \ref{subsec:BBHsInDetBandsAtForm} for details. The vertical dashed and dashed-dotted lines correspond to $1 \, \Hz$ and $10 \, \Hz$ characteristic GW frequencies, respectively, separating the regions detectable by space-based and Earth-based GW detectors.
  \label{fig:fGW0Dist} } 
\end{figure}

\subsection{Peak GW frequency at the time of binary formation}
\label{subsec:BBHsInDetBandsAtForm}
 
 The top panel in Figure \ref{fig:fGW0Dist} shows the distribution of $f_{\rm GW,0}$, the peak GW frequency at the time the binary forms due to GW capture, for binaries with $z \lesssim 0.99$ without restricting to the subpopulation possibly detectable by aLIGO/AdV/KAGRA. We find that the $f_{\rm GW,0}$ distribution $P(f_{\rm GW,0})$ ranges between remarkably high values between $\sim 1 - 70 \, \Hz$, $\sim 41 - 56 \%$ of the binaries form above $10 \, \Hz$ for $0.5 \leqslant p_0 \leqslant 0.9$ and the considered $\mathcal{F}_{\rm BH}$ models ($\sim 45 \%$ for $p_0 = 0.5$ and $ \beta = 2$). Note that $\sim 58 \%$ of systems are above $10 \, \Hz$ in the case of an extreme mass-segregated cusp model with $p_0 = 1.5$ and $ \beta = 2$, as for the young stars in the Galactic Center (Appendix \ref{subsec:GNs_BHPops}). The region between $1 \, \Hz$ and $10 \, \Hz$ is accessible with the proposed deci-Hertz detectors (TianQin and \mbox{(B-)DECIGO)} and in part by higher-generation Earth-based detectors: Einstein Telescope \citealt{Punturoetal2010,Hildetal2011}; LIGO Voyager \citealt{Adhikarietal2017}; Advanced LIGO Plus \citealt{Milleretal2015}; Cosmic Explorer \citealt{Abbottetal2017d}).\footnote{Note that LISA will not be sensitive to this GW source population since a negligible ($\lesssim 7 \%$) fraction of binaries form in the LISA band ($\lesssim 1 \Hz$; \citealt{Robsonetal2019}).} These findings indicate that joint multi-band GW observations with advanced (or higher generation) detectors and deci-Hertz space-based detectors may have the capability to observe the full GW signal from formation to merger for this GW source population, which exhibits an initial series of repeated bursts turning into a continuous waveform in time domain \citep{KocsisLevin2012,Loutrel2020b}.
 
 The middle panel in Figure \ref{fig:fGW0Dist} shows $P(f_{\rm GW,0})$ as seen by aLIGO, and $\sim 55 - 63 \%$ of the binaries form above $10 \, \Hz$ for $0.5 \leqslant p_0 \leqslant 0.9$ and the considered $\mathcal{F}_{\rm BH}$ models ($\sim 69 \%$ when $p_0 = 1.5$ and $ \beta = 2$) with $e_0 \gtrsim 0.95$ as discussed in Sections \ref{subsec:ResMC_SingleGN}. The results are similar to that obtained without restricting to the detectable population. The difference is due to the fact that aLIGO is more sensitive to sources with lower $z$, where GW redshift is less significant, which leads to somewhat higher frequencies. Similar trends apply to the subpopulations detectable by AdV/KAGRA, but in this case $P(f_{\rm GW,0})$ shift toward somewhat higher frequencies as the detector is sensitive to sources at lower redshifts than aLIGO (Section \ref{subsec:DLdistribGNs}) due to its different sensitivity curve. Indeed, $\sim 63 - 75 \%$ ($\sim 79 \%$ ) and $\sim 60 - 71 \%$ ($74 \%$) of the binaries form above $10 \, \Hz$ for $0.5 \leqslant p_0 \leqslant 0.9$ and the considered $\mathcal{F}_{\rm BH}$ models ($\sim 79 \%$  and $\sim 74 \%$ when $p_0 = 1.5$ and $ \beta = 2$) for AdV and KAGRA, respectively.
 
 We find that more massive binaries form with lower $f_{\rm GW,0}$ as shown in the bottom panel in Figure \ref{fig:fGW0Dist}. This characteristic of GW capture BBHs is not evident as more massive binaries form with lower $\rho_{\rm p0}$ (Section \ref{subsec:ResMC_SingleGN}) and $f_{\rm GW,0} \propto M_{\rm tot,z}^{-1} \rho_{\rm p0}^{-3/2}$. This indicates a negative correlation between $f_{\rm GW,0}$ and the (redshifted) mass-dependent parameters.\footnote{Specifically for aLIGO detections, the correlation coefficient $r_{\rm S}$ (Section \ref{subsec:CorrBBHparams}) between $M_{\rm tot}$ ($M_{\rm tot,z}$) and $f_{\rm GW,0}$ ranges between $-0.188 \leqslant r_{\rm S} \leqslant -0.041$ ($-0.179 \leqslant r_{\rm S} \leqslant -0.029$) depending on $p_0$ and $\mathcal{F}_{\rm BH}$.} The bottom panel in Figure \ref{fig:fGW0Dist} shows that the peak of the $f_{\rm GW,0}$ distribution is at $\sim 15 \,\Hz$ for the heaviest BBHs and $\sim 30 \, \Hz$ for the lightest BBHs in the BBH population detectable by aLIGO. The reason is that $\rho_{\rm p0}$ decreases systematically with $M_{\rm tot}$ by a factor of 3 between $M_{\rm tot} = 10 - 100 \, \Msun$ (Figure \ref{fig:ParamDistGWcBBHs_MassRanges}), implying that the overall variation in the peak $f_{\rm GW,0}$ is within a factor of $\sim 3$ for different mass mergers.
 
 Single-single GW captures in GNs form at a frequency band much higher than binary populations in GCs or in galactic fields. The value of $f_{\rm GW,0}$, if measured, may be used to distinguish different astrophysical merger channels. \citet{Samsingetal2020} found that $P({\rm log}_{10}(f_{\rm GW,0}))$ peaks at different locations in GCs for mergers per unit volume in the local Universe: roughly about $0.001 \, \Hz$, $0.1 \, \Hz$, and $2 \, \Hz$ for 2--body, single--single GW capture, and 3-body mergers, respectively. To compare apples-with-apples, we generate mock catalogs of GW capture BBHs merging in GNs per unit volume in the local Universe for various $p_0$ and $\mathcal{F}_{\rm BH}$ as prescribed in Section \ref{subsec:SetupMCaux}, and in this case $P({\rm log}_{10}(f_{\rm GW,0}))$ has a systematically higher peak between $\sim 16 - 35  \, \Hz$ depending on $p_0$ and $\mathcal{F}_{\rm BH}$. This indicates that by measuring $f_{\rm GW,0}$, multi-band GW observations with the advanced GW detectors (or higher generation detectors) and proposed deci-Hertz space-based detectors may distinguish this GW source population from those formed in GCs or in galactic fields. Indeed, the $f_{\rm GW,0}$ distribution  carries direct information on the characteristic relative velocity between compact objects in the host environment through $\rho_{\rm p0}$ because $\rho_{\rm p0} \propto (\eta / w^2)^{-2/7}$ \citep{Gondanetal2018b,GondanKocsis2019} and $f_{\rm GW,0} \propto \rho_{\rm p0}^{-3/2}$.

\begin{figure*}
    \centering
    \includegraphics[width=75mm]{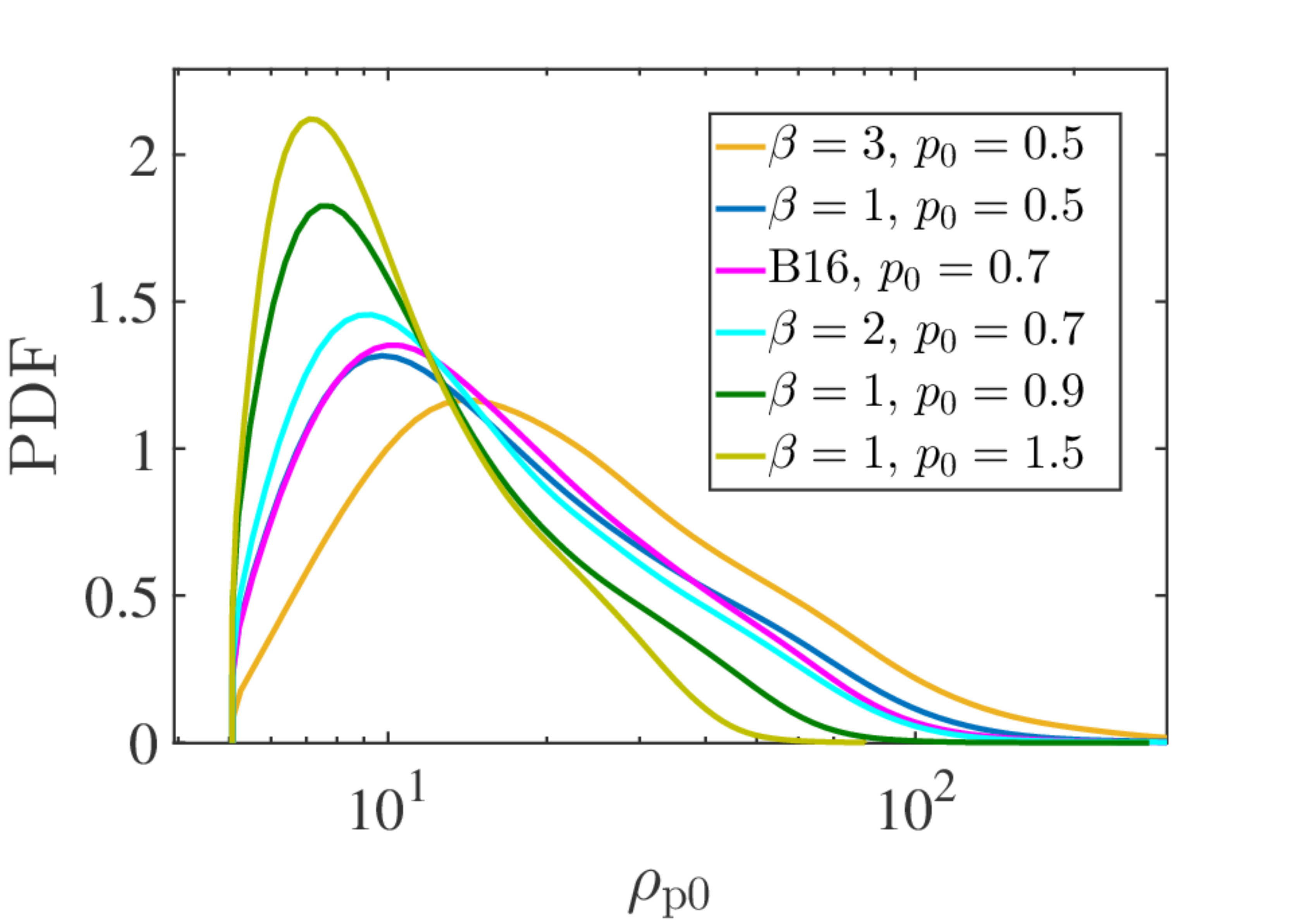}
    \includegraphics[width=75mm]{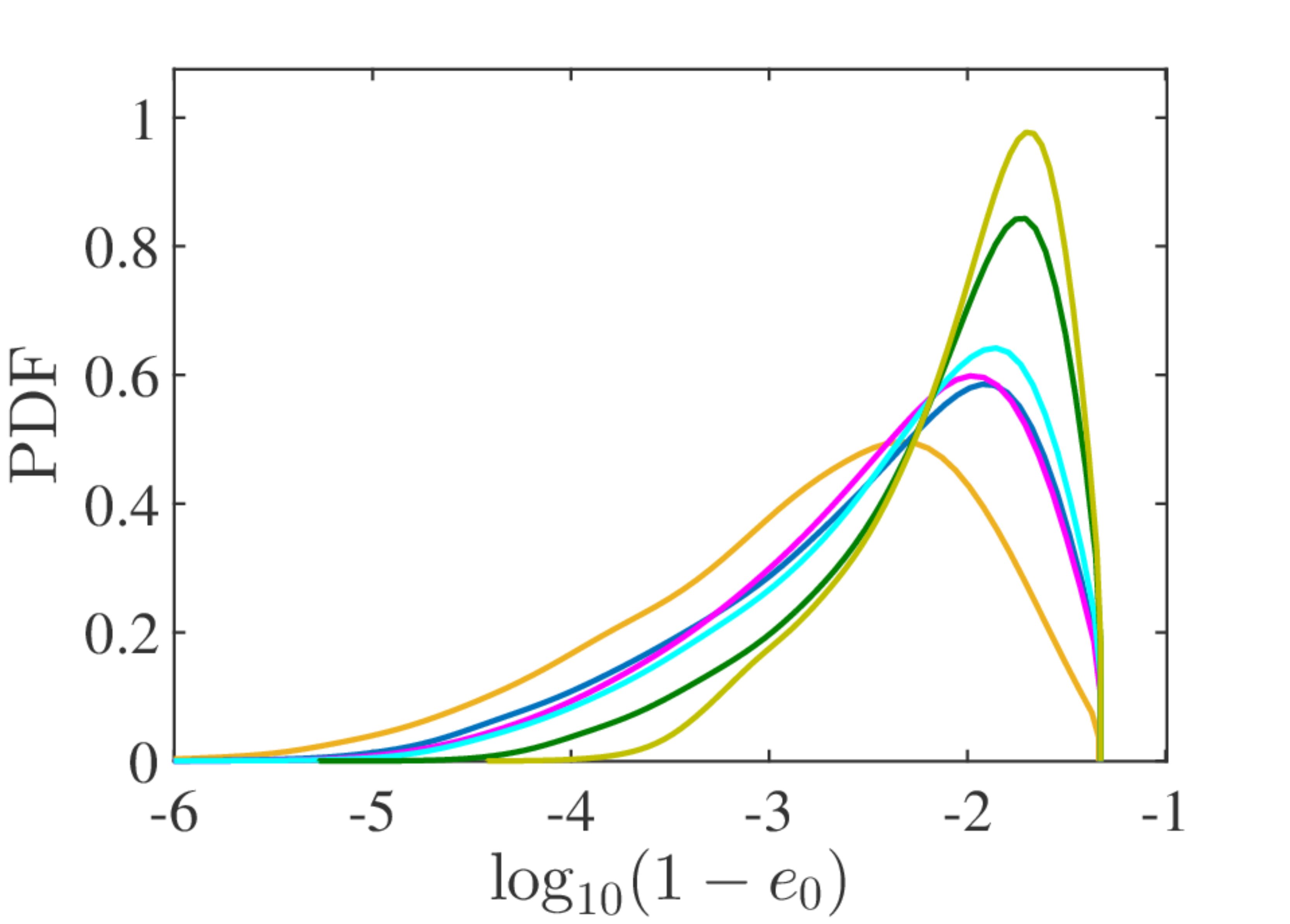}
    \\
    \includegraphics[width=75mm]{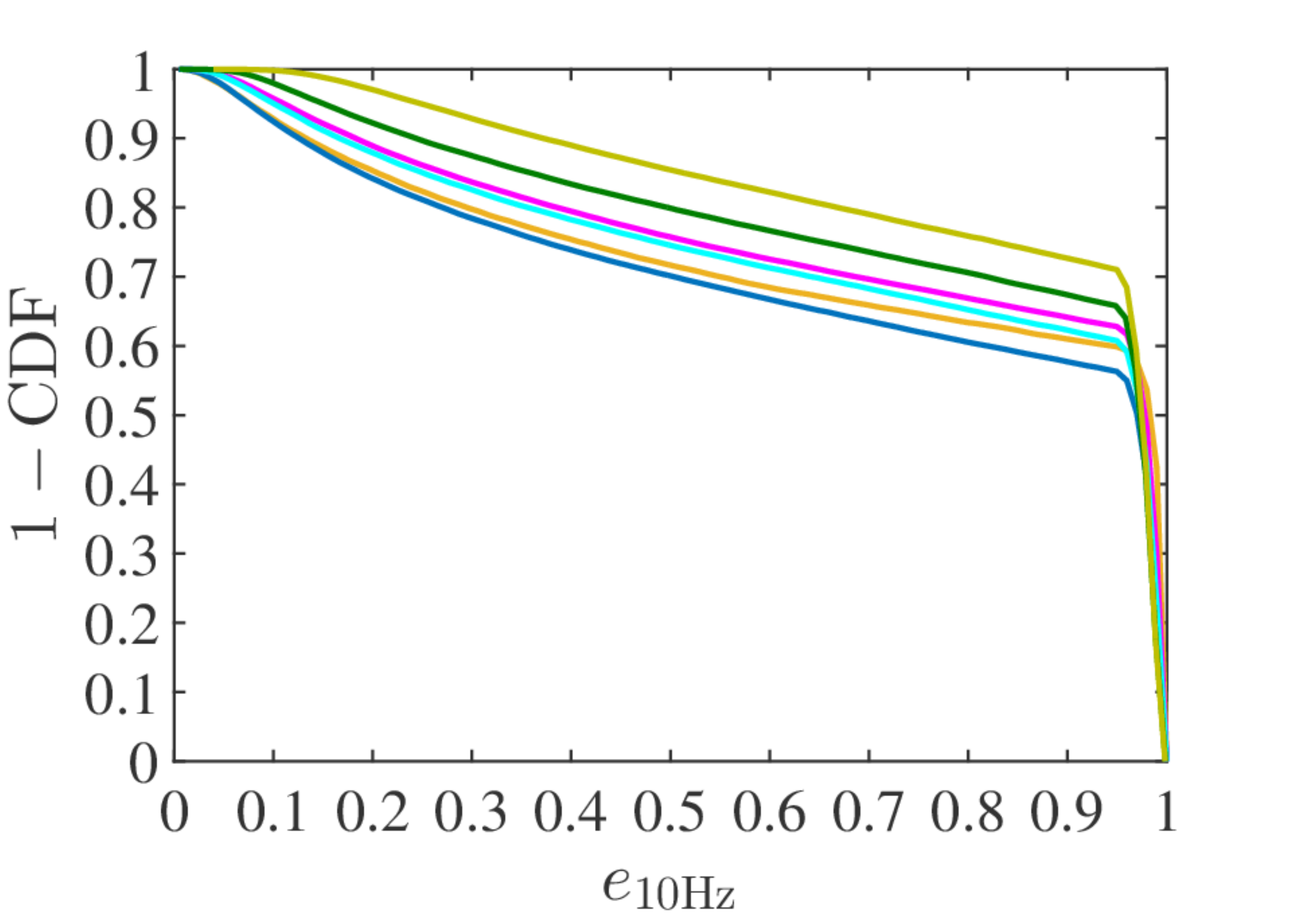}
    \includegraphics[width=75mm]{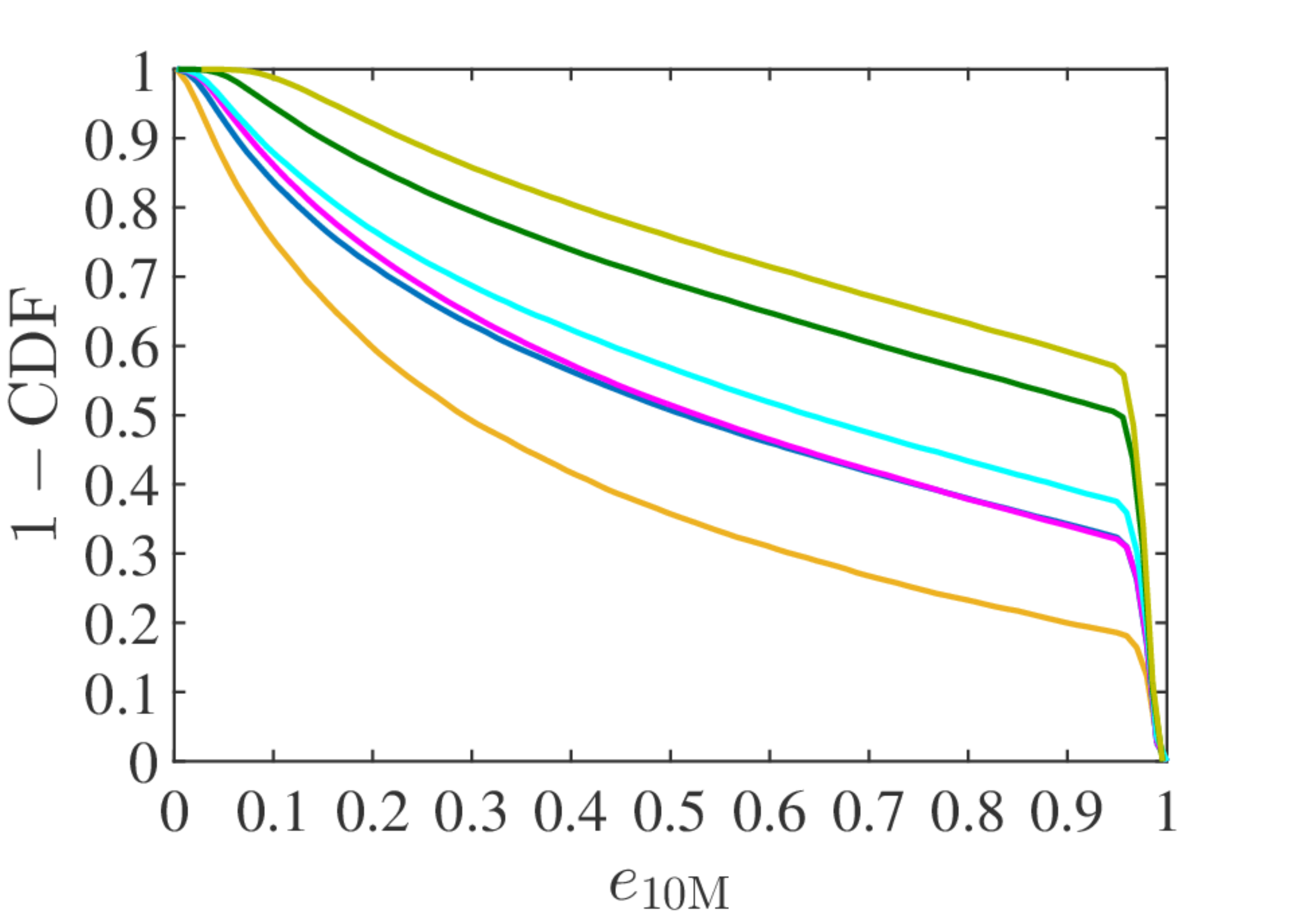} 
    \\
    \includegraphics[width=75mm]{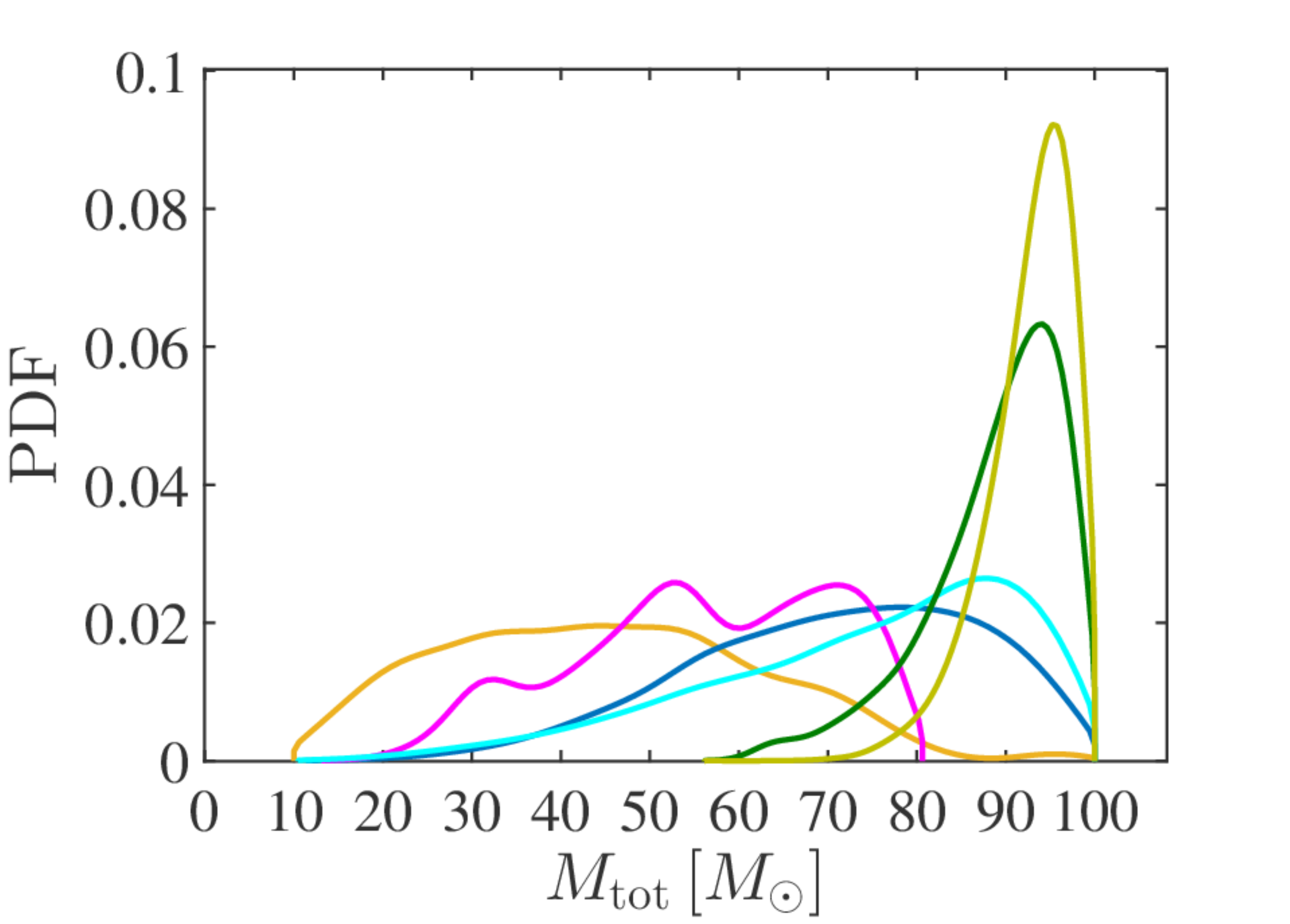}
    \includegraphics[width=75mm]{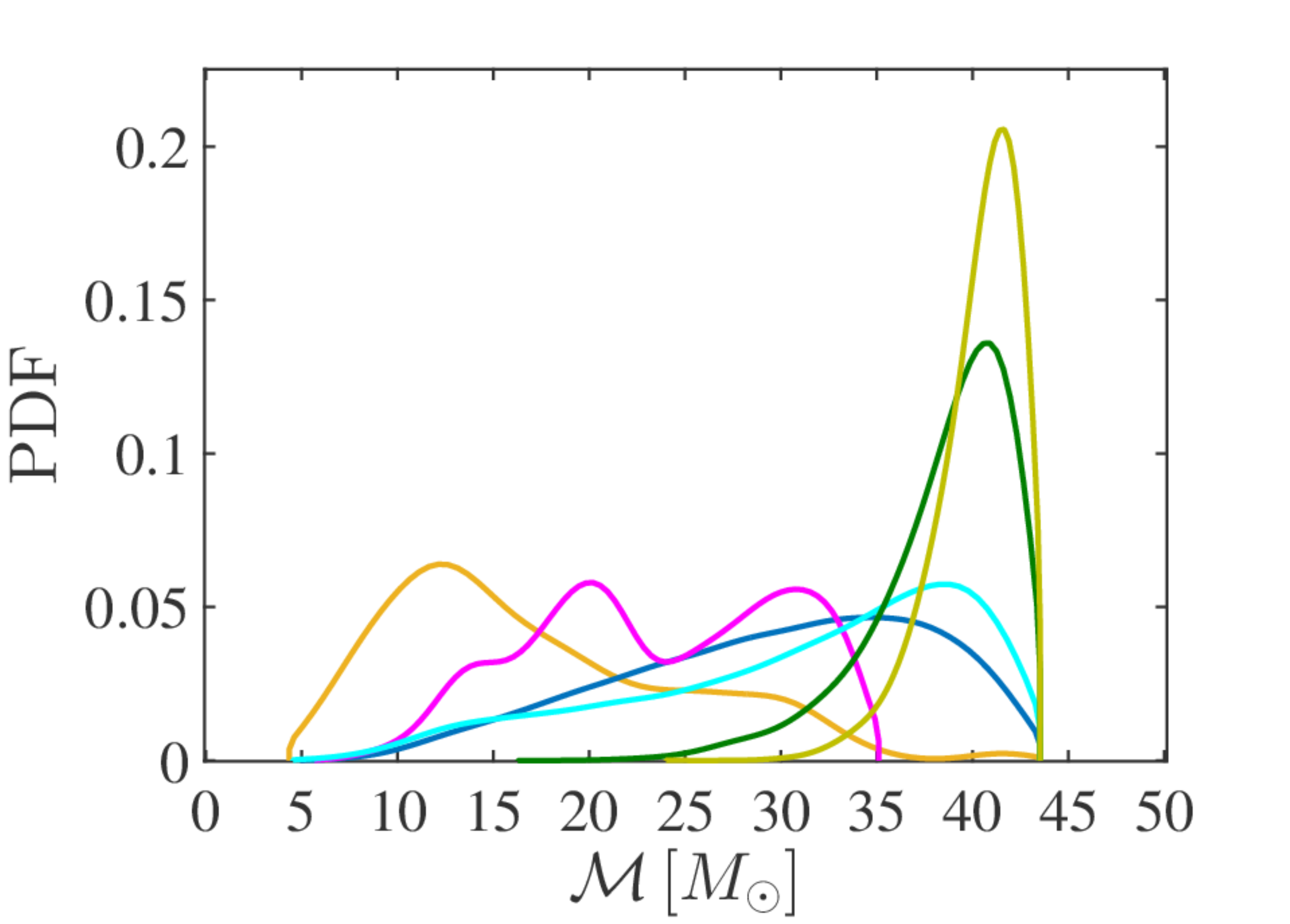}
    \\
    \includegraphics[width=75mm]{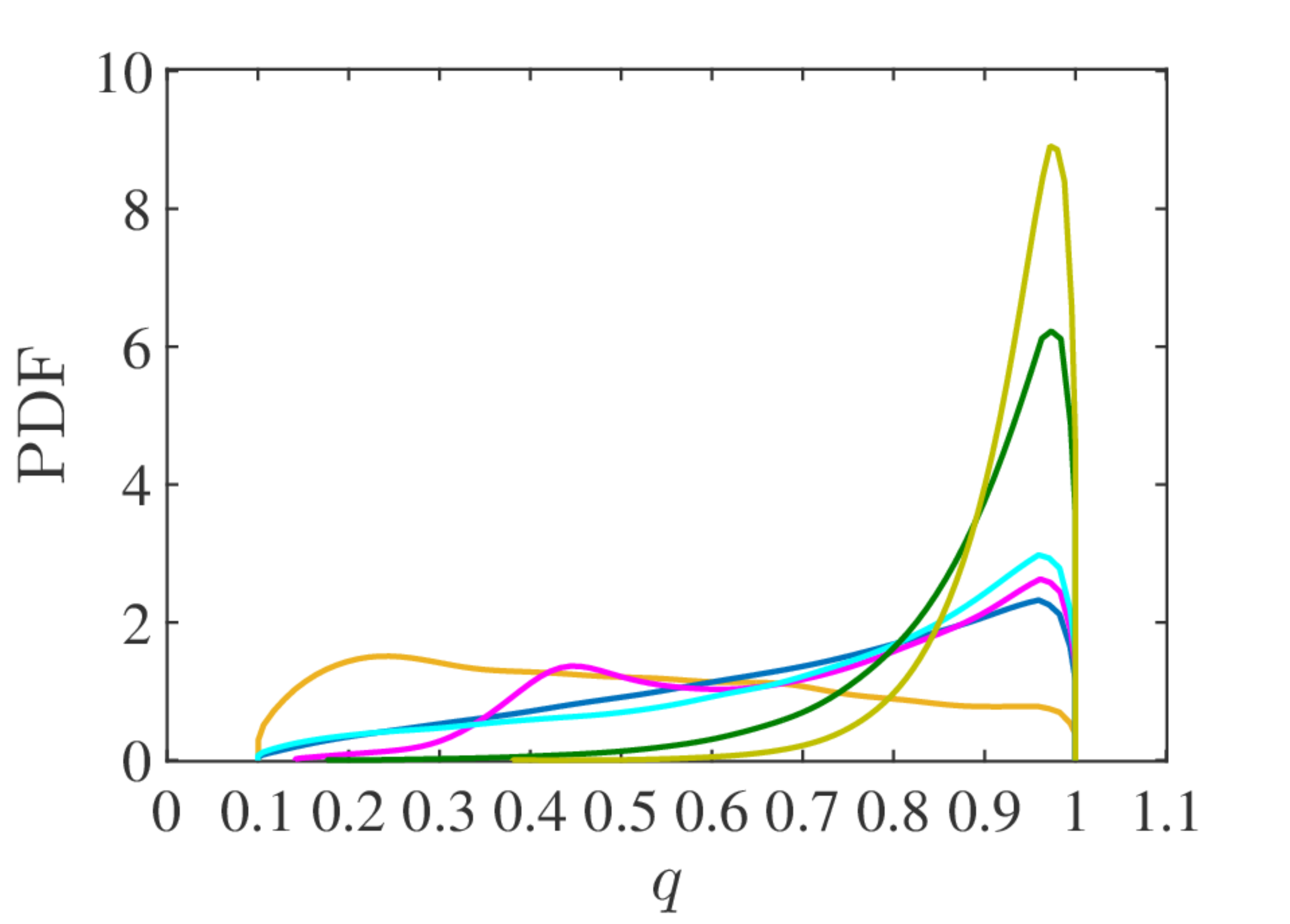}
    \includegraphics[width=75mm]{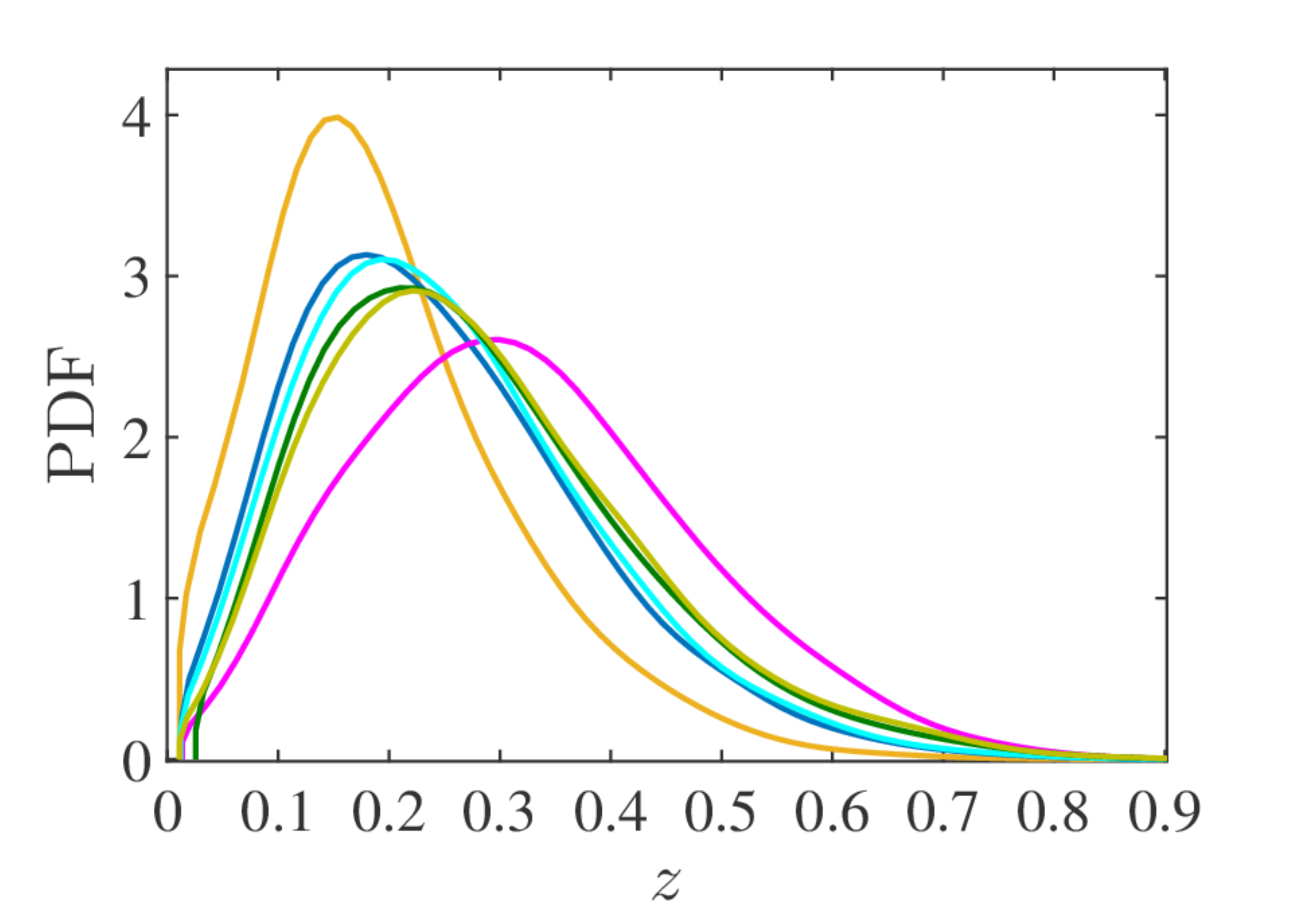}
  \caption{Distributions of binary parameters with $\rm S/N > 8$ for the inspiral phase with aLIGO at design sensitivity for single--single GW capture BBH sources in isotropic GNs assuming $5 \, \Msun \leqslant \ m_{\rm BH} \leqslant 50 \, \Msun$: initial dimensionless pericenter distance $\rho_{\rm p0}$ (row 1, left), initial orbital eccentricity $e_0$ (row 1, right), the $1 - {\rm CDF}$ of $e_{\rm 10 Hz}$ (row 2, left) and of $e_{\rm 10 M}$ (row 2, right), total mass $M_{\rm tot}$ (row 3 left), chirp mass $\mathcal{M}$ (row 3, right), mass ratio $q$ (row 4, left), and redshift $z$ (row 4, right). \label{fig:ParamDistGWcBBHs_Volume_Limits} } 
\end{figure*}

\begin{figure*}
    \centering
    \includegraphics[width=75mm]{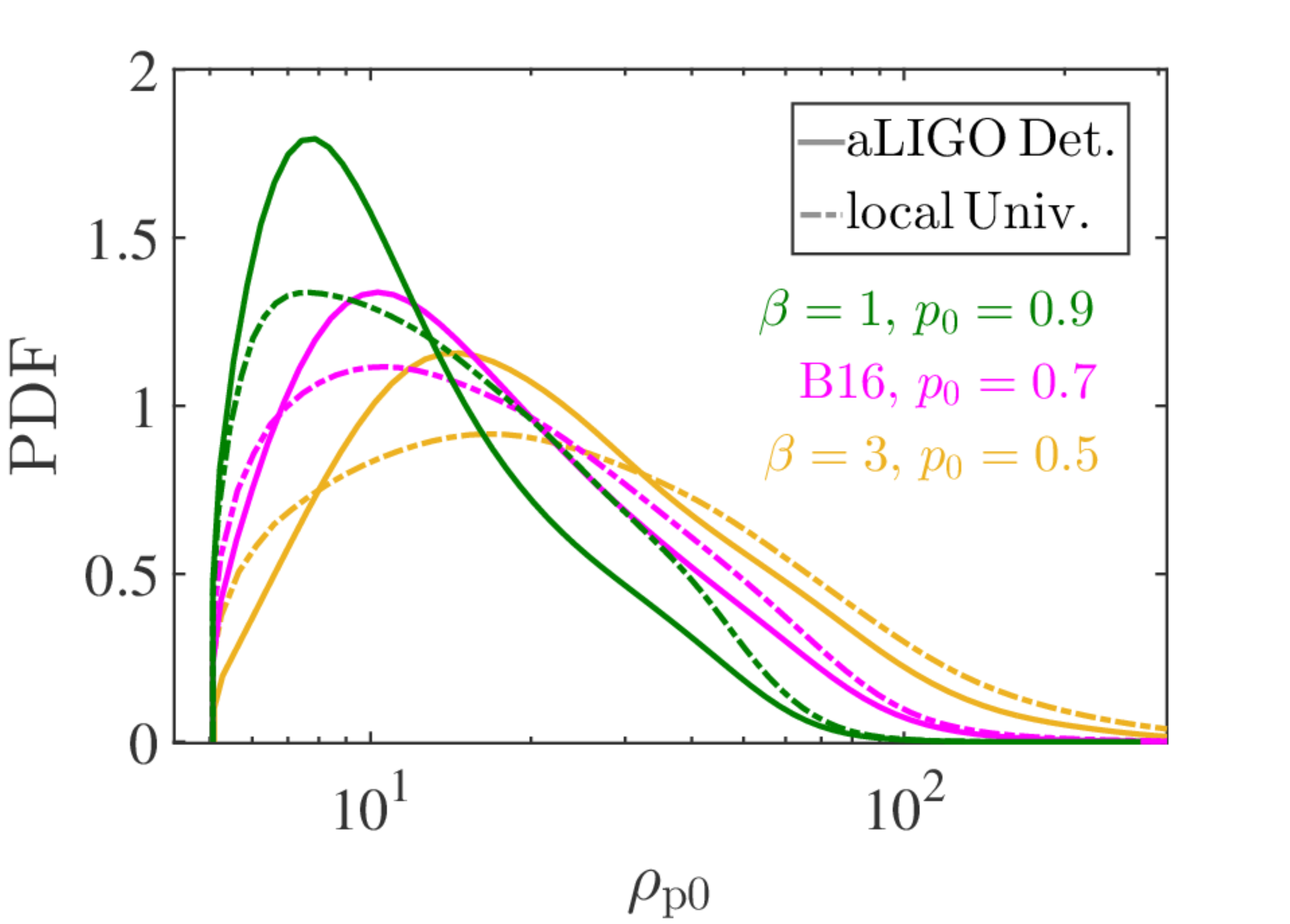}
    \includegraphics[width=75mm]{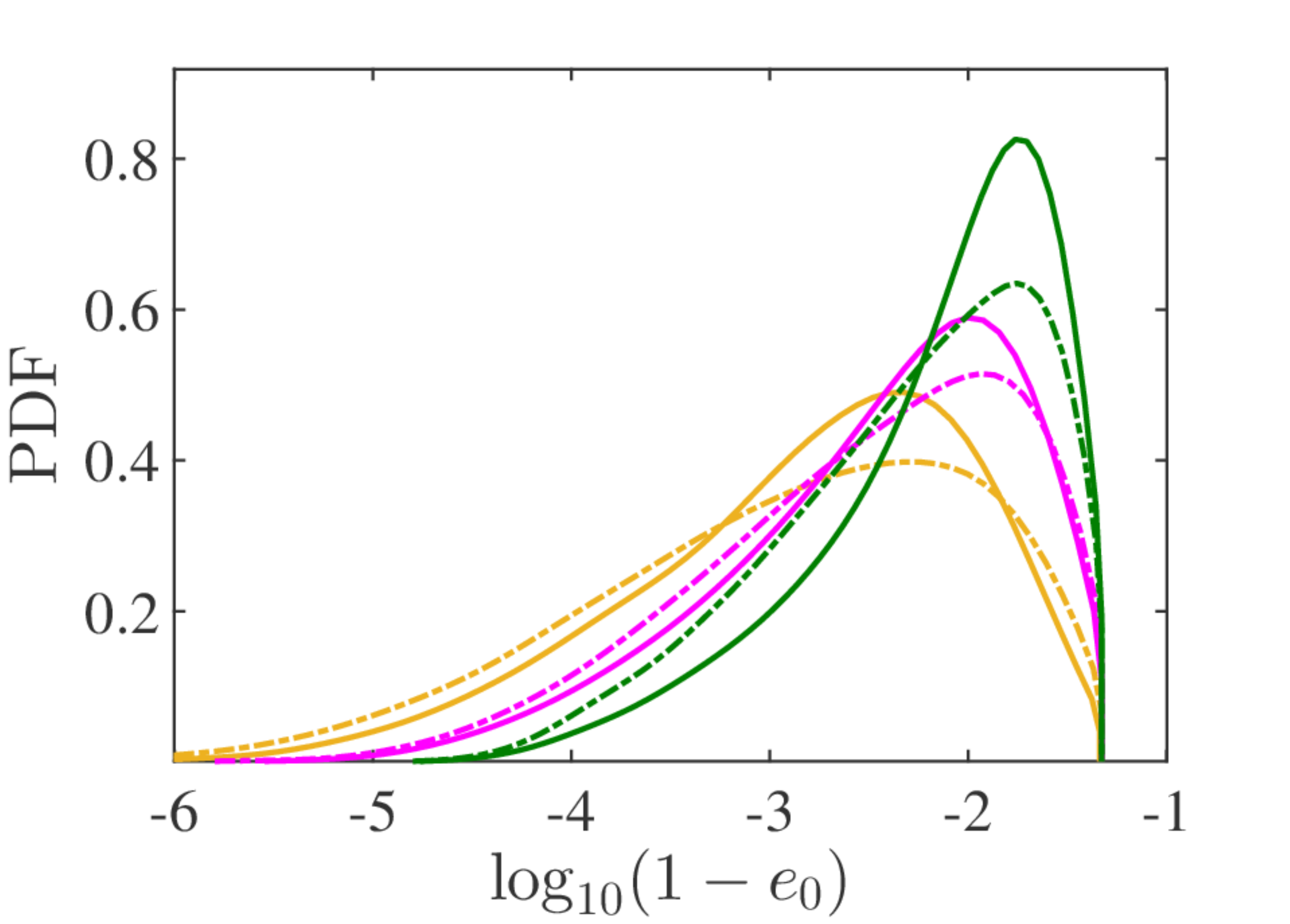}
    \\
    \includegraphics[width=75mm]{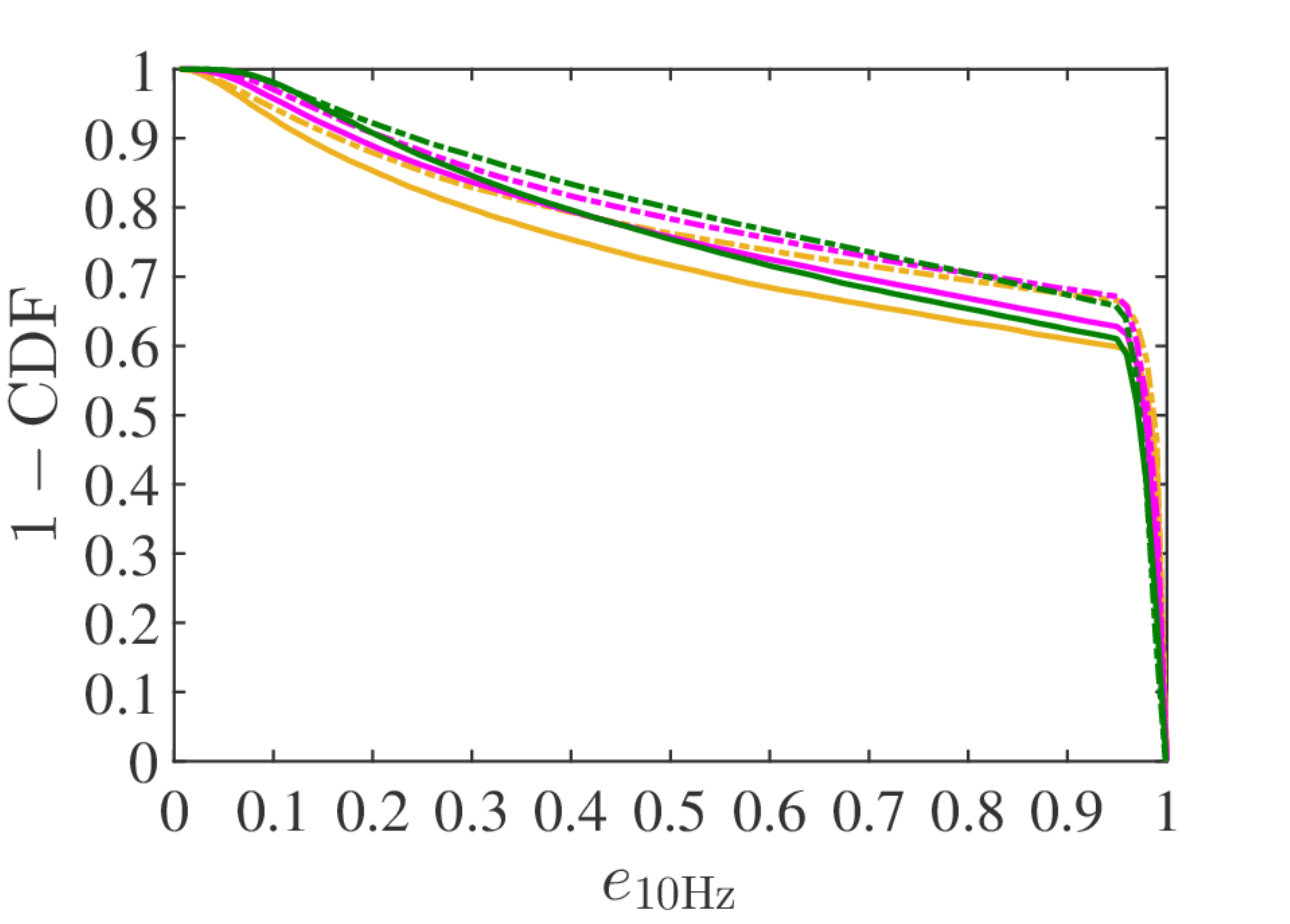}
    \includegraphics[width=75mm]{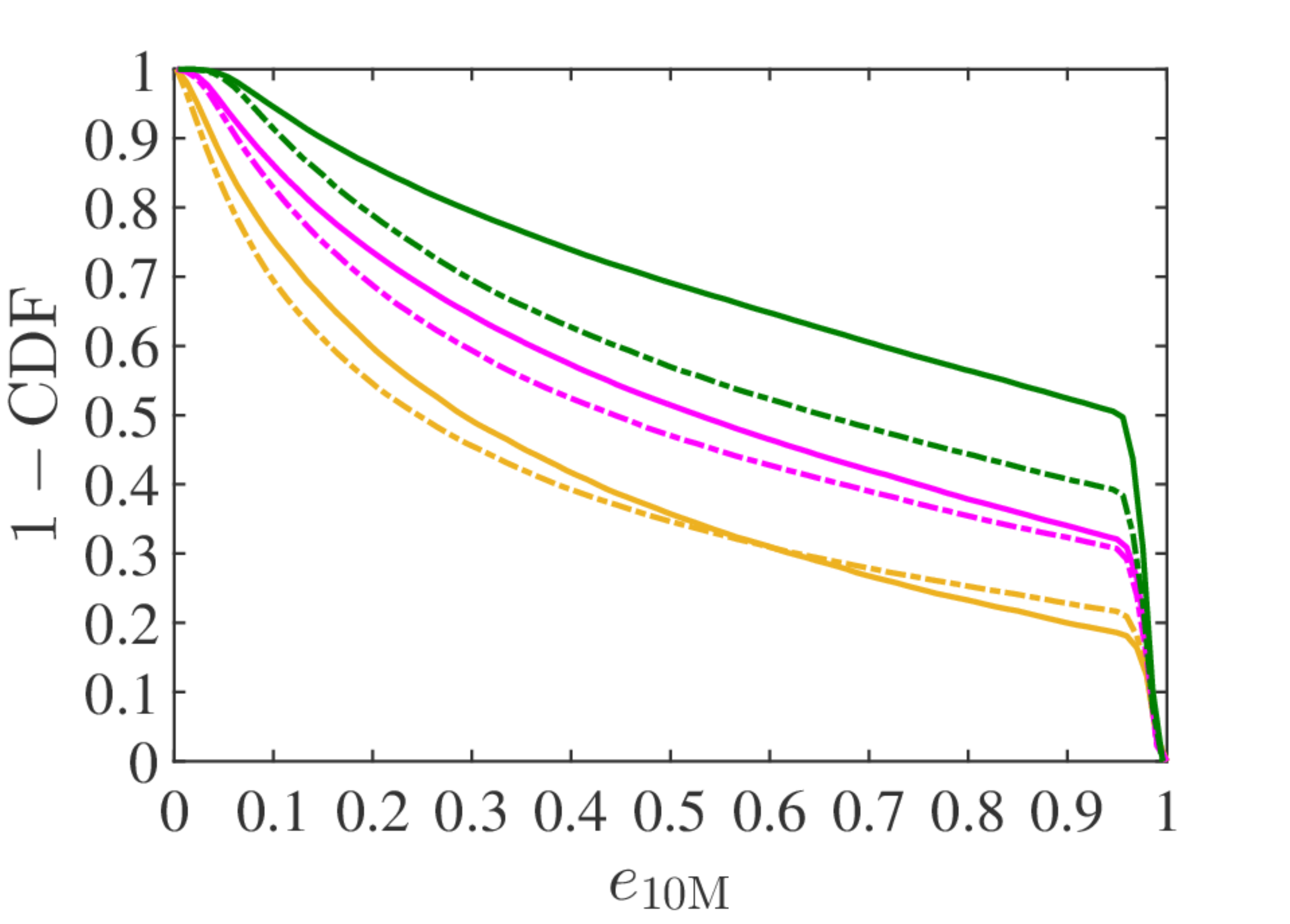} 
    \\
    \includegraphics[width=75mm]{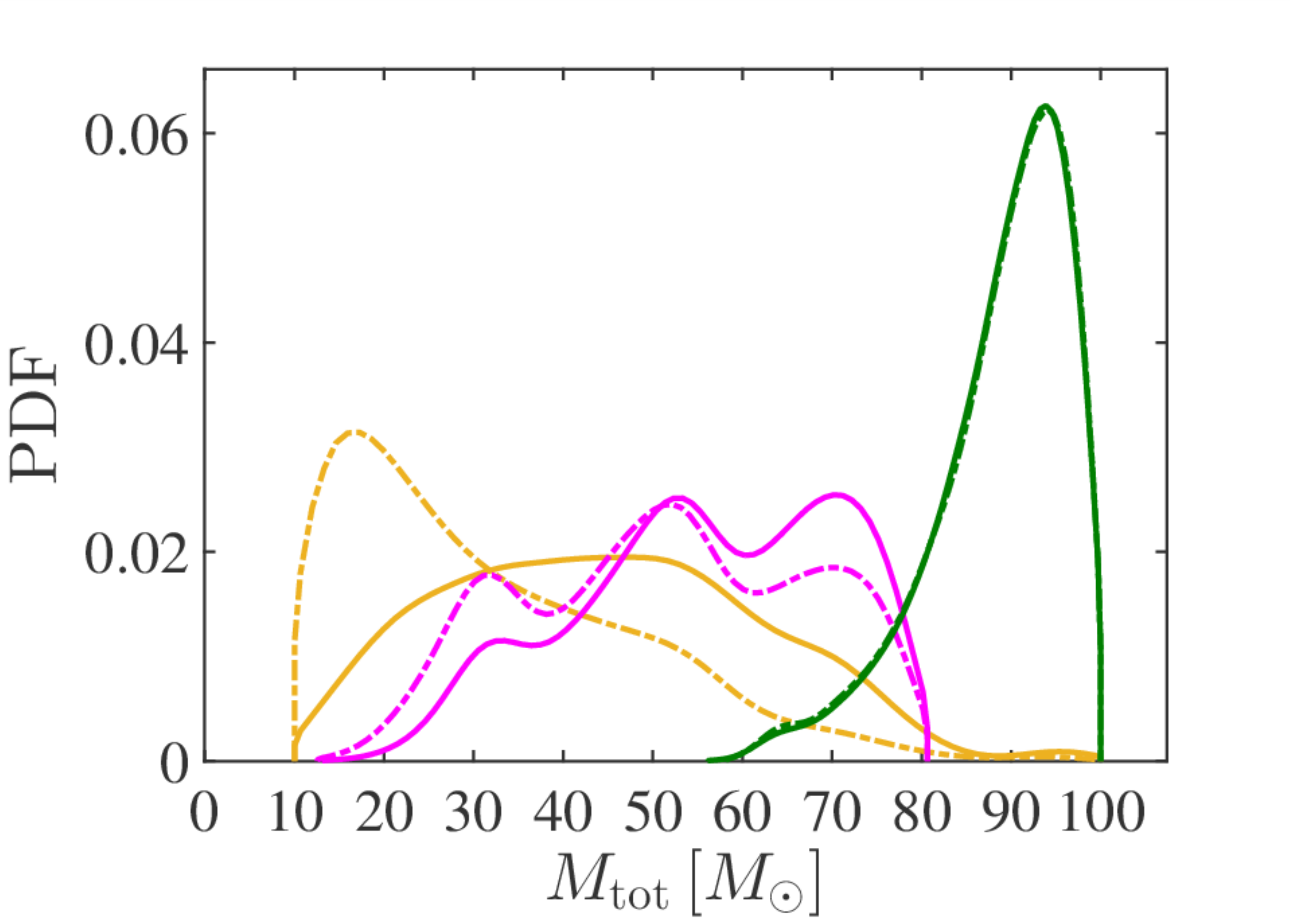}
    \includegraphics[width=75mm]{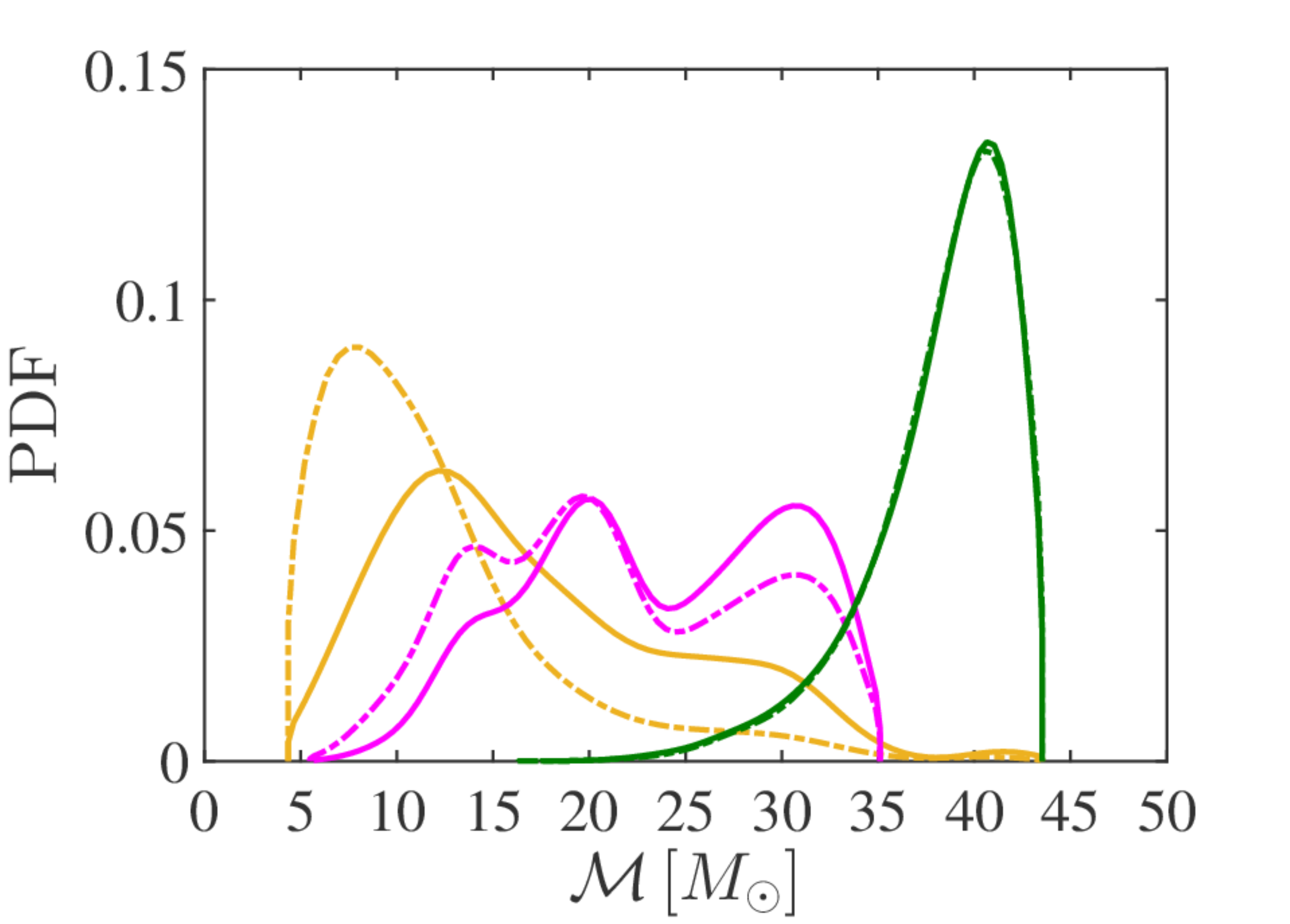}
    \\
    \includegraphics[width=75mm]{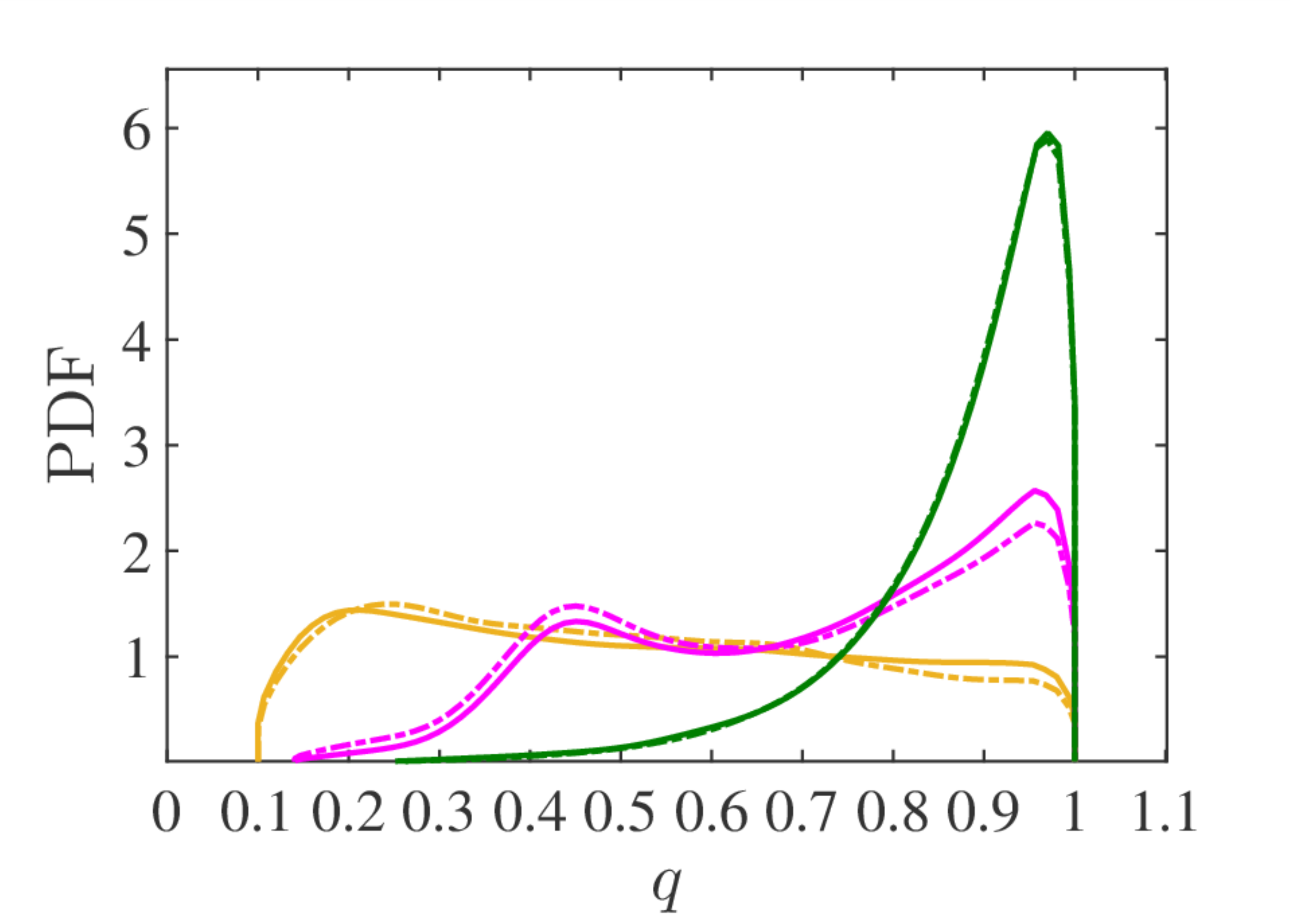}
  \caption{The impact of observational bias on the distributions of binary parameters for various examples as in Figure \ref{fig:ParamDistGWcBBHs_Volume_Limits} (labelled in the top left panel): initial dimensionless pericenter distance $\rho_{\rm p0}$ (row 1, left), initial orbital eccentricity $e_0$ (row 1, right), the $1 - {\rm CDF}$ of $e_{\rm 10 Hz}$ (row 2, left) and of $e_{\rm 10 M}$ (row 2, right), total mass $M_{\rm tot}$ (row 3 left), chirp mass $\mathcal{M}$ (row 3, right), and mass ratio $q$ (row 4). \label{fig:Impact_ObsBias_ParamDists} } 
\end{figure*}

\begin{figure*}
    \centering
    \includegraphics[width=75mm]{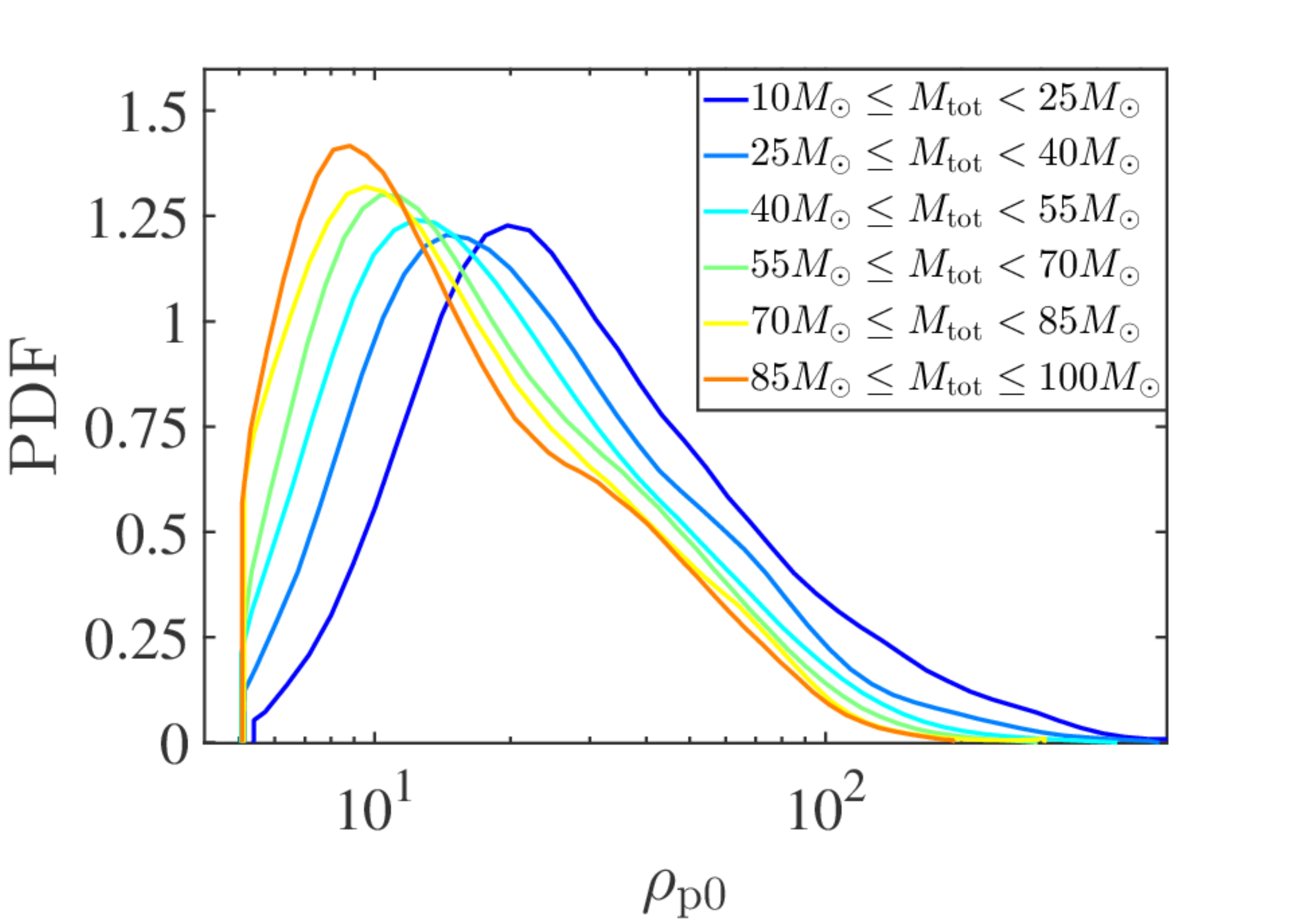}
    \includegraphics[width=75mm]{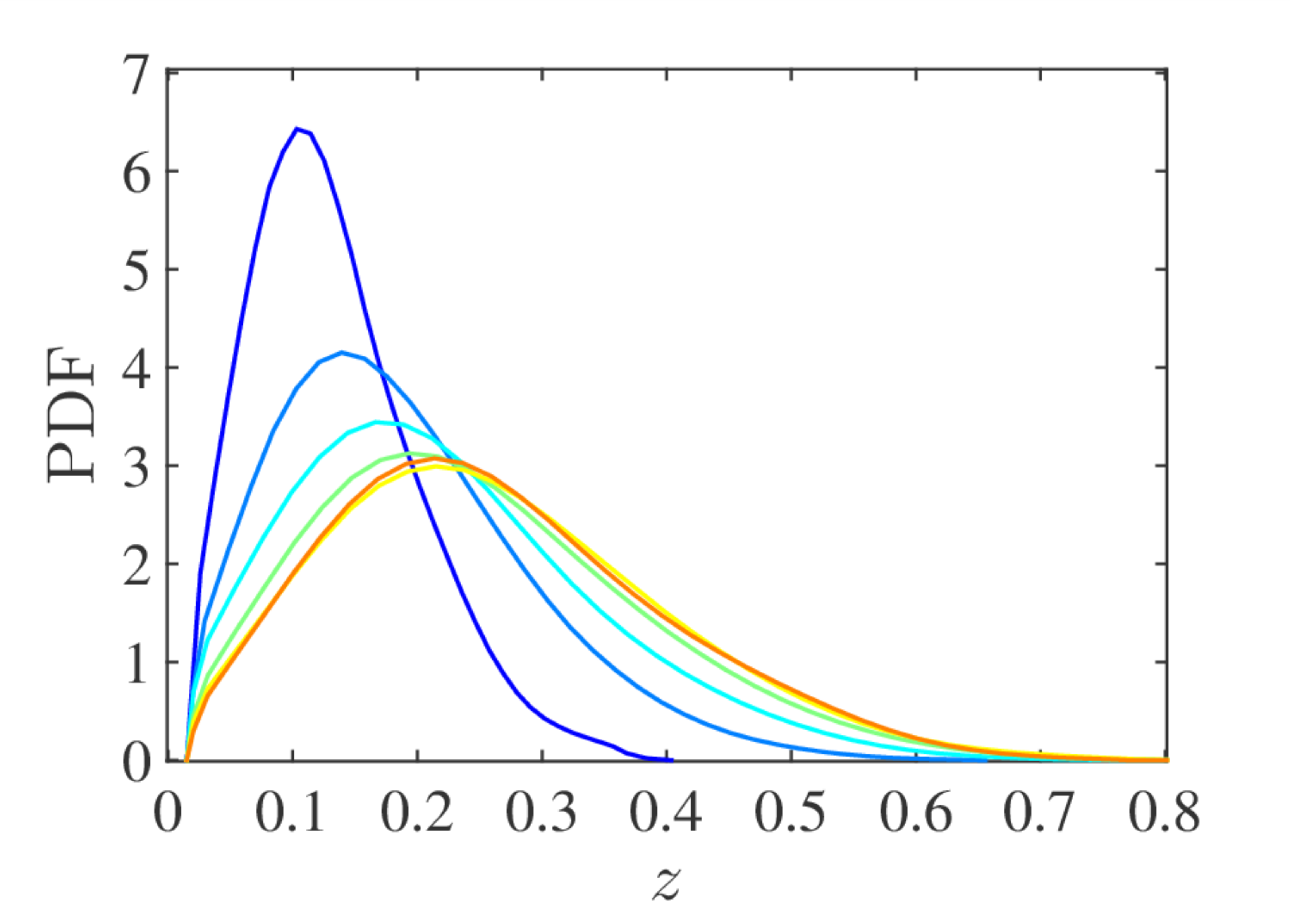}
    \\
    \includegraphics[width=75mm]{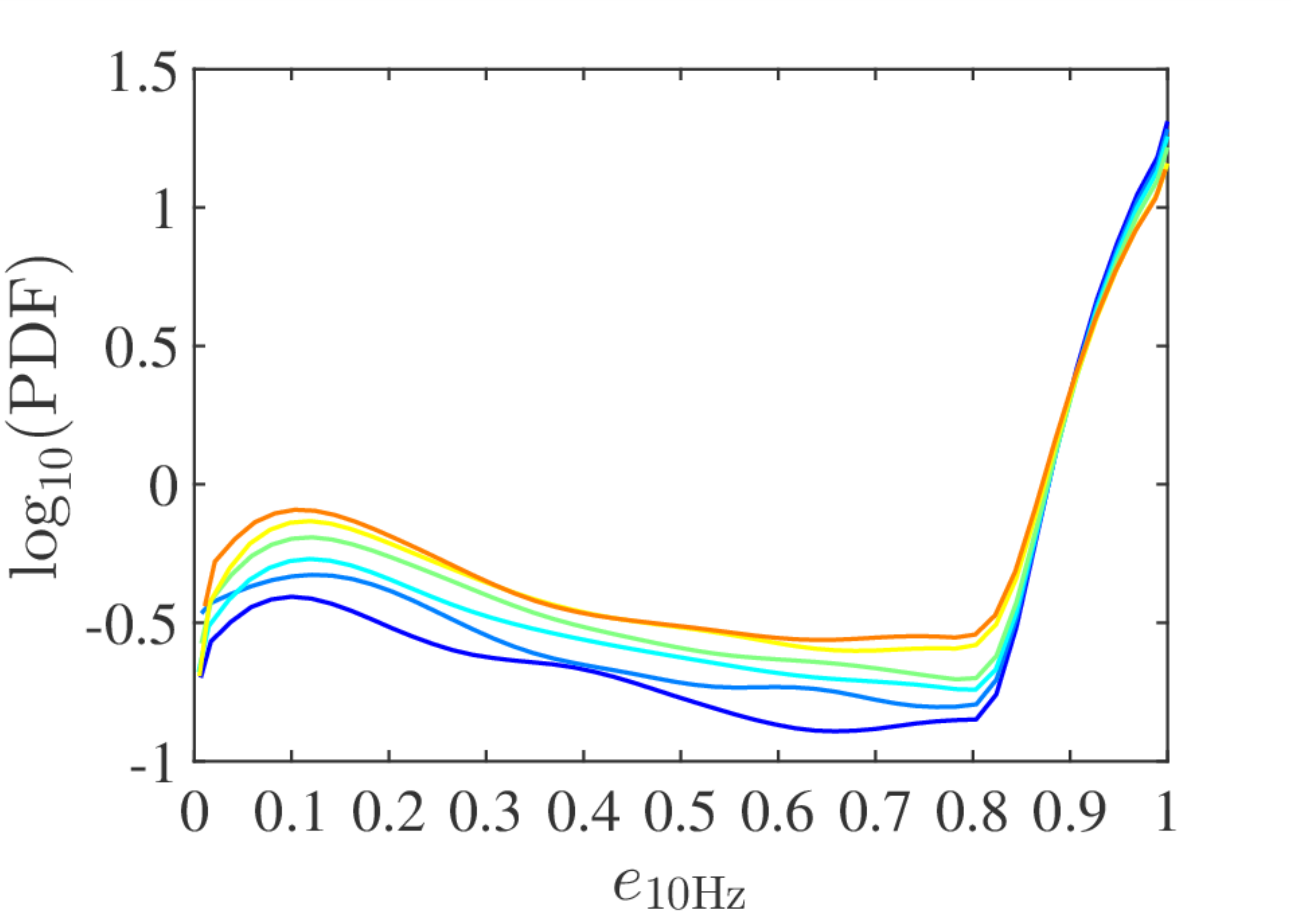}
    \includegraphics[width=75mm]{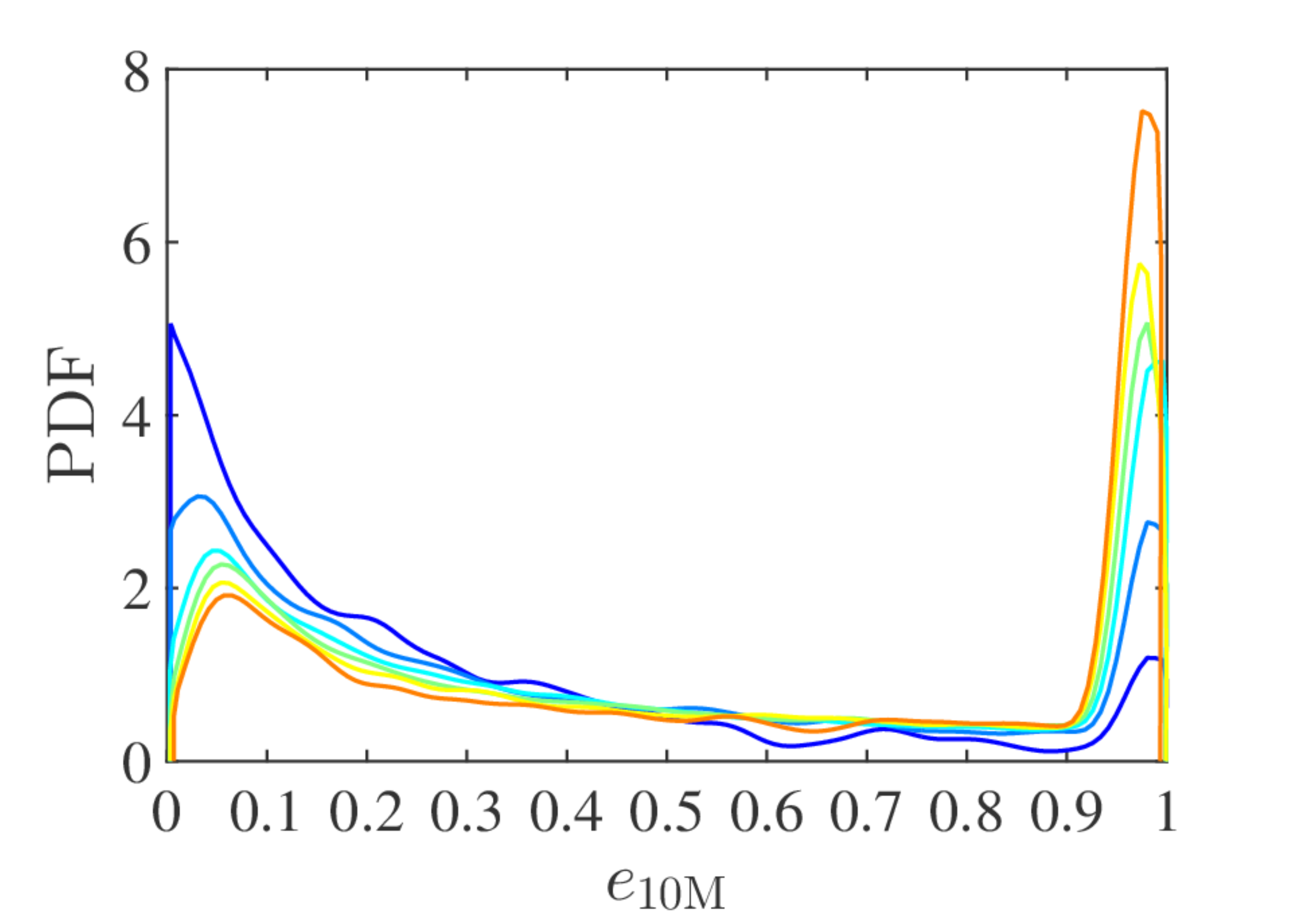}
    \\
    \includegraphics[width=75mm]{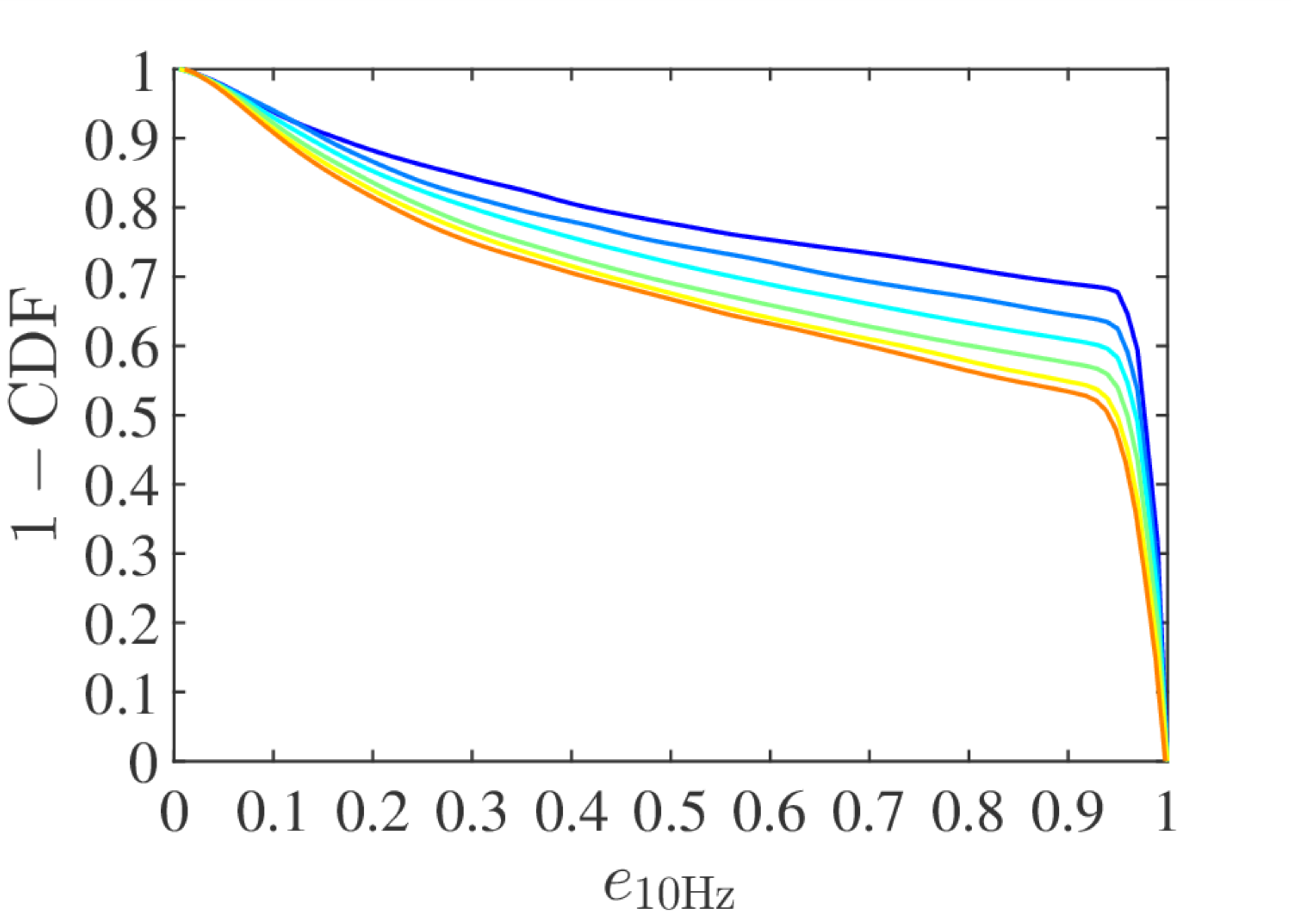}
    \includegraphics[width=75mm]{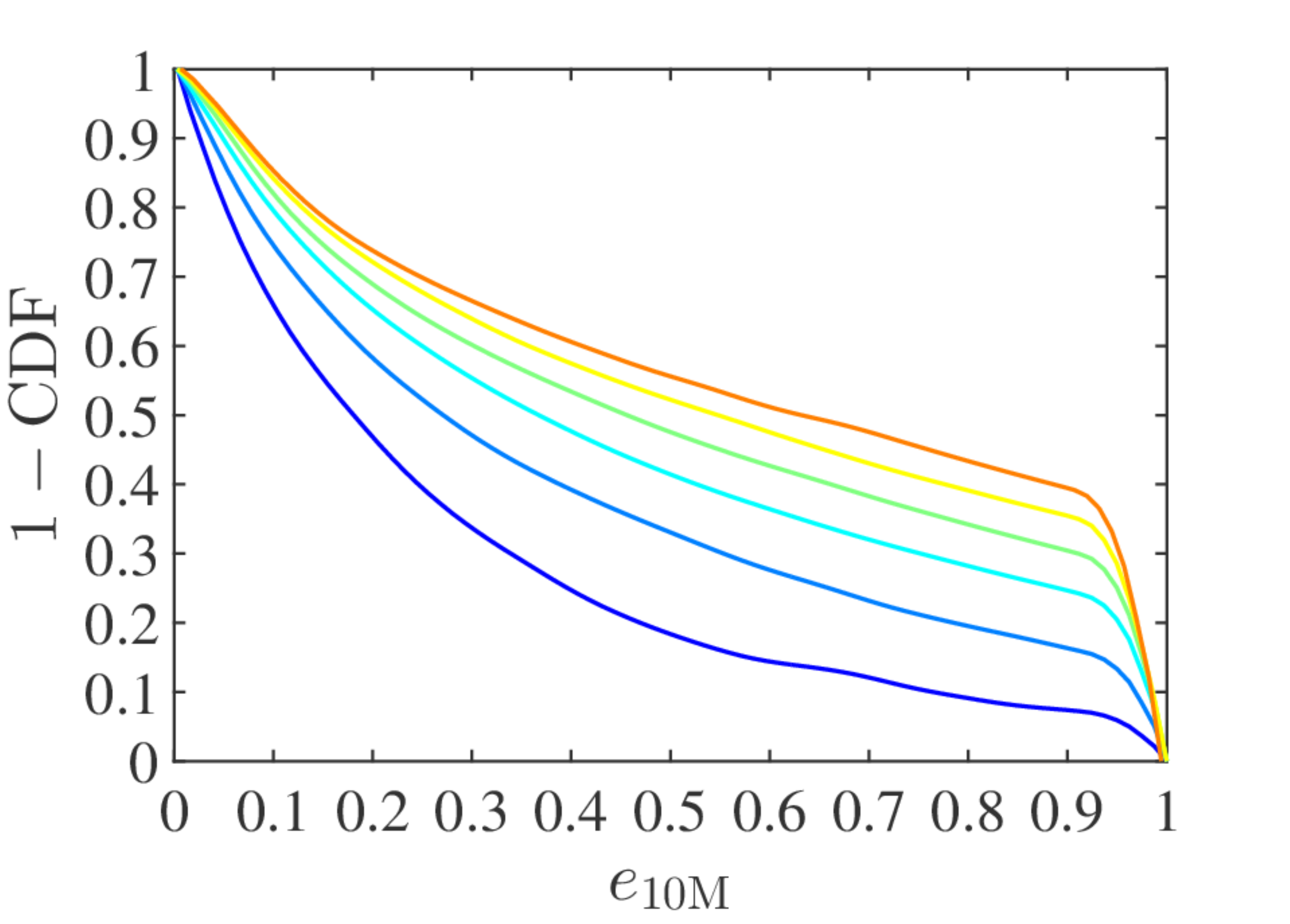}
    \\
    \includegraphics[width=75mm]{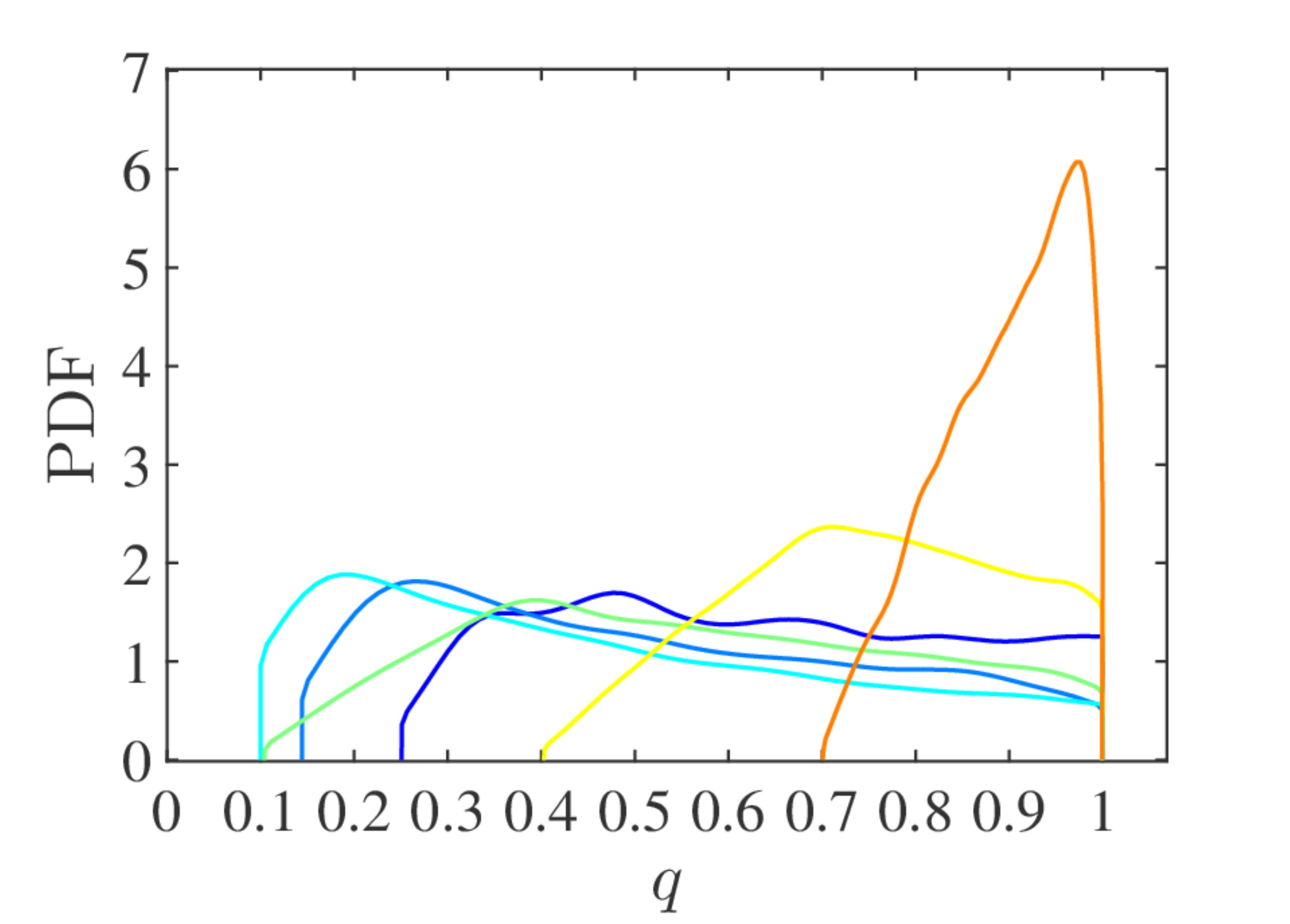}
  \caption{Distributions of binary parameter in various mass ranges in the fiducial model ($\beta = 2$ and $p_0 = 0.5$) for aLIGO detections as in Figure \ref{fig:ParamDistGWcBBHs_Volume_Limits}, but showing the initial dimensionless pericenter distance $\rho_{\rm p0}$ (row 1 left), redshift $z$ (row 1, right), residual eccentricity $e_{\rm 10 Hz}$ (row 2, left), and residual eccentricity $e_{\rm 10 M}$ (row 2, right). The panels in the 3rd row shows $1-$CDF corresponding to the row 2 panels, the fraction of detected sources higher than $e$. The last row accounts for the mass ratio $q$.
  \label{fig:ParamDistGWcBBHs_MassRanges} }
\end{figure*}

\subsection{Distributions of binary parameters for Advanced LIGO detections}
\label{subsec:DistaLIGOdet}
 
 The distributions of physical parameters as seen by aLIGO are presented in Figure \ref{fig:ParamDistGWcBBHs_Volume_Limits}, and the impact of observational bias on the parameter distributions are illustrated in Figure \ref{fig:Impact_ObsBias_ParamDists}. As seen, the results are significantly different from the detector-independent merger rate density in the local Universe.
 \begin{itemize}
     \item The $\rho_{\rm p_0}$ distribution is tilted toward lower values in comparison to the distributions of mergers per unit volume in the local Universe, and peaks between $\rho_{\rm p_0} \sim 7 - 15$ depending on the level of mass segregation and the underlying BH mass function. The reason for the tilt is that binaries with lower $\rho_{\rm p0}$ are generally detected to larger distances (Figure \ref{fig:HorizonDistGWcBBHs}).
     
     \item The $e_0$ distribution is also skewed toward lower values compared to that for the local Universe, which arises from the fact that observational bias favors high-mass binaries with low $\rho_{\rm p0}$ that form with lower $e_0$ in single GNs. However, $e_0$ remains high ($0.95 \lesssim$) for the detectable population.
     
     \item The eccentricity distribution at $10 \, \Hz$ is weakly sensitive to the observational bias. We find that more than $92\%$ have $e_{\rm 10 Hz} > 0.1$ independently of the BH mass function and the mass--segregation parameters and $\sim 60 - 71 \%$ of detectable GW capture BBH events have $e_{\rm 10 Hz} >0.8$ for $0.5 \leqslant p_0 \leqslant 0.9$ and the considered $\mathcal{F}_{\rm BH}$ models. Note that the corresponding fractions are $99 \%$ and $76 \%$ for an extreme mass-segregated model with $p_0 = 1.5$ and $ \beta = 2$, as for the young stars in the Galactic Center (Appendix \ref{subsec:GNs_BHPops}). Similarly, more than $93 \% $ ($\sim 66-79 \%$) and $93 \%$ \mbox{($\sim 62-73 \%$)} of detectable binaries have $e_{\rm 10 Hz} > 0.1$ ($e_{\rm 10 Hz} > 0.8$) as seen by AdV and KAGRA, respectively. Furthermore, these fractions are $99 \%$ ($85 \%$) and $98 \%$ ($79 \%$) for the considered extreme mass-segregated model for AdV and KAGRA, respectively.
     
     \item The $e_{\rm 10 M}$ distribution is skewed toward higher eccentricities due to observational bias, which originates from the one-to-one correspondence between $\rho_{\rm p0}$ and $e_{\rm 10 M}$ (Section \ref{subsec:Evolution}). As seen in Figure \ref{fig:ParamDistGWcBBHs_Volume_Limits}, $e_{\rm 10 M}> 0.1$ and \mbox{$e_{\rm 10 M} > 0.8$} has a probability of $\sim 75 - 95\%$ ($98 \%$) and $\sim 23 - 56 \%$ ($63 \%$), respectively, for $0.5 \leqslant p_0 \leqslant 0.9$ and the considered $\mathcal{F}_{\rm BH}$ models (when $p_0 = 1.5$ and $ \beta = 2$). Similarly, AdV and KAGRA detect $\sim 74 - 96 \%$ ($\sim 27 - 67 \%$) and $\sim 71 - 93 \%$ ($\sim 24 - 62 \%$) of mergers with $e_{\rm 10 M}> 0.1$ ($e_{\rm 10 M} > 0.8$), respectively, while these fractions are $74 \%$ ($99 \%$) and $69 \%$ ($98 \%$) for the considered extreme mass-segregated model, respectively.
     
     \item The $M_{\rm tot}$ and $\mathcal{M}$ distributions are skewed to higher masses in comparison to the distributions of mergers per unit volume in the local Universe due to the larger detection distance (Figure \ref{fig:HorizonDistGWcBBHs}). For instance, in the case of standard mass segregation ($p_0 = 0.5$), $P(M_{\rm tot})$ peaks at $M_{\rm tot}\sim 3m_{\rm BH, min}$ and at $1 - 1.6\,m_{\rm BH\max}$ for an underlying BH mass function of $m_{\rm BH}^{-3}$ and $m_{\rm BH}^{-1}$ for the mergers per unit volume in the local Universe (e.g. Figure \ref{fig:Impact_ObsBias_ParamDists}), respectively, while the corresponding detected distributions peak at $0.8 \, m_{\rm BH, max}$ and \mbox{$\sim 1.6\,m_{\rm BH, max} \, \Msun$} (Figure \ref{fig:ParamDistGWcBBHs_Volume_Limits}), respectively.\footnote{These results are based on MC simulations with $m_{\rm BH, max} = 50 \, \Msun$ and $100 \, \Msun$ cases.} The detected distributions and distributions of mergers per unit volume in the local Universe are practically the same in the strongly mass-segregated limit ($p_0 \gtrsim 0.8$). The reason is that the BBH population principally consists of those systems that are preferentially selected by the observational selection effect.
     
     \item The mass ratio distribution $P(q)$ is weakly skewed toward equal masses, which arises from the fact that observational bias preferentially selects binaries with mass ratios closer to unity. Furthermore, $P(q)$ is closer to a uniform distribution for $p_0 \simeq 0.5 - 0.7$.
     
     \item We find that the $z$ distribution of detectable mergers $P(z)$ is not very sensitive to $p_0$ and the BH mass function. The vast majority of binaries form with $z \lesssim 0.7$, and $P(z)$ peaks between $\sim 0.15$ and $\sim 0.3$ depending on $p_0$ and $\mathcal{F}_{\rm BH}$. As a consequence, the detection completeness is also not very sensitive to these parameters. One example is shown for the detection completeness together with the intrinsic distribution of GNs in the detection volume in Figure \ref{fig:PDF_Dlum}.
 \end{itemize}
 
 Comparing these results with that obtained for other eccentric formation channels, we conclude that the single--single GW capture BBH merger channel in GNs produces the largest fraction of binaries that form in the aLIGO/AdV/KAGRA band with $e_{\rm 10 Hz} > 0.1$ and the highest fraction of highly eccentric merging binaries.
 
 The GW capture process and mass segregation lead to relations between various physical parameters of the merging binaries in single GNs, e.g. mergers with higher masses or higher mass ratios have lower $\rho_{\rm p0}$ and $e_0$ (see \citealt{Gondanetal2018a}, Figure 4  and Equations 49-53 therein). Observational bias modifies the relations between $\{ \rho_{\rm p0}, q, M_{\rm tot}, D_{\rm L} \}$ since it preferentially selects more massive binaries with low $\rho_{\rm p0}$ and equal masses (Figure \ref{fig:HorizonDistGWcBBHs}). To understand these trends, we present the distributions of several binary parameters in various mass ranges for aLIGO detections in Figure \ref{fig:ParamDistGWcBBHs_MassRanges}. The negative correlation between $\rho_{\rm p0}$ and $M_{\rm tot}$ identified for merging populations in single GNs \citep{Gondanetal2018a} also holds for future aLIGO/AdV/KAGRA detections. In addition, this indicates a positive correlation between $e_{\rm 10 M}$ and $M_{\rm tot}$ for aLIGO/AdV/KAGRA detections due to the one-to-one correspondence between $\rho_{\rm p0}$ and $e_{\rm 10 M}$ (Section \ref{subsec:Evolution}). We also find a positive correlation between $z$ (or equivalently $D_{\rm L}$) and $M_{\rm tot}$ since more massive binaries are detected to larger distances (Figure \ref{fig:HorizonDistGWcBBHs}). Remarkably, a negative correlation is obtained between $e_{\rm 10 Hz}$ and $M_{\rm tot}$ for detectable populations, while this cannot be observed in the case of single GNs \citep{Gondanetal2018a}. We investigate possible correlations among various binary parameters systematically in the next subsection.

\setlength{\tabcolsep}{3.5pt}
 
\begin{table*}
\centering  
   \begin{tabular}{@{}cc|ccccccccc}
     \multicolumn{11}{c}{Spearman's Rank Correlation Coefficients} \\
     \hline
      $p_0$ & $\mathcal{F}_{\rm BH}$ & $\rho_{\rm p0} - M_{\rm tot}$ & $\rho_{\rm p0} - M_{\rm tot,z}$ & $\rho_{\rm p0} - q$ & $M_{\rm tot} - q$ & $M_{\rm tot,z} - q$ & $\rho_{\rm p0} - D_{\rm L}$ & $M_{\rm tot} - D_{\rm L}$ & $M_{\rm tot,z} - D_{\rm L}$ & $q - D_{\rm L}$ \\
     \hline\hline
     $0.5$ & $\beta = 1$ & $-0.154$ & $-0.203$ & $-0.080$ & $0.631$ & $0.581$ & $-0.176$ & $0.135$ & $0.499$ & $0.124$  \\
     $0.5$ & $\beta = 2$ & $-0.189$ & $-0.214$ & $-0.053$ & $0.491$ & $0.473$ & $-0.146$ & $0.255$ & $0.462$ & $0.131$  \\
     $0.5$ & $\beta = 3$ & $-0.262$ & $-0.272$ & $-0.036$ & $0.385$ & $0.346$ & $-0.135$ & $0.374$ & $0.434$ & $0.265$  \\
     $0.5$ &     B16     & $-0.159$ & $-0.168$ & $-0.029$ & $0.422$ & $0.405$ & $-0.119$ & $0.184$ & $0.474$ & $0.075$  \\
     $0.7$ & $\beta = 1$ & $-0.102$ & $-0.187$ & $-0.069$ & $0.593$ & $0.555$ & $-0.202$ & $0.054$ & $0.610$ & $0.048$  \\
     $0.7$ & $\beta = 2$ & $-0.151$ & $-0.201$ & $-0.091$ & $0.616$ & $0.641$ & $-0.187$ & $0.146$ & $0.528$ & $0.079$  \\
     $0.7$ & $\beta = 3$ & $-0.232$ & $-0.243$ & $-0.060$ & $0.497$ & $0.456$ & $-0.179$ & $0.251$ & $0.519$ & $0.209$  \\
     $0.7$ &     B16     & $-0.131$ & $-0.143$ & $-0.051$ & $0.488$ & $0.433$ & $-0.121$ & $0.165$ & $0.504$ & $0.049$  \\
     $0.9$ & $\beta = 1$ & $-0.059$ & $-0.151$ & $-0.037$ & $0.466$ & $0.412$ & $-0.215$ & $0.010$ & $0.771$ & $0.012$  \\
     $0.9$ & $\beta = 2$ & $-0.081$ & $-0.176$ & $-0.064$ & $0.517$ & $0.496$ & $-0.191$ & $0.021$ & $0.692$ & $0.026$  \\
     $0.9$ & $\beta = 3$ & $-0.126$ & $-0.221$ & $-0.086$ & $0.552$ & $0.524$ & $-0.183$ & $0.120$ & $0.603$ & $0.106$  \\
     $0.9$ &     B16     & $-0.085$ & $-0.114$ & $-0.061$ & $0.543$ & $0.526$ & $-0.154$ & $0.057$ & $0.651$ & $0.038$  \\
     $1.5$ & $\beta = 2$ & $-0.040$ & $-0.027$ & $-0.020$ & $0.282$ & $0.255$ & $-0.204$ & $0.013$ & $0.783$ & $0.011$  \\
   \end{tabular} 
   \caption{Spearman's Rank Correlation Coefficients for GN models with different BH mass functions ($m_{\rm BH}^{-\beta}$ and B16) and different mass--segregation parameters $p_0$ (Equations \ref{eq:n(r)app} and \ref{eq:pmbh}) between the measurable parameters $\{ \rho_{\rm p0}, q, M_{\rm tot}, M_{\rm tot,z}, D_{\rm L} \}$ for binaries with $\rm S/N > 8$ for the inspiral phase with aLIGO at design sensitivity.  \label{tab:correlation1}}
\end{table*}

\begin{table*}
\centering  
   \begin{tabular}{@{}cc|cccccccccc}
     \multicolumn{12}{c}{Spearman's Rank Correlation Coefficients} \\
     \hline
      $p_0$ & $\mathcal{F}_{\rm BH}$ & $e_{\rm 10 Hz} - \rho_{\rm p0}$ & $e_{\rm 10 Hz} - M_{\rm tot}$ & $e_{\rm 10 Hz} - M_{\rm tot,z}$ & $e_{\rm 10 Hz} - q$ & $e_{\rm 10 Hz} - D_{\rm L}$ & $e_{\rm 10 M} - \rho_{\rm p0}$ & $e_{\rm 10 M} - M_{\rm tot}$ & $e_{\rm 10 M} - M_{\rm tot,z}$ & $e_{\rm 10 M} - q$ & $e_{\rm 10 M} - D_{\rm L}$  \\
     \hline\hline
     $0.5$ & $\beta = 1$ & $-0.638$ & $-0.164$ & $-0.125$ & $-0.131$ & $-0.028$ & $-0.942$ & $0.121$ & $0.155$ & $0.050$ & $0.177$  \\
     $0.5$ & $\beta = 2$ & $-0.594$ & $-0.230$ & $-0.212$ & $-0.187$ & $-0.035$ & $-0.966$ & $0.160$ & $0.179$ & $0.024$ & $0.145$  \\
     $0.5$ & $\beta = 3$ & $-0.522$ & $-0.258$ & $-0.250$ & $-0.226$ & $-0.085$ & $-0.979$ & $0.243$ & $0.268$ & $-0.012$ & $0.139$  \\
     $0.5$ &     B16     & $-0.570$ & $-0.187$ & $-0.174$ & $-0.028$ & $-0.138$ & $-0.977$ & $0.142$ & $0.167$ & $0.039$ & $0.111$  \\
     $0.7$ & $\beta = 1$ & $-0.572$ & $-0.123$ & $-0.052$ & $-0.099$ & $0.038$ & $-0.929$ & $0.069$ & $0.160$ & $0.038$ & $0.206$  \\
     $0.7$ & $\beta = 2$ & $-0.546$ & $-0.205$ & $-0.163$ & $-0.170$ & $0.015$ & $-0.940$ & $0.108$ & $0.181$ & $0.011$ & $0.185$  \\
     $0.7$ & $\beta = 3$ & $-0.468$ & $-0.219$ & $-0.194$ & $-0.206$ & $-0.032$ & $-0.965$ & $0.209$ & $0.281$ & $-0.024$ & $0.174$  \\
     $0.7$ &     B16     & $-0.498$ & $-0.155$ & $-0.132$ & $-0.110$ & $-0.016$ & $-0.944$ & $0.124$ & $0.162$ & $0.025$ & $0.128$  \\
     $0.9$ & $\beta = 1$ & $-0.472$ & $-0.066$ & $-0.034$ & $-0.054$ & $0.070$ & $-0.918$ & $0.035$ & $0.187$ & $0.018$ & $0.219$  \\
     $0.9$ & $\beta = 2$ & $-0.446$ & $-0.093$ & $-0.057$ & $-0.070$ & $0.047$ & $-0.927$ & $0.049$ & $0.194$ & $-0.017$ & $0.198$  \\
     $0.9$ & $\beta = 3$ & $-0.412$ & $-0.201$ & $-0.144$ & $-0.168$ & $0.017$ & $-0.934$ & $0.079$ & $0.302$ & $-0.041$ & $0.186$  \\
     $0.9$ &     B16     & $-0.455$ & $-0.124$ & $-0.084$ & $-0.095$ & $0.014$ & $-0.921$ & $0.060$ & $0.154$ & $0.013$ & $0.155$  \\
     $1.5$ & $\beta = 2$ & $-0.341$ & $-0.041$ & $-0.032$ & $-0.036$ & $0.025$ & $-0.892$ & $0.017$ & $0.202$ & $-0.010$ & $0.205$  \\
   \end{tabular} 
   \caption{Spearman's Rank Correlation Coefficients for different mass functions and mass--segregation parameters as in Table \ref{tab:correlation1} but for the eccentricity parameters. \label{tab:correlation2} }
\end{table*}

\subsection{Correlations among binary parameters}
\label{subsec:CorrBBHparams}
 
 Here we investigate possible correlations among various measurable parameters describing the GW capture sources in GNs, including mass-dependent parameters together with initial orbital parameters and $\{ D_{\rm L}, e_{\rm 10 Hz}, \, e_{\rm 10 M} \}$.
 
 We measure the strength and direction of the possible monotonic association between binary parameters $X$ and $Y$ using the Spearman correlation coefficient test \citep{Spearman1904} since this test does not carry any assumptions about the distribution of the data. Spearman's rank correlation coefficient $r_{\rm S}$ is defined as the Pearson correlation coefficient (i.e., the degree of the relationship between linearly related parameters) between the rank variables ${\rm rg}_X$ and ${\rm rg}_Y$ as
\begin{equation}   \label{eq:rS}
   r_{\rm S} = \frac{ {\rm cov}({\rm rg}_X , {\rm rg}_Y) }{ \sigma_{{\rm rg}_X} \sigma_{{\rm rg}_Y} } \, 
\end{equation}
 where ${\rm cov}({\rm rg}_X , {\rm rg}_Y)$ is the covariance of the rank variables, and $\sigma_{{\rm rg}_X}$ and $\sigma_{{\rm rg}_Y}$ are the standard deviations of the rank variables. For a sample size of $N_{\rm samp}$, the data sets $X_i$ and $Y_i$ are converted to ranks ${\rm rg}_{X_i}$ and ${\rm rg}_{Y_i}$ as follows. For the binary parameter $X$ (or $Y$), we assign a rank of 1 to the highest value, a rank of $N_{\rm samp}$ to the lowest value, and any other intermediate values of $X_i$ are ranked in ascending order between 1 and $N_{\rm samp}$. In case of identical values in $X_i$ or in $Y_i$, we take the average of the ranks that they would have otherwise occupied because there is no way of knowing which element should be ranked with a higher index. $r_S$ can take values from $-1$ to $+1$, where $r_{\rm S} = 0$ indicates that there is no tendency for $Y$ to either increase or decrease when $X$ increases, while $r_{\rm S} = 1$ ($r_{\rm S} = -1$) means that any two pairs of data values $( X_i, Y_i )$ and $(X_j, Y_j )$ that $X_i - X_j$ and $Y_i - Y_j$ always have the same (opposite) sign. Furthermore, a positive (negative) value corresponds to an increasing (decreasing) monotonic trend between $X$ and $Y$; and the closer $r_{\rm S}$ is to zero, the weaker the monotonic trend between $X$ and $Y$.
 
 For any two binary parameters $X$ and $Y$, we investigate whether the obtained $r_{\rm S}$ value is convergent. As the generated mock catalogs contain $N_{\rm samp} \approx 1.5 \times 10^5 - 3.3 \times 10^5$ binaries (Appendix \ref{subsec:SetupMC}), we randomly draw $k \times 10^3$ pairs from the paired data $\{ (X_1, Y_1), \ldots , (X_{N_{\rm samp}}, Y_{N_{\rm samp}}) \}$ with $k = \{1, 2, \ldots , [N_{\rm samp} / 10^3] \} $, where the bracket $[ \, ]$ denotes the floor function. We compute the Spearman's rank correlation coefficient for each paired data sample $r_{\rm S,k}$, then use the Cauchy criterion to test the convergence of the sequence $\{ r_{\rm S,k} \}$. For each $r_{\rm S}$, we also carry out a statistical null-hypothesis test using Student's t-distribution\footnote{We use Student's t-distribution as it approximates well the distribution of $r_{\rm S}$ in the zero-correlation case for large sample sizes.} to construct the p-value with the null hypothesis statement that there is no monotonic association between the two binary parameters $X$ and $Y$. For the $r_{\rm S}$ values presented in Tables \ref{tab:correlation1} and \ref{tab:correlation2}, the corresponding $r_{\rm S,k}$ sequences are convergent and the obtained p-values are below $10^{-8}$. Note that $r_{\rm S}$ between a binary parameter $X$ and the parameters $q$ and $\eta$ are the same because there is a one-to-one correspondence between $q$ and $\eta$.
 
 Tables \ref{tab:correlation1} and \ref{tab:correlation2} show the correlation coefficients between various parameters for detections with a single aLIGO detector. We identify the following correlations.
 \begin{itemize}
     \item The MC analysis shows that $e_{\rm 10 M}$ and $\rho_{\rm p0}$ are perfectly anticorrelated, as expected, since there is a one-to-one correspondence between these parameters (Section \ref{subsec:Evolution}). Furthermore, we also identify a strong anticorrelation between $e_{\rm 10 Hz}$ and $\rho_{\rm p0}$ because binaries with higher $\rho_{\rm p0}$ systematically enter a given frequency band with lower residual eccentricity (Equation \ref{eq:f_GW}).
     
     \item Initial dimensionless pericenter distance $\rho_{\rm p0}$ and $M_{\rm tot}$ are anticorrelated as are $e_{\rm 10 Hz}$ and $M_{\rm tot}$, while $e_{\rm 10M}$ and $M_{\rm tot}$ are correlated, which are also apparent in Figure \ref{fig:ParamDistGWcBBHs_MassRanges}. We find the same trends for other mass-dependent parameters such as $\mathcal{M}$ and $\mu$ and the corresponding redshifted masses as well.
     
     \item Source distance $D_{\rm L}$ and $M_{\rm tot}$ are correlated, since $D_{\rm hor}$ is higher for higher masses up to \mbox{$M_{\rm tot,z} \sim 90-250 \, \Msun$} as shown in Figure \ref{fig:HorizonDistGWcBBHs}. Similarly, $\mathcal{M}$ and $\mu$ are also correlated with $D_{\rm L}$ from the same reason such as the redshifted mass parameters.
     
     \item There is a strong correlation between $q$ and $M_{\rm tot}$ ($\mathcal{M}$ and $\mu$) meaning that the mass ratio is closer to unity for heavier BBHs.
 \end{itemize}
 We find the same correlations for AdV and KAGRA detections.
 
 The exceptionally strong correlation between $e_{\rm 10M}$ and $\rho_{\rm p0}$ imply that the measurement of $e_{\rm 10M}$ may be used to infer $\rho_{\rm p0}$, which is useful to put constraints on models of the host environment of the source, particularly its velocity dispersion or escape speed; see Section \ref{subsec:VelDist_Host} for details. Furthermore, because of the obtained strong correlation between $e_{\rm 10 Hz}$ and $\rho_{\rm p0}$, $e_{\rm 10 Hz}$ may also be used to constrain the host environment as shown in Section \ref{subsec:VelDist_Host}.
 
 The correlation between $e_{\rm 10M}$ and mass parameters imply that heavier BHs reside in higher velocity dispersion environments, which arises due to mass segregation. Thus, a correlation between $e_{\rm 10M}$ and mass parameters are a smoking gun signature of sources in GNs.
 
 Finally, we put an order of magnitude estimate on the measurement accuracy of $e_{\rm 10M}$, $\Delta e_{\rm 10 M}$, in future observations based on \citet{GondanKocsis2019}, who among others determined the measurement accuracy of orbital eccentricity at several stages of the binary evolution for the aLIGO--AdV--KAGRA detector network at design sensitivity for the eccentric inspiral phase. We estimate $\Delta e_{\rm 10 M}$ for two scenarios: when BBHs form below $\rho_{\rm p0} = 10$ and in this case $e_{\rm 10 M}: = e_0$ by definition (i.e. $e_{\rm 10 M} \gtrsim 0.95$; Section \ref{subsec:Evolution}), and when BBHs reach $\rho_{\rm p} = 10$ with $e_{\rm 10 M}$. In the former scenario, $\Delta e_{\rm 10 M}$ is in the range $(10^{-3} - 10^{-2}) \times (100 \, {\rm Mpc} / D_{\rm L} )$ depending on $(m_A, m_B)$ and $\rho_{\rm p0}$, which based on Figure 5 in \citet{GondanKocsis2019} and the facts that the vast majority of BBHs form with $\rho_{\rm p0} \lesssim 100$ and that the most massive BBHs form with a few tens of $\rho_{\rm p0}$ (Section \ref{subsec:DistaLIGOdet}). This result holds for $\sim 20 - 50 \%$ of detectable mergers depending on $p_0$ and $\mathcal{F}_{\rm BH}$, and for $\sim 60 \%$ of detectable systems for the considered extreme mass-segregated model (Figure \ref{fig:ParamDistGWcBBHs_Volume_Limits}). In the later scenario, we estimate $\Delta e_{\rm 10 M}$ as follows. First, we note that the measurement error of eccentricity gradually decreases as the orbit shrinks \citep{GondanKocsis2019} and that $e_{\rm 10 M} \lesssim e_{\rm 10 Hz}$, thus $\Delta e_{\rm 10 M} \lesssim \Delta e_{\rm 10 Hz}$. Therefore, we estimate the measurement error of $e_{\rm 10 M}$ with that of $e_{\rm 10 Hz}$ because $\Delta e_{\rm 10 M}$ has not yet been investigated in the literature. We find that the vast majority of detectable BBHs have $0.05 \lesssim e_{\rm 10 Hz} \lesssim 0.95$ (Figure \ref{fig:ParamDistGWcBBHs_Volume_Limits}),  accordingly, the corresponding $\Delta e_{\rm 10 Hz}$ and thereby $\Delta e_{\rm 10 M}$ are in the range $(10^{-4} - 10^{-2}) \times (100 \, {\rm Mpc} / D_{\rm L} )$, which mainly depend on $e_{\rm 10 Hz}$ and secondly on the component masses; Figure 9 in \citet{GondanKocsis2019}. This result holds for the remaining detectable population, i.e. $\sim 50 - 20 \%$ depending on $\{ p_0, \mathcal{F}_{\rm BH} \}$ and $\sim 40 \%$ for the considered extreme mass-segregated model (Figure \ref{fig:ParamDistGWcBBHs_Volume_Limits}). In summary, $\Delta e_{\rm 10 M}$ is at the level of $(10^{-4} - 10^{-2}) \times (100 \, {\rm Mpc} / D_{\rm L} )$. Note that the reconstruction accuracy of $e_{\rm 10 M}$ may improve significantly by taking into account the merger and ringdown phases, similar to the case of quasi-circular binaries (e.g., \citealt{AjithBose2009}).

\subsection{Escape speed of the host environment based on eccentricity}
\label{subsec:VelDist_Host}

 For single--single GW capture BBH events at a given location, $\rho_{\rm p0}$ is uniformly distributed between 0 and  $\rho_{\rm p0, \max}$, where
 \begin{equation}\label{eq:rhop0max_vesc}
     \rho_{\rm p0, \max} = 
 \left(\frac{85\pi\eta}{24 \sqrt{2}}\right)^{2/7} v_{\rm esc}^{-4/7} 
 \simeq 118 \, (4\eta)^{2/7} \left(\frac{v_{\rm esc}}{100\,\rm km/s}\right)^{-4/7} \, ,
 \end{equation}
 and $v_{\rm esc}$ is the escape speed \citep{OLearyetal2009}. $v_{\rm esc} = \sqrt{2}\sigma$ in a Keplerian potential, where $\sigma$ is the circular velocity or the velocity dispersion. Similarly, for GW captures during strong binary--single interactions \citep{Rodriguezetal2018,Rodriguezetal2018b,Samsing2018}, $\rho_{{\rm p}0}$ is also expected to be approximately uniformly distributed accoridng to Equation \eqref{eq:rhop0max_vesc}, but here $v_{\rm esc}$ is replaced by the velocity of the binary components preceding the encounter $v_{\rm orb}$, 
\begin{equation} \label{eq:rhop0-v}
    \rho_{\mathrm{p}0,\max} = \left(\frac{85\pi\eta}{24 \sqrt{2}}\right)^{2/7} v_{\rm orb}^{-4/7} \simeq 118 \, (4\eta)^{2/7} \left(\frac{v_{\rm orb}}{100\,\rm km/s}\right)^{-4/7} \, .
\end{equation}
 However, $v_{\rm orb}$ cannot be arbitrary. In globular clusters, it is typically larger than the value for the hard-soft boundary, and smaller than $\sqrt{ 24 }\, v_{\rm esc} $ \citep{Samsing2018}, because otherwise binary-single interactions would have either disrupted the binary or ejected it from the cluster. Eccentric mergers are most common among binaries with the maximum orbital velocities, i.e when $v_{\rm orb} = \sqrt{ 24 }\, v_{\rm esc} $\,. In the following, we will use Equation \eqref{eq:rhop0max_vesc} to express $v_{\rm esc}$ for single-single scattering for given orbital eccentricity, understanding that similar results hold for binary-single interaction sources if replacing $v_{\rm esc}\rightarrow \sqrt{24}v_{\rm esc}$.
 
 To find the maximum velocity for a given $\rho_{\rm p0}$, $v_{\rm esc}$ can be expressed by $\eta$ and $\rho_{\mathrm{p}0}$ as
\begin{equation}  \label{eq:v-rhop0eta}
   v_{\rm esc} = 
    \left(\frac{85\pi\eta}{24 \sqrt{2}}\right)^{1/2} \rho_{\mathrm{p}0}^{-7/4}
   \simeq
   133 \, \rm km/s \times (4\eta)^{1/2} \left(\frac{\rho_{\mathrm{p}0}}{100}\right)^{-7/4}  \, .
\end{equation} 
 Note that the parameters $\{\rho_{\rm p0}, \eta, v_{\rm esc} \}$ may be measured for eccentric systems with the aLIGO--AdV--KAGRA detector network at design sensitivity \citep{Gondanetal2018a,GondanKocsis2019}.
 
 To set a relation between $v_{\rm esc}$ and the eccentricities $e_{\rm 10M}$ and $e_{\rm 10Hz}$, we use the leading-order orbital evolution equation \citep{Peters1964}
\begin{equation}\label{eq:rhoe}
 \rho_\mathrm{p} (e) = \frac{c_0}{M_\mathrm{tot}}\frac{ e^{12/19}}{ 1+e }
 \left( 1+ \frac{ 121 }{ 304 } e^2 \right)^{ \frac{870}{2299} } \, ,
\end{equation}
 where $c_0/ M_{\rm tot}$ may be expressed with $e_0$ and $\rho_{\rm p0}$, by solving Equation \eqref{eq:rhoe} for $\rho_{\rm p} = \rho_{\rm p0}$ and $e = e_0 \sim 1$.\footnote{Both single--single GW capture BBHs (e.g., Section \ref{subsec:ResMC_SingleGN}) and BBHs forming through binary--single interactions (e.g., \citealt{Samsingetal2014}) are expected to form with $e_0 \sim 1$.} For further use, we rewrite this equation to a more appropriate form as
\begin{equation}  \label{eq:rho_ee0rhop0}
  \rho_{\rm p} = \rho_{\rm p0} \frac{ h(e) }{ h(1) } \, ,  \quad h(x) = \frac{ x^{12/19} }{ 1 + x } \left( 1 + \frac{121}{304} x^2 \right)^{ \frac{870}{2299} } \, .
\end{equation}
 The relation between $v_{\rm esc}$ and $e_{\rm 10M}$ can be given by first setting $\rho_{\rm p} = 10$ and $e = e_{\rm 10 M}$ in Equation \eqref{eq:rho_ee0rhop0}, then substituting $\rho_{\rm p0}$ by Equation \eqref{eq:rhop0-v}, and finally expressing $v_{\rm esc}$ in terms of the remaining parameters as
 \begin{align}    \label{eq:vesc_e10M}
  v_{\rm esc} & = \left( \frac{ 85\pi \eta }{ 24\sqrt{2} } \right)^{1/2} \rho_{\rm p}^{-7/4} \left(\frac{h(e)}{h(1)}\right)^{7/4} \nonumber\\
  & \simeq 2 \times 10^4 \, \rm km/s \times \,(4\eta)^{1/2} [h(e_{\rm 10 M})]^{7/4}\, ,
\end{align}
 where $[h(e_{\rm 10 M})]^{7/4} \approx e_{\rm 10 M}^{21/19}$ for $e_{\rm 10 M} \ll 1$. For $e_{\rm 10 Hz}$, we first substitute Equation \eqref{eq:rho_ee0rhop0} into Equation \eqref{eq:f_GW}, set $e = e_{\rm 10 Hz}$ and $f_{\rm GW} = 10\,\mathrm{Hz}$, then substitute $\rho_{\rm p0}$ by Equation \eqref{eq:rhop0-v}. Finally, we express $v_{\rm esc}$ in terms of the remaining parameters as
 \begin{align}  \label{eq:vesc_e10Hz}
  v_{\rm esc} & = \left(\frac{85\pi \eta}{24\sqrt{2}}\right)^{1/2} [(1+e)^{0.3046}\pi M_{\rm tot,z} f_{\rm GW}]^{7/6} \left(\frac{h(e)}{h(1)} \right)^{7/4} \nonumber \\ & \simeq 596.6 \, \rm km/s \times (4\eta)^{1/2} \left(\frac{M_{\rm tot,z}}{10\Msun}\right)^{7/6} 
  g(e_{\rm 10 \Hz})\, ,
\end{align}
 where $g(e_{\rm 10 Hz}) = [h(e_{\rm 10 Hz})]^{7/4} (1 + e_{\rm 10 Hz})^{0.3554}$. Similar to the case of $v_{\rm esc}(e_{\rm 10 M})$, $g(e_{\rm 10 Hz}) \approx e_{\rm 10 Hz}^{21/19}$ for $e_{\rm 10 Hz} \ll 1$.

\begin{figure}
    \centering
    \includegraphics[width=84mm]{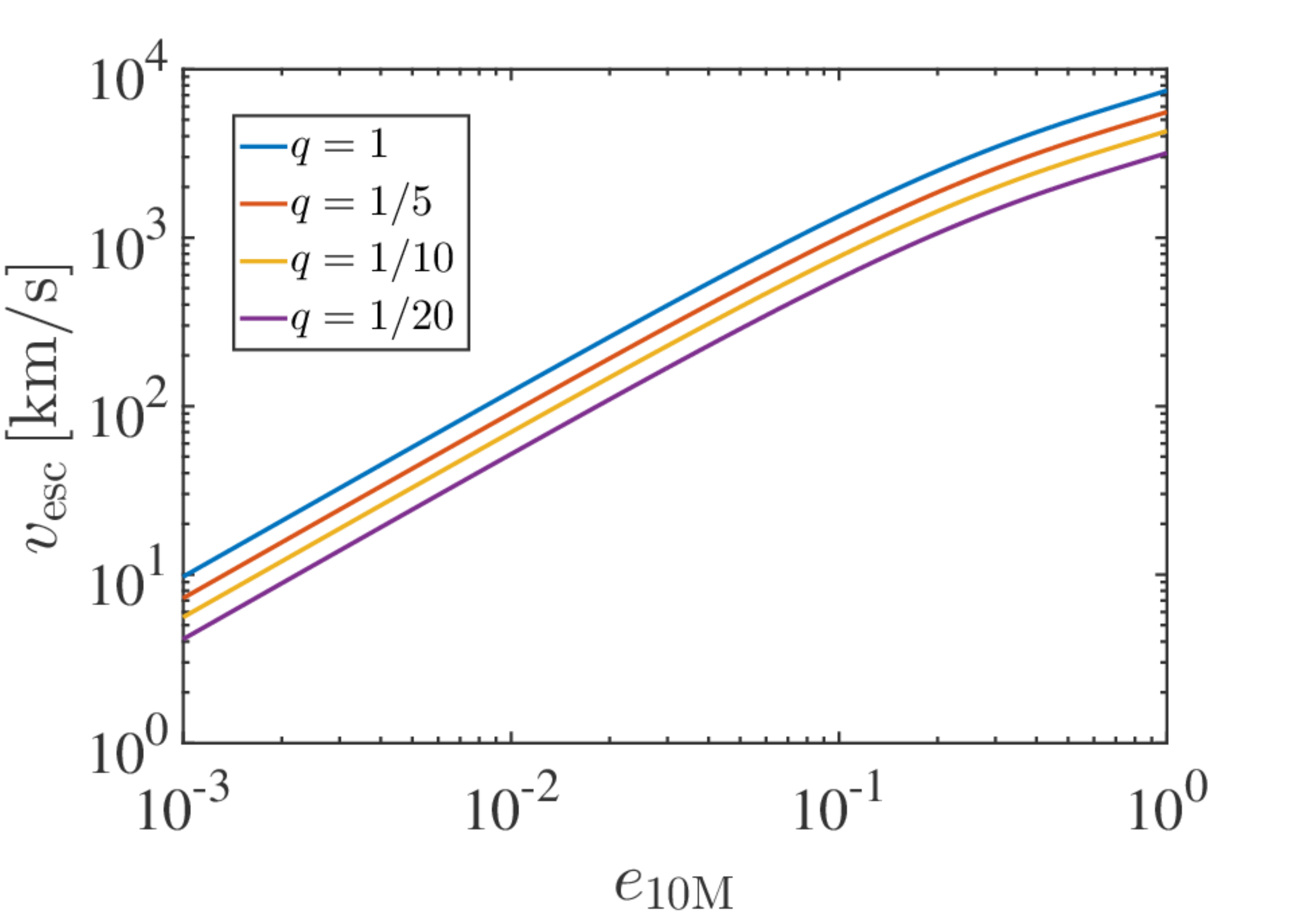}
    \\
    \includegraphics[width=84mm]{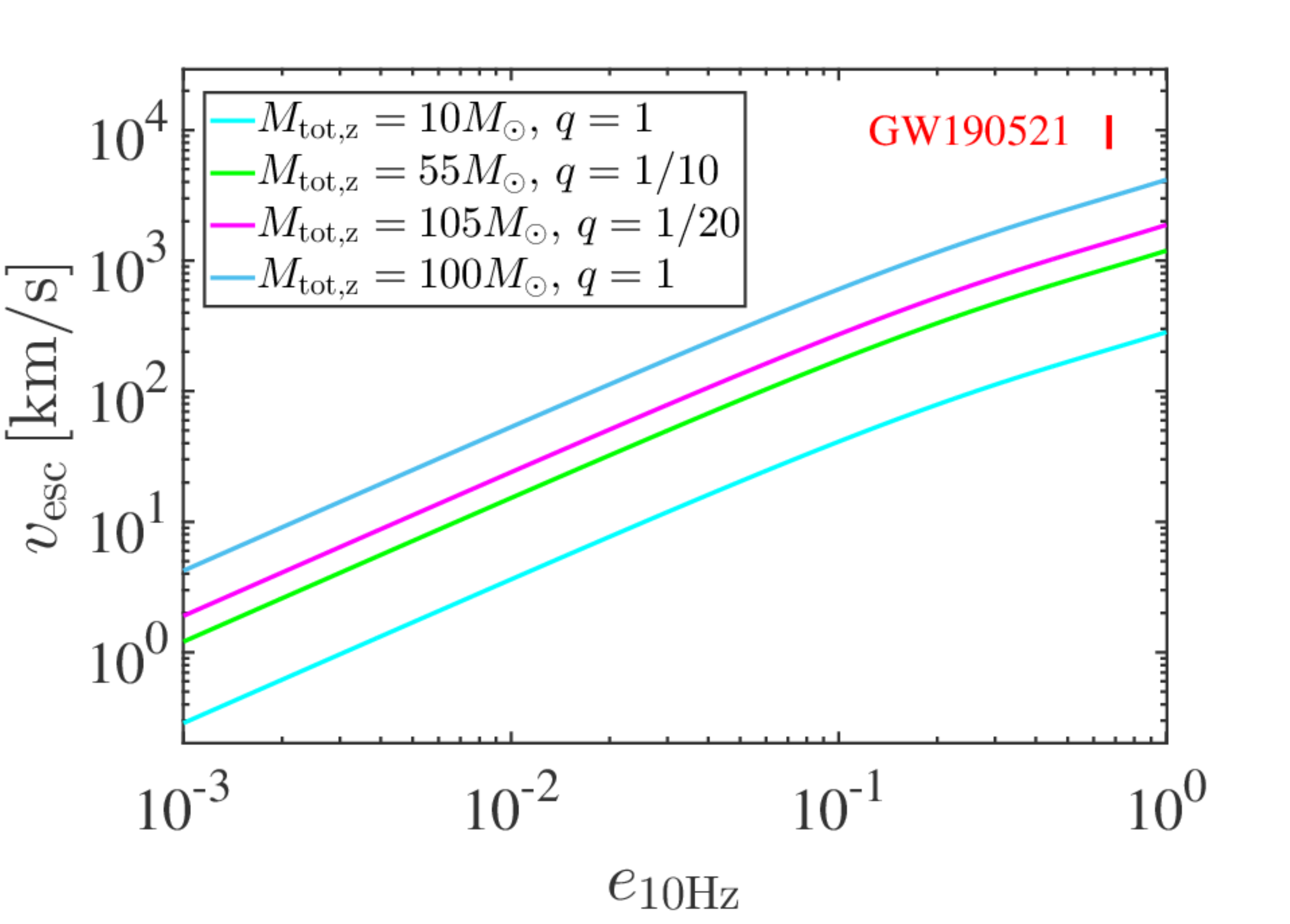}
  \caption{The estimated escape speed of the host environment $v_{\rm esc}$ as a function of $e_{\rm 10 M}$ (top panel) and $e_{\rm 10 Hz}$ (bottom panel) for the single--single GW capture process in GNs. $v_{\rm esc}(e_{\rm 10M})$ and $v_{\rm esc}(e_{\rm 10 Hz})$ are calculated using Equations (\ref{eq:vesc_e10M}) and (\ref{eq:vesc_e10Hz}), respectively, and examples are labelled in the figures. Based on the inferred parameters of GW190521 such as $\{ m_A, m_B, z, e_{\rm 10Hz} \}$ (Table 1 in \citealt{Gayathrietal2020}), we find that $v_{\rm esc}$ ranges between $\sim 7.5 \times 10^3 \, \rm km/s - 1.2 \times 10^4 \, \rm km/s$ (red vertical region at $e_{\rm 10 Hz} = 0.67$) if GW190521 formed through the single--single GW capture process.  \label{fig:vorb_vs_ecc}   } 
\end{figure}

 Examples for $v_{\rm esc}(e_{\rm 10M})$ and $v_{\rm esc}(e_{\rm 10 Hz})$ are displayed in Figure \ref{fig:vorb_vs_ecc}. As seen, the single--single GW capture process leads to binaries with higher $e_{\rm 10 M}$ and $e_{\rm 10 Hz}$ in host environments with higher escape speeds. 
 
 Note that the introduced method will only give a crude estimate for the true escape speed for single sources.

%% file: Sections/Summary_Conclusions.tex
\newpage

\section{Discussion and Conclusions} 
\label{sec:SummAndConc} 
 
 We extended the studies by \citet{OLearyetal2009,Gondanetal2018b}, who investigated GW capture BBH populations in single GNs, by including the 1PN corrections for the formation and evolution of these BBHs and accounted for the effect of DF. We found that these corrections do not modify significantly the distributions of binary parameters in single GNs (Section \ref{subsec:ResMC_SingleGN}), thus, the conclusions of those papers continue to hold for single GNs under more general assumptions. We highlight the most important of these next.
\begin{itemize}
 \item In single GNs, single--single GW capture BBHs typically form with high eccentricities ($0.95 \lesssim e_0 \lesssim 0.9999$) and low orbital separations ($5 \, M_{\rm tot } \lesssim r_{\rm p0} \lesssim 100 \, M_{\rm tot}$), where the distribution of initial orbital separation drops off quickly beyond $\sim 10 - 20 \, M_{\rm tot}$ depending on the level of mass segregation and the BH mass function. Furthermore, the vast majority of binaries ($92 \% \lesssim $) retain eccentricity beyond $0.1$ when forming in or entering the aLIGO/AdV/KAGRA band, and $\sim 59 - 74 \%$ of them have an eccentricity larger than $0.8$ depending on the parameters of the BH population. Specifically for an extreme mass-segregated model referring to that of young stars in the Galactic Center, the corresponding fractions are $97 \% \lesssim $ and $\sim 67 - 71 \%$. The merging BBHs retain a significant eccentricity even at the point when the dimensionless pericenter distance reaches $\rho_{\rm p} = 10$ or when they for at $\rho_{\rm p0} < 10$, the majority ($\sim 71 - 94 \% $) of these binaries have $e_{\rm 10M} > 0.1$ and a non-negligible fraction ($\sim 21 - 42 \%$) of them have $e_{\rm 10 M} > 0.8$. Similarly, $91 - 95 \%$ and $46 - 50 \%$ for the considered extreme mass-segregated model.
 
 \item In single GNs, the distributions of binary parameters are weakly sensitive to the BH formation rate in the centers of galaxies for relaxed GNs (Section \ref{subsec:MergRates}) as long as they populate GNs and form a relaxed distribution by $z\sim 1$. Furthermore, we found that the distributions do not depend significantly on $\Msmbh$ (Section \ref{subsec:ResMC_SingleGN}).
 
 \item However, not surprisingly, the $M_{\rm tot}$ and $q$ distributions in single GNs are sensitive to the underlying mass function and the level of mass--segregation. More mass segregated clusters or more top-heavy BH mass functions lead to more massive merging binaries with somewhat lower $\rho_{\rm p0}$ and with a preference for equal-mass systems. In particular, the distribution of $q$ is roughly uniform and the typical $M_{\rm tot}$ is roughly $m_{\rm BH, max}$ for standard mass segregation ($p_0 \simeq 0.5 - 0.6$) with an $\propto m_{\rm BH}^{-2}$ BH mass function. For strongly mass segregated systems with $p_0 \gtrsim 0.8$, the GW capture binaries form with a strong preference for the maximum component masses, favoring equal masses (see Figures \ref{fig:ParamDist_GN_p0Dep} and \ref{fig:ParamDist_GN_betaDep}).
\end{itemize}

 We have extended previous studies to obtain the distribution of physical parameters describing the binaries as seen by aLIGO/AdV/KAGRA at design sensitivity, by taking into account the detection range for different masses, eccentricity, source direction, inclination, and polarization angle. Our main results for the detectable GW capture BBH populations can be summarized as follows:
  
\begin{itemize}
   \item We find that $\sim 41-58\%$ of single-single GW capture BBHs in GNs form above $10 \, \Hz$ depending on the level of mass--segregation and the underlying BH mass function, allowing aLIGO/AdV/KAGRA to observe these binaries from formation to merger (Figure \ref{fig:fGW0Dist}). The rest of the sources form outside the advanced GW detectors' frequency band. Higher-generation detectors such as Advanced LIGO Plus, Einstein Telescope, LIGO Voyager, and Cosmic Explorer will observe a larger fraction of sources at formation, but ultimately the deci-Hertz instruments such as (B-)DECIGO and TianQin may be best suited to observe the formation of all other GW capture events. LISA will not be able to detect this source population.
   
   \item For GW capture BBHs merging in GNs in the local Universe, the distribution of peak GW frequency at binary formation $f_{\rm GW,0}$ is between $1 \, \Hz$ and $100 \, \Hz$, and their $\log f_{\rm GW,0}$ distribution peaks at \mbox{$ \sim 16 - 35 \, \Hz$} depending on the level of mass--segregation and the BH mass function (Section \ref{subsec:BBHsInDetBandsAtForm}). The location of the peak is higher than other eccentric source populations such as 2-body, single--single GW capture, or 3-body GW mergers in GCs \citep{Samsingetal2020}, indicating that BBH mergers in the GW capture merger channel in GNs may be disentangled by measuring the peak GW frequency at binary formation.
   
   \item Observational bias shifts the distributions of $\rho_{\rm p0}$ toward \mbox{$\rho_{\rm p0} \sim 5-17$}, $M_{\rm tot}$ and $\mathcal{M}$ shift toward higher masses, and $q$ weakly toward equals-mass binaries. The distribution of $e_{\rm 10 Hz}$ is also weakly affected by observational bias, while the $e_{\rm 10 M}$ distribution is skewed toward higher eccentricities. Examples are depicted in Figure \ref{fig:Impact_ObsBias_ParamDists}.
   
   \item We find that more than $92 \%$ of binaries have eccentricity beyond $0.1$ when forming in or entering the aLIGO/AdV/KAGRA band, and $\sim 60 - 79 \%$ of them have an eccentricity larger than $0.8$ for aLIGO/AdV/KAGRA detections and for $0.5 \leqslant p_0 \leqslant 0.9$ and the considered $\mathcal{F}_{\rm BH}$ models (Section \ref{subsec:DistaLIGOdet}). Note that somewhat higher fractions characteize the considered extreme mass-segregated model. Furthermore, at the point when the pericenter reaches $10 \, M_{\rm tot}$ the eccentricity satisfies $e_{\rm 10M} > 0.1$ and $e_{\rm 10M} > 0.8$ for $\sim 71-95\%$ and $\sim 23 - 67 \%$ of the sources, respectively, for aLIGO/AdV/KAGRA detections. Somewhat higher fractions characterize the considered extreme mass-segregated model for both $e_{\rm 10M}$ than the presented maximum values. Obtained results make the single-single GW capture BBH merger channel in GNs the most eccentric source population among the currently known stellar-mass BH merger channels in the Universe.
   
   \item We find a significant correlation between the eccentricities $\{ e_{\rm 10 Hz}, e_{\rm 10M} \}$ and the total binary mass or the chirp mass together with reduced mass, which arise due to mass segregation in these systems (Section \ref{subsec:CorrBBHparams}). Furthermore, there is an exceptionally strong correlation between $e_{\rm 10M}$ and $\rho_{\rm p0}$. Thereby, the measurement of $e_{\rm 10M}$ may be used to put constraints on the host environment of the source as $\rho_{\rm p0}$ gives an estimate on the magnitude of the escape speed for single--single GW captures (and similarly for GW captures during strong binary--single interactions; Section \ref{subsec:VelDist_Host}); see examples in Figure \ref{fig:vorb_vs_ecc}. Similarly, $e_{\rm 10 Hz}$ may also be used to constrain the escape speed in the host environment because of the strong correlation between $e_{\rm 10 Hz}$ and $\rho_{\rm p0}$. We also find a correlation between source mass and mass ratio (higher $M_{\rm tot}$ yields more equal-mass sources) and between source mass and distance (Section \ref{subsec:CorrBBHparams}).
 \end{itemize}
 
 We worked under the assumption that (i) the steepness of the BH number density profile is constant in radius for each BH species and that (ii) each BH species has a spherically symmetric number density profile around an SMBH. Note, however, that both assumptions may be violated to some degree, primarily for the heaviest BHs, by the following effects. (i) In the innermost region, where the heavy BHs outnumber the otherwise more common lighter objects, the heavy BHs will move towards the standard weak mass segregation regime $n\propto r^{-1.75}$  \citep{Vasiliev2017}. Furthermore, the number density profile of each BH species by mass could become shallower at small radii due to the mutual scattering of BHs into the loss cone (e.g., \citealt{AlexanderPfuhl2014}). However, for a mass-segregated profile, the rate of mergers scales weakly with radius as $d\Gamma /d\ln r \propto r^{11/14-2p_0 m/m_{\rm max}}$ i.e. $r^{-0.2}-r^{-0.4}$ for $p_0\sim 0.5-0.6$ and $m\sim m_{\rm max}$, but $d\Gamma /d\ln r \propto r^{-2.1}$ for the extreme case of $p_0=1.5$. Thus, the merger rates and parameter distributions are highly sensitive to the very inner regions only for the extreme values of $p_0>1$. In those cases, if the density profiles of the heavy BH species become shallower, the $\rho_{\rm p0}$ and $e_0$ distributions shift toward higher values, while the mass-dependent distributions shift toward lower masses with a stronger preference for unequal-mass systems; see Figure \ref{fig:ParamDist_GN_p0Dep}. Furthermore, the $e_{\rm 10 M}$ distribution is expected to shift toward lower values as well. However, $e_{\rm 10Hz}$ would not be affected much as $P(e_{\rm 10Hz})$ depends weakly on the steepness of the BH number density profile. The predicted merger rate densities of the heaviest BHs \citep{OLearyetal2009,KocsisLevin2012,Tsang2013,RasskazovKocsis2019} may also decrease. (ii) On the other hand, heavier objects segregate to a more flattened distribution due to vector resonant relaxation \citep{SzolgyeneKocsis2018,Szolgyenetal2021}. This mechanism increases the BH number density, especially the heavy ones, close to the SMBH, thereby increasing the merger rate of heavy BBHs \citep{KocsisLevin2012}.
 
 In conclusion, our findings for the binary parameter distributions and correlations among binary parameters may be useful to discriminate this formation channel from others in the observed sample of mergers and identify binaries that may have formed in galactic nuclei via the singe--single GW capture mechanism.
 
 The parameters of GW190521 were inferred under the assumption that it was a quasi-circular BBH coalescence \citep{Abbottetal2020b}, and results showed evidence for a BH in the mass gap predicted by pair-instability pulsation supernova theory. A list of astrophysical processes which populate the mass gap in which GW190521 could have formed are summarized in \citet{Abbottetal2020b}, and recent studies have proposed alternative pathways to produce GW190521-like BBH mergers (e.g.,  \citealt{Belczynski2020b,DeLucaetal2020,Farrelletal2020,Fishbachetal2020,Fragioneetal2020b,Kremeretal2020,Kritosetal2020}). Most recently, \citet{Gayathrietal2020} and \citet{RomeroShaw2020} have pointed out that GW190521 may be interpreted as a highly eccentric source. Given the reconstructed masses ($m_A = m_B = 102 ^{+7}_{-11} \, \Msun$) together with reconstructed residual eccentricity at $10 \, \Hz$ $e_{\rm 10 Hz} = 0.67$ and redshift $z = 0.35 ^{+0.16}_{-0.09}$ for this source (Table 1 in \citealt{Gayathrietal2020}), the initial pericenter distance is approximately $\rho_{\rm p0} \simeq 8.9 - 10.7$ if the initial eccentricity is close to 1 as in the GW capture channel ($0.95 \lesssim e_0 \lesssim 0.9999$).\footnote{We use Equations \eqref{eq:f_GW} and \eqref{eq:drhopde_3p5PN} together with the conditions $e_{\rm 10 Hz} = 0.67$ and $0.95 \leqslant e_0 \leqslant 0.9999$ to determine limits for $\rho_{\rm p0}$ over the possible values of both $M_{\rm tot}$ and $z$.} Furthermore, using the inferred parameters $\{ e_{\rm 10 Hz}, q, M_{\rm tot,z} \}$, the estimated escape speed of the host environment ranges between $\sim 7.5 \times 10^3 \, \rm km/s - 1.2 \times 10^4 \, \rm km/s$ (Figure \ref{fig:vorb_vs_ecc}). These parameters may be directly compared with the expectations for the single--single GW capture mechanism derived in this paper:
 \begin{itemize}
   \item The reconstructed parameter set \{$\rho_{\rm p0} \simeq 8.9 - 10.7$, $M_{\rm tot,z} \sim 229 - 329 \, \Msun$, $q = 1$\} lies close to the sweet--spot where the GW inspiral detection is most sensitive for a single aLIGO detector ($M_{\rm tot,z} \sim 90 - 250 \, \Msun$, $\rho_{\rm p0} \sim 5 - 10$, $q = 1$, see Figure \ref{fig:HorizonDistGWcBBHs}).\footnote{Note, however, that these values correspond to the detector's design sensitivity, which has not been reached in O3. Given the higher noise at low frequencies, the actual sweet spot for O3 was at a somewhat lower mass.}
   
   \item The predicted mass distribution of the GW capture sources is strongly tilted towards high masses for single--single GW capture sources, even if the host population is bottom heavy. The peak of the total binary mass distribution is between $m_{\rm max, BH}$ and $1.5 \, m_{\rm max, BH}$ for the fiducial choice of mass-segregation and it is at $\sim 1.8 \, m_{\rm max, BH}$ for strongly mass-segregated systems, where $m_{\rm max, BH}$ is the maximum mass in the system (Figure \ref{fig:ParamDistGWcBBHs_MassRanges} which assumed $m_{\rm max, BH} = 50 \, \Msun$). The fact that the first eccentric source has been observed to be the heaviest seen to date, probably a 2nd or higher generation merger, is consistent with the single--single GW capture scenario. Note, however, that while the merger remnants of single--single GW capture events are typically retained in galactic nuclei due to the SMBH potential, the likelihood of repeated mergers is low in this merger channel alone as long as the BH distribution is isotropic. Repeated mergers are possible in cases where the distribution is flattened in a disk either by resonant relaxation \citep{SzolgyeneKocsis2018} or gaseous processes \citep{Tagawaetal2020}. Alternatively, the high-mass BHs could have formed in and delivered to the galactic nucleus by globular clusters \citep{OLearyetal2016,Rodriguezetal2019}. The results of this paper indicate that if such high mass BHs exist in galactic nuclei, they will be the ones most commonly observed to be undergoing single--single GW capture events.
   
   \item The reconstructed $\rho_{\rm p0} \simeq 8.9 - 10.7$ and $q = 1$ also lie close to the peak of the predicted distribution for single--single GW capture events in GNs regarding the mass-dependent characteristics of corresponding distributions as shown in Figure \ref{fig:ParamDistGWcBBHs_MassRanges}.
   
   \item The estimated escape speed of the host environment $\sim 7.5 \times 10^3 \, \rm km/s - 1.2 \times 10^4 \, \rm km/s$ also supports a GN origin of GW190521. Note that it is a crude estimate of the true escape speed of the host environment (Sections \ref{subsec:VelDist_Host}).
   
   \item The overall BBH merger rate observationally constrained by the LIGO-Virgo Collaboration using their second GW transient catalogue \citep{Abottetal2020e} to the range \mbox{$\mathcal{R}_{\rm BBH} = 15.3 - 38.2 \, \Gpc ^{-3} \yr ^{-1}$} \citep{Abottetal2020_O3population}, and the estimated rate of mergers similar to GW190521 is $0.08 \, \Gpc ^{-3} \yr ^{-1}$ \citep{LIGOcollOthers2021}. The luminosity distance of the single potentially eccentric source is $D_{\rm L} \sim 1.8 - 2.9\, \Gpc$ detected during 6 months of the O3a observing run \citep{Gayathrietal2020}, which implies a very crude estimate of the rate of highly eccentric mergers of $\mathcal{R}_{\rm ecc} \simeq (0.5 \, \yr)^{-1}(4\pi D_{\rm L}^3/3)^{-1} = 0.02 - 0.08 \, \Gpc ^{-3} \yr ^{-1}$. The theoretical rate estimates for single--single GW captures have been shown to be exponentially sensitive to the mass--segregation parameter \citep{RasskazovKocsis2019}. The measured value of $\mathcal{R}_{\rm ecc}$ is consistent with the single-single GW capture channel for moderately strongly mass-segregated cusps with $p_0 = 0.85-0.93$ and a BH mass function $m^{-2.35}$ (see Figure 5 in \citealt{RasskazovKocsis2019}). In this case, the radial number density distribution of the heaviest BHs is between $r^{-2.35}$ and $r^{-2.43}$.
 \end{itemize}
 
 In conclusion, the reconstructed total mass, mass ratio, eccentricity, along with the estimated initial pericenter distance and eccentric merger rate are all fully consistent with the single--single GW capture scenario for GW190521. 
 
 Recently, \citet{Samsingetal2020b} and \citet{Tagawaetal2020c} have shown that the AGN merger channel also produces highly eccentric sources with similar parameters as GW190521. In this case, we find that the escape speed of the host environment ranges between $\sim 1.5 \times 10^3 \, \rm km/s - 2.5 \times 10^3 \, \rm km/s$, supporting the assumption of an AGN origin and a hierarchical--merger scenario as well (see also \citealt{GerosaFishbach2021} and references therein).\footnote{see Section \ref{subsec:VelDist_Host} and the inferred parameters of GW190521 (Table 1 in \citealt{Gayathrietal2020}).} The single--single GW channel however produces on average higher eccentricities, particularly close to the merger (e.g. at a pericenter of $10 M_{\rm tot}$) in comparison to the AGN merger channel. As the number of detections is expected to increase rapidly in the coming years with the further development of GW detectors, these highly eccentric source populations may be statistically disentangled and will represent different peaks in a multimodal distribution, possibly with different sets of intrinsic correlations among the merging binaries' parameters.

%% file: Sections/Data_Availability.tex
\section*{Data Availability}
 
 The data underlying this article will be shared on reasonable request to the corresponding author.

%% file: Sections/Acknowledgments.tex
\section*{Acknowledgments}

 We thank the anonymous referee for constructive comments that helped improve the quality of the paper. We thank Hiromichi Tagawa and Chris Belczynski for useful discussions. We are grateful to M\'aria Kolozsv\'ari for help with logistics and administration related to the research.  L\'aszl\'o Gond\'an acknowledges support from the \mbox{{\'U}NKP-18-3} New National Excellence Programs of the Ministry of Human Capacities. This work received founding from the European Research Council (ERC) under the European Union's Horizon 2020 Programme for Research and Innovation ERC-2014-STG under grant agreement No. 638435 (GalNUC).

%% file: Appendices/Appendix_Properties_StellarPops.tex
\section{Stellar Populations in Relaxed Galactic Nuclei}
\label{sec:StellPopsProp}
 
 We introduce our assumptions for the BH and light stellar (MS, WD, and NS) populations in GNs in Appendices \ref{subsec:GNs_BHPops} and \ref{subsec:GNs_LightStarPops}, respectively.

\subsection{Black hole population}
\label{subsec:GNs_BHPops}  
 
 Recent observations found that there is a large variation in the metallicity for stars in the Galactic center, the metallicity ranges from sub-solar ($\sim 10 \% \, Z_{\odot}$) to super-solar \mbox{($\sim 500 \% \, Z_{\odot}$)} metallicity, and more than 90\% of stars have metallicities between $\sim Z_{\odot} - 500 \% \, Z_{\odot}$ (e.g., \citealt{Doetal2015,Rydeetal2016,Feldmeieretal2017,Richetal2017,Doetal2018,NoguerasLaraetal2018,Doetal2020}). Stars with metallicity beyond the solar value are expected to collapse to low-mass BHs with masses between $\sim 5-15 \, \Msun$; see e.g. Figure 5 in \citet{Belczynskietal2016}. Additionally, DF can also bring such low-mass BHs into the GN \citep{MiraldaEscudeGould2000,AntoniniMerritt2012,RasskazovKocsis2019} from the high metallicity inner Central bulge  ($Z_{\odot} \lesssim $) stars (e.g., \citealt{Majewskietal2017,Rojasetal2017,Garciaetal2018}). We set the lower limit of the BH mass range of interest to be $m_{\rm BH, min} = 5 \, \Msun$.\footnote{Additionally, results of X-ray observations also suggest a lower limit of $\sim 5 \, \Msun$ for BH masses \citep{Bailynetal1998,Ozeletal2010,Farretal2011,Belczynskietal2012}.} Sub-solar metallicity stars constitute $7\%$ of stars in the GN. These stars may have arrived in the Galactic center as a result of the infall of massive stellar systems into the Galactic center, either Galactic star clusters  \citep{ArcaSeddaetal2015,ArcaSeddaetal2018,ArcaSedda2020b} and extra-galactic dwarf galaxies \citep{ArcaSedda2020b}. For these sub-solar metallicity stars, both pair-instability supernova and pair-instability pulsation are expected to produce BH with masses up to $\sim 50 \, \Msun$ in the stellar-mass range (and $m_{\rm BH} \gtrsim 130 \, \Msun$ in the intermediate-mass range; see e.g. \citealt{Belczynskietal2016,Belczynskietal2016c,Marchantetal2016,FishbachHolz2017,Mapellietal2017,Marchantetal2018,Farmeretal2019,Stevensonetal2019,Woosley2019}).\footnote{Under special circumstances, BHs may form up to $\sim 60 - 70 \, \Msun$ from metal-poor stars \citep{SperaMapelli2017,Woosley2017,Giacobboetal2018,Mapellietal2020} and high-metallicity stars \citep{Belczynskietal2020}, and up to $\sim 80 \, \Msun$ from intermediate-metallicity stars \citep{Limongietal2018}.} Nuclear star clusters are expected to form, at least partially, through the infall of massive stellar systems via DF \citep{Tremaineetal1975,CapuzzoDolcetta1993}, thus sub-solar metallicity stellar populations may commonly occur in other GNs. Therefore, we set the maximum possible BH mass to be \mbox{$m_{\rm BH, max} = 50 \, \Msun$}.\footnote{Note that the mass gap may be filled by multiple generations of mergers since the hierarchical merger process is more efficient in denser and heavier environments such as nuclear star clusters or accretion disks of active galactic nuclei (e.g., \citealt{MillerLauburg2009,Fishbachetal2017,GerosaBerti2017,Christianetal2018,McKernanetal2018,Rodriguezetal2018,GerosaBerti2019,Yangetal2019,Tagawaetal2020b}).}
 
 Since the mass distribution of BHs in GNs is still not well-understood, we consider two types of BH mass distribution models in the present work: a power-law multi-mass distribution and a multi-mass distribution given by a population synthesis method. In the case of the power-law model, the mass distribution of BHs normalized to unity is
\begin{equation}  \label{eq:BHmassFunc}
  \mathcal{F}_{\rm BH} = \frac{ (1-\beta) \, m_{\rm BH}^{-\beta} }{ m_{\rm BH,max}^{1-\beta} - m_{\rm BH,min}^{1-\beta} } \, ,
\end{equation} 
 where we assume $\beta \in \{ 1, 2, 3\}$. This model is motivated by recent LIGO--Virgo detections \citep{Abbottetal2016a,Abbottetal2019c,Abbottetal2019}, results of numerical simulations for the steady-state mass distribution of stellar populations around a SMBH \citep{AlexanderHopman2009}, and that low-mass BHs may dominate BH populations in GNs because of the high-metallicity environment in GNs. We also carry out MC simulations for a BH mass distribution obtained in a population synthesis study \citep{Belczynskietal2016} for single  $10 \% \, Z_{\odot}$ metallicity stars. We refer to this model as B16 throughout the paper. In this case, the BH mass distribution ranges between $m_{\rm BH,min} \sim 5 \, \Msun$ and $m_{\rm BH,max} \sim 40 \, \Msun$, and represents a more complicated distribution of BH masses. We set the fiducial power--law mass distribution exponent to be $\beta = 2$. The evolution of $\mathcal{F}_{\rm BH}$ across cosmic time due to DF is discussed in Appendix \ref{sec:ImpactDF_GWcBBHs}.
 
 Recall that the equilibrium number density distribution in Equation \eqref{eq:n(r)} can be parameterized using Equation \eqref{eq:alpha_p} as
 \begin{equation}  \label{eq:n(r)app}
  n(r,m) =  C_{\rm frac} n_{\rm inf} \mathcal{F}(m) \left( \frac{r}{r_{\rm max}} \right)^{-(3/2)-p(m)} \, . 
\end{equation}
 Here, $p(m)$ can be approximated for BHs as
\begin{equation}\label{eq:pmbh}
  p(m_{\rm BH}) = p_0 \frac{ m_{\rm BH} }{ m_{\rm BH, max} }  \, ,
\end{equation} 
 where the mass--segregation parameter $p_0$ is a constant for relaxed GNs and ranges between $0.5 - 0.6$ based on Fokker--Planck simulations \citep{OLearyetal2009}. This implies a density profile as steep as $\alpha_{\rm BH} = 2 - 2.1$ for the heaviest BHs. However, this profile may be steeper due to the binary star disruption by the SMBH's tidal field ($\alpha_{\rm BH} \lesssim 2.25$; \citealt{FragioneSari2018}), star formation ($\alpha_{\rm BH} \lesssim 2.3$ if $m_{\rm BH, max} = 30 \, \Msun$; \citealt{AharonPerets2016}), and when the heavy objects are rare ($\alpha_{\rm BH} \lesssim 2.75$; \citealt{AlexanderHopman2009}). To study the effects of the cusp slope on the distributions of binary parameters, we assume that $p_0$ is between $0.5$ and $0.9$, i.e. $\alpha_{\rm BH} \leqslant 2.4$. Here we set a somewhat higher $p_0$ than that was obtained in \citet{AharonPerets2016} because we consider a higher $m_{\rm BH, max}$. The fiducial mass--segregation parameter is set to be $p_0 = 0.5$.
 
 Observations showed evidence for warped disks of young massive stars in the Galactic Center with a radial surface density profile of \mbox{$\propto r^{-2}$} \citep{Paumardetal2006,Bartkoetal2009,Luetal2009}, which is equivalent to a 3D density profile of $\propto r^{-3}$. Using different mass distributions in isotropic steady-state Fokker--Planck models, \citet{Keshetetal2009} showed that the density profile can be as steep as $\alpha_{\rm BH} = 3$, which corresponds to $p_0 = 1.5$. To capture the effect of extreme mass segregation, we also consider a model with $p_0 = 1.5$. Furthermore, observational studies suggest an initial mass function with $1.7 \lesssim \beta \lesssim 2.35$ for the massive young cluster in the Galactic Center region (e.g., \citealt{Bartkoetal2010,Luetal2013}). Therefore, assuming that the BH mass function follows the initial mass function of stars, we adopt $\beta = 2$ to the extreme mass--segregation cusp model in MC simulations.
 
 As many as $\sim 20,000$ BHs are predicted to settle into the Galactic center as a result of DF \citep{Morris1993,MiraldaEscudeGould2000,Gebhardtetal2002,Freitagetal2006,HopmanAlexander2006}. The number of segregated BHs is proportional to the mass of the SMBH through the infall rate of BHs \citep{MiraldaEscudeGould2000}, we,  therefore, approximate the present-day total number of BHs in a GN with an SMBH of mass $\Msmbh$ in its center as 
\begin{equation}  \label{eq:NtotBHappr}
   N_{\rm BH} \approx 20,000 \times \frac{ \Msmbh }{M_{\rm SgrA^*}} \, .
\end{equation}
 Here, $M_{\rm SgrA^*} \approx 4.28 \times 10^6 \, \Msun$ is the recently estimated mass of the SMBH in the Galactic center, \mbox{Sagittarius A$^{*}$} \citep{Gillessenetal2017}. The impact of DF on $N_{\rm BH}$ is discussed in Appendix \ref{sec:ImpactDF_GWcBBHs}.
 
 BHs infall into the SMBH close to it due to GW emission within a much shorter timescale than their local two--body relaxation timescale. This leads to the removal of BHs, which creates a loss-cone around the SMBH. Because of that, the number density distribution of BHs exhibits a cut off at an inner radius of $r_{\rm min}$. For $r > r_{\rm min}$, we assume that BHs are continuously replenished from the outside, due to relaxation to maintain a steady-state density. Following \citet{Gondanetal2018b}, we calculate $r_{\rm min}$ as the radius where the GW inspiral timescale becomes shorter than the local two-body relaxation timescale; see Appendix A in \citealt{Gondanetal2018b} for details. Note that results for $r_{\rm min}$ are robust and only weakly sensitive to the assumptions on the mass distributions of light star populations (MSs, WDs, and NSs) because $r_{\rm min} \propto [ n_{\rm tot}(r) \langle M^2 \rangle (r) ]^{-2/11}$. Finally, the innermost radius of the GN at which BHs with masses $m_A$ and $m_B$ can form GW capture BBHs can be defined as
\begin{equation}  \label{eq:rmin_AB}
  r_{\rm min}^{A,B} := {\rm max}\left[ r_{\rm min}(m_A), r_{\rm min}(m_B) \right] 
\end{equation}
 \citep{Gondanetal2018b}. Thus we assume that the BHs span the range
\begin{equation}  \label{eq:Cond_r}
  r_{\rm min}^{A,B} \leqslant r \leqslant r_{\rm max} \, ,
\end{equation}
 where $r_{\max}$ is the SMBH radius of influence. We discuss the impact of DF on $r_{\rm min}^{A,B}$ in Appendix \ref{sec:ImpactDF_GWcBBHs}.
 
 We obtain the parameter $C_{\rm frac}$ in Equation \eqref{eq:n(r)app} for the BH population as follows. We express the number of BHs in two ways: using Equation \eqref{eq:NtotBHappr}, and by calculating the enclosed number of BHs within $r_{\rm max}$ using Equation \eqref{eq:n(r)app} as\footnote{$n_{\rm BH}(r)$ is a smooth and strictly monotonically increasing function and $r_{\rm min} \ll r_{\rm max}$ \citep{Gondanetal2018b}, thus the lower bound for the integration in Equation \eqref{eq:Nenclosed_BH} can be substituted with zero.}
\begin{equation}  \label{eq:Nenclosed_BH}
  N_{\rm BH}(< r_{\rm max}) = \int _0 ^{ r_{\rm max} } 4 \pi r^2 n_{\rm BH}(r) dr \, .
\end{equation}
 $C_{\rm frac}$ is determined numerically by solving $N_{\rm BH} = N_{\rm BH} (< r_{\rm max} )$.

\subsection{Population of light stars}
\label{subsec:GNs_LightStarPops} 
 
 Light stellar components such as MSs, WDs, and NSs influence $r_{\rm min}$ (Appendix \ref{subsec:GNs_BHPops}) and also occasionally participate in binary--single interactions in which case we conservatively discard the binary from the MC sample (Section \ref{subsec:BBHevaporation}).
 
 Star formation can be very different deep in the potential of an SMBH from that of the field \citep{AlexanderHopman2009,Bartkoetal2010}, hence the mass distributions of MSs, WDs, and NSs are not well known in GNs. \citet{Gondanetal2018b} showed that $r_{\rm min}$ is weakly sensitive to the mass distributions of these stellar populations. Therefore, as a simplicity, we follow \citet{HopmanAlexander2006} and adopt single mass MS, WD, and NS populations with component masses $1 \, \Msun$, $0.6 \, \Msun$, and $1.4 \, \Msun$, respectively.\footnote{Accordingly, mass distributions for the MS, WD, and NS populations are $\mathcal{F}_{\rm MS} \propto \delta(1 \, \Msun)$, $\mathcal{F}_{\rm WD} \propto \delta(0.6 \, \Msun)$, and \mbox{$\mathcal{F}_{\rm NS} \propto \delta(1.4 \, \Msun)$}, respectively.}
 
 Relaxation of multi-mass stellar populations around SMBHs has two limiting cases depending on the fraction ratio of heavy objects in the population \citep{AlexanderHopman2009}: weak and strong mass-segregation in which heavy stars are relatively common and rare, respectively. Light stars form density cusps with $3/2 \leqslant \{ \alpha_{\rm MS}, \alpha_{\rm WD}, \alpha_{\rm NS} \} \leqslant 7/4$ in both cases, where, according to mass segregation, lower $\alpha$ values correspond to lower masses. We therefore consider $\{ \alpha_{\rm MS}, \alpha_{\rm WD}, \alpha_{\rm NS} \}$ in the range $[3/2, \, 7/4]$, and set the fiducial values to be $3/2$.
 
 Following \citet{Gondanetal2018b}, we adopt the number fraction ratios of MSs, WDs, and NSs as $1 : 0.1 : 0.01$ obtained for continuous star-forming populations \citep{Alexander2005}, and find that
\begin{equation}
  C_{\rm MS} = 1, \quad C_{\rm WD} = 0.1 \frac{3 - \alpha_{\rm MS}}{3 - \alpha_{\rm WD}}, \quad C_{\rm WD} = 0.01  \frac{3 - \alpha_{\rm MS}}{3 - \alpha_{\rm NS}} \, ;
\end{equation}
 see Equation (12) in \citet{Gondanetal2018b}.

%% file: Appendices/Appendix_Rates_1PN.tex
\section{Analytic Estimates of Differential Merger Rate Distributions}
\label{sec:Rates_1PN}
 
 We derive the radial distributions of GW capture BBH mergers with fixed component masses taking into account 1PN order effects in Section \ref{subsec:RadDist_1PN}, and similarly determine the component-mass-dependent merger rate in Section \ref{subsec:MergRate_1PN}.

\subsection{Radial distribution for fixed component masses}
\label{subsec:RadDist_1PN}
 
 We start by recalling that the differential merger rate distribution of GW capture BBHs as a function of component masses $(m_A, m_B)$ and radial distance $r$ from the SMBH can be given as
\begin{align}  \label{eq:DiffMergRate}
  \frac{\partial^3 \Gamma_{\rm 1GN}}{\partial m_A \partial m_B \partial r} 
 & \approx  N_{\rm BH}^2 C_{r}(m_A) C_{r}(m_B) r^{- \alpha_{\rm BH}(m_A)- \alpha_{\rm BH}(m_B)} 
\nonumber \\
 & \times 4 \pi^2 r^2 v_{\rm circ} \mathcal{F}(m_A) \mathcal{F}(m_B) \left\{ b_{\rm max}^2 - b_{\rm min}^2  \right\}
\end{align}
 \citep{Gondanetal2018b}. Here, $v_{\rm circ} = \sqrt{\Msmbh/r}$ is the circular velocity at radius $r$, both $b_{\rm max}$ and $b_{\rm min}$ are evaluated at $w = v_{\rm circ}$, and $C_{r}$ as a function of $m_{\rm BH}$ is given as
\begin{equation}  \label{eq:CrmBH}
  C_{r}(m_{\rm BH}) = \frac{ 1 }{4\pi} \frac{ 3-\alpha_{\rm BH}(m_{\rm BH}) }{r_\mathrm{max}^{ 3-\alpha_{\rm BH}(m_{\rm BH}) } -
  [r_\mathrm{min} (m_{\rm BH})]^{ 3-\alpha_{\rm BH}(m_{\rm BH})} } \, .
\end{equation}
 
 The radial distribution of BBHs with fixed component masses $P_{AB}(r)$ can be given by first fixing $m_A$ and $m_B$ in Equation \eqref{eq:DiffMergRate} then normalizing it over the radius range $r \in [r_{\rm min}^{A,B}, r_{\max}]$. In order to account for additional 1PN terms in $P_{AB} (r)$, one can express $b_{\rm max}$ as $b_{\rm max} = b_{\rm max,L} ( 1 + Q_{\rm 1PN} w^{4/7} )$, where $b_{\rm max,L}$ is the leading-order term and $Q_{\rm 1PN}$ is given as (Equation \ref{eq:bmax})
\begin{equation}  \label{eq:Q_1PN}
  Q_{\rm 1PN} = \left( \frac{ 5763 - 3220 \eta }{ 3400 } \right) \left( \frac{ 3 }{ 340 \pi \eta }\right)^{2/7} \, .
\end{equation}
 Accordingly, $P_{AB}(r)$ can be separated into two part as
\begin{align}  \label{eq:P_ABr}
  P_{AB}(r) & \propto 4 \pi^2 r^2 v_{\rm circ} \left\{ b_{\rm max,L}^2 - b_{\rm min}^2 \right\}
  \nonumber
  \\
  & + 4 \pi^2 r^2 v_{\rm circ} \left\{ 2 b_{\rm max,L}^2 Q_{\rm 1PN} v_{\rm circ}^{ 4/7 } + b_{\rm max,L}^2 Q_{\rm 1PN}^2 v_{\rm circ}^{ 8/7 } \right\} \, ,
\end{align}
 where the second row accounts for the 1PN order correction in $b_{\rm max}$. Finally, after substituting $\{ b_{\rm max,L}, b_{\rm min}, v_{\rm circ}\}$, $P_{AB}(r)$ can be expressed as
\begin{align}  \label{eq:PDF_r}
  P_{AB}(r) & = C_{\rm N} \left[ \frac{ c_0 r^{-\frac{3}{14} - p_0 \frac{m_A + m_B}{ m_{\rm BH,max} }}} { \Msmbh ^
  {11/14}} - \frac{ 16 \, r^{-\frac{1}{2} - p_0
 \frac{m_A + m_B}{m_{\rm BH,max}}}}{ \Msmbh ^{1/2}} \right]
 \nonumber
 \\
 & + c_0 C_{\rm N} 
 \left[ \frac{2 Q_{\rm 1PN} r^{-\frac{1}{2} - p_0 \frac{m_A + m_B}{ m_{\rm BH,max} }} }{ \Msmbh ^
  {1/2}} - \frac{ Q_{\rm 1PN}^2 r^{-\frac{11}{14} - p_0
 \frac{m_A + m_B}{m_{\rm BH,max}}}}{ \Msmbh ^{3/14}} \right]
\end{align} 
 where $c_0 = (340 \pi \eta /3)^{2/7}$. Note that binaries form closer to the SMBH due to the adopted PN correction terms, and the impact of these terms on the radial distribution of binaries systematically increases with both $p_0$ and $M_{\rm SMBH}$.

\subsection{Component-mass-dependent merger rate}
\label{subsec:MergRate_1PN}
 
 The component-mass-dependent merger rate can be given by marginalizing Equation \eqref{eq:DiffMergRate} over radius $r$ as
\begin{align}  \label{APg2}
  \frac{\partial^2 \Gamma_{\rm 1GN}}{\partial m_A \partial m_B}
 & \approx  N_{\rm BH}^2 C_{rA} C_{rB} \mathcal{F}(m_A) \mathcal{F}(m_B)
 \nonumber \\
 & \times \int_{r_\mathrm{min}^{A,B}}^{r_\mathrm{max}} d r \, 4 \pi^2 r^{2 - \alpha_{\rm BH}(m_A) - \alpha_{\rm BH}(m_B)} v_{\rm circ} \left\{ b_{\rm max}^2 - b_\mathrm{min}^2\right\} \, ,
\end{align}
 where again $b_{\rm max}$ and $b_{\rm min}$ are evaluated at $w = v_{\rm circ}$. The collisional term ($\propto b_{\rm min}^2$) can be neglected because its contribution to $\partial^{2} \Gamma_{\rm 1GN} / \partial m_A \partial m_B $ is negligible even for $b_{\rm max} := b_{\rm max,L}$ \citep{Gondanetal2018b}. Similar to the $P_{AB}(r)$ case (Section \ref{subsec:RadDist_1PN}), we separately derive $\partial^{2} \Gamma_{\rm 1GN} / \partial m_A \partial m_B $ for the leading order and 1PN order correction terms as
\begin{align}  \label{eq:Rate_mAmB_Gen}
  \frac{\partial^2 \Gamma_{\rm 1GN}}{\partial m_A \partial m_B}
 & \approx C_{\rm P} \int_{r_\mathrm{min}^{A,B}}^{r_\mathrm{max}} d r \, 4 \pi^2 r^{2 - \alpha_{\rm BH}(m_A) - \alpha_{\rm BH}(m_B)} v_{\rm circ} b_{\rm max,L}^2 
 \nonumber 
 \\
 & + C_{\rm P} \int_{r_\mathrm{min}^{A,B}}^{r_\mathrm{max}} d r \, 8 \pi^2 r^{2 - \alpha_{\rm BH}(m_A) - \alpha_{\rm BH}(m_B)} b_{\rm max,L}^2 Q_{\rm 1PN} v_{\rm circ}^{ 11/7 }
 \nonumber 
 \\
 & + C_{\rm P} \int_{r_\mathrm{min}^{A,B}}^{r_\mathrm{max}} d r \, 4 \pi^2 r^{2 - \alpha_{\rm BH}(m_A) - \alpha_{\rm BH}(m_B)} b_{\rm max,L}^2 Q_{\rm 1PN}^2 v_{\rm circ}^{ 15/7 }
\end{align}
 Here, the $C_{\rm P}$ prefactor is given as
\begin{align}  \label{eq:CP_Pref}
  \nonumber
  C_{\rm P} & =  N_{\rm BH}^2 C_{rA} C_{rB} \mathcal{F}(m_A) \mathcal{F}(m_B) 
  \nonumber \\
  & = \frac{ 9 m_\mathrm{BH,max}^2 - 6 p_0 m_\mathrm{BH,max} M_\mathrm{tot}
 + 4 p_0^2 \mu M_\mathrm{tot} }{ 64 \pi^2 m_\mathrm{BH,max}^2 } 
 \nonumber 
 \\
 & \times N_{\rm BH}^2 \mathcal{F}_{\rm BH}(m_A) \mathcal{F}_{\rm BH}(m_B)
 \, r_\mathrm{max} ^{ - 3 + \frac{ p_0 M_\mathrm{tot} }{ m_\mathrm{BH,max} } } \, ,
\end{align}
 where we used the condition $r_{\rm min}^{A,B} \ll r_{\rm max}$ to simplify the resulting expression. By evaluating the integrals in Equation \eqref{eq:Rate_mAmB_Gen}, we get that
\begin{align}  \label{eq:rate_mAmB}
 \frac{\partial^2 \Gamma_{\rm 1GN}}{\partial m_A \partial m_B} 
 & \approx \frac{ 4 \pi^2 c_0 C_{\rm P} M_{\rm tot}^2 }{ \Msmbh^{11/14} }  
  \frac{ r_\mathrm{max}^{ \frac{11}{14} - \frac{ p_0 M_\mathrm{tot} }{ m_\mathrm{BH,max} }
  } - \left( r_\mathrm{min}^{A,B} \right) ^{ \frac{11}{14} - \frac{ p_0 M_\mathrm{tot}}{ m_\mathrm{BH,max} }
  } }{\frac{11}{14} - \frac{ p_0 M_\mathrm{tot} }{ m_\mathrm{BH,max} } } 
  \nonumber
  \\
  & + \frac{ 8 \pi^2 c_0 C_{\rm P} M_{\rm tot}^2 Q_{\rm 1PN} }{ \Msmbh^{1/2} }  
  \frac{ r_\mathrm{max}^{ \frac{1}{2} - \frac{ p_0 M_\mathrm{tot} }{ m_\mathrm{BH,max} }
  } - \left( r_\mathrm{min}^{A,B} \right) ^{ \frac{1}{2} - \frac{ p_0 M_\mathrm{tot}}{ m_\mathrm{BH,max} }
  } }{\frac{1}{2} - \frac{ p_0 M_\mathrm{tot} }{ m_\mathrm{BH,max} } }
  \nonumber
  \\
  & + \frac{ 4 \pi^2 c_0 C_{\rm P} M_{\rm tot}^2 Q_{\rm 1PN}^2 }{ \Msmbh^{3/14} }  
  \frac{ r_\mathrm{max}^{ \frac{3}{14} - \frac{ p_0 M_\mathrm{tot} }{ m_\mathrm{BH,max} }
  } - \left( r_\mathrm{min}^{A,B} \right) ^{ \frac{3}{14} - \frac{ p_0 M_\mathrm{tot}}{ m_\mathrm{BH,max} }
  } }{\frac{3}{14} - \frac{ p_0 M_\mathrm{tot} }{ m_\mathrm{BH,max} } } \, .
\end{align}
 Note that the additional PN terms mildly change the gradient of $\partial^{2} \Gamma_{\rm 1GN} / \partial m_A \partial m_B $ obtained in leading order, thereby their impact on the distributions of binary parameters is not significant.

%% file: Appendices/Appendix_EvolBBHparams.tex
\section{Impact of Dynamical Friction on Stellar Populations in Galactic Nuclei}
\label{sec:ImpactDF_GWcBBHs}
 
 DF is predominantly effective in bringing high-mass objects such as BHs into the GN \citep{MiraldaEscudeGould2000,RasskazovKocsis2019}. Therefore, we first determine the corresponding changes in the BH population, then investigate how DF influences the light stellar populations (MS, WD, NSs) and MC sample sizes of GW capture BBHs assigned to single GNs.
 
 Generally, the BH mass function is expressed as the initial value times a dimensionless parameter factor $\zeta$ that accounts for the accumulation of objects due to DF (\mbox{Section \ref{subsec:MergRates}}) as
\begin{equation}  \label{eq:NtotntotInitZeta}
  \frac{d N_{\rm BH} }{ d m_{\rm BH} } (t_{\rm DF}) = \frac{d N_{\rm BH,init} }{ d m_{\rm BH} }  \, \zeta (t_{\rm DF}) \, .
\end{equation}
 Quantities with the subscript "init" account for their values $12 \, \Gyr$ ago. We focus on the $t_{\rm DF}$ dependence of both $\mathcal{F}_{\rm BH}$ and $N_{\rm BH}$ in a single GN across cosmic time in this section, therefore, as a simplicity, we denote only this parameter in the arguments of $\mathcal{F}_{\rm BH}$, $N_{\rm BH}$, and $\zeta$. Using the general relation between the mass distribution and the normalized mass distribution $dN / dm = N \mathcal{F}$, Equation \eqref{eq:NtotntotInitZeta} can be rewritten as
\begin{equation}  \label{eq:RelNBHfBHZeta}
  N_{\rm BH} (t_{\rm DF}) \, \mathcal{F}_{\rm BH}(t_{\rm DF}) = N_{\rm BH, init} \, \mathcal{F}_{\rm BH, init} \, \zeta(t_{\rm DF}) \, ,
\end{equation}
 which sets the following relation between any two times $ t_{\rm DF}$ and $t_{\rm DF}'$:
\begin{equation}  \label{eq:RelNBHfBHZetaGen}
  N_{\rm BH} (t_{\rm DF}) \, \mathcal{F}_{\rm BH}(t_{\rm DF}) = N_{\rm BH} (t_{\rm DF}') \, \mathcal{F}_{\rm BH}(t_{\rm DF}') \frac{ \zeta(t_{\rm DF}) }{ \zeta(t_{\rm DF}') }  \, .
\end{equation}
 Thus,
\begin{align}  \label{eq:NBH_tDF}
 N_{\rm BH} (t_{\rm DF}) & = N_{\rm BH} (t_{\rm DF}') \, \int _{ m_{\rm BH, min} } ^{ m_{\rm BH, max} }  dm_{\rm BH}  \, \mathcal{F}_{\rm BH}(t_{\rm DF}') \, \frac{ \zeta(t_{\rm DF}) }{ \zeta(t_{\rm DF}') } \, ,
 \\
  \label{eq:FBH_tDF}
  \mathcal{F}_{\rm BH}(t_{\rm DF}) & = \frac{ \mathcal{F}_{\rm BH}(t_{\rm DF}') \, \frac{ \zeta(t_{\rm DF}) }{ \zeta(t_{\rm DF}') } }{ \int _{ m_{\rm BH, min} } ^{ m_{\rm BH, max} } dm_{\rm BH} \, \mathcal{F}_{\rm BH}(t_{\rm DF}') \, \frac{ \zeta(t_{\rm DF}) }{ \zeta(t_{\rm DF}') } }  \, .
\end{align}
 These equations can be parameterized with redshift instead of DF time by changing variable from $t_{\rm DF}$ to $z$ using Equation \eqref{eq:tDF_vs_z}. Accordingly, $N_{\rm BH}$ and $\mathcal{F}_{\rm BH}$ at redshift $z$ can be expressed with their values at redshift $z'$ as
\begin{align}  \label{eq:NBH_z}
 N_{\rm BH} (z) & = N_{\rm BH} (z') \, \int _{ m_{\rm BH, min} } ^{ m_{\rm BH, max} }  dm_{\rm BH}  \, \mathcal{F}_{\rm BH}(z') \, \frac{ \zeta(z) }{ \zeta(z') } \, ,
 \\
  \label{eq:FBH_z}
  \mathcal{F}_{\rm BH}(z) & = \frac{ \mathcal{F}_{\rm BH}(z') \, \frac{ \zeta(z) }{ \zeta(z') } }{ \int _{ m_{\rm BH, min} } ^{ m_{\rm BH, max} } dm_{\rm BH} \, \mathcal{F}_{\rm BH}(z') \, \frac{ \zeta(z) }{ \zeta(z') } }  \, .
\end{align}
 
 First, we use Equations \eqref{eq:NBH_z} and \eqref{eq:FBH_z} to investigate the impact of DF on both $N_{\rm BH}$ and $\mathcal{F}_{\rm BH}$. For this purpose, we compute $N_{\rm BH} (z) / N_{\rm BH} (0)$ and $\mathcal{F}_{\rm BH}(z)$ over the redshift range of interest $z \in [0,1]$ (Section \ref{subsec:DLdistribGNs}) for various values of $\Msmbh$ (Section \ref{subsec:SMBH_MassRange}) and for the two BH formation models (Section \ref{subsec:MergRates}) choosing $\mathcal{F}_{\rm BH}(z=1)$ among the two $\mathcal{F}_{\rm BH}$ models. We find (i) similar results for $\mathcal{F}_{\rm BH}(0)$ and $\mathcal{F}_{\rm BH}(1)$, (ii) that $N_{\rm BH}$ at $z = 1$ is $\sim 80 - 92\%$ of the present day $N_{\rm BH}$ depending mostly on $\Msmbh$, (iii) that $N_{\rm BH}$ systematically increases with decreasing $z$ due to the continuous inflow of BHs inside the GN across cosmic time, (iv) and that $N_{\rm BH}$ is more than $\sim 84 - 93 \%$ of the present-day $N_{\rm BH}$ between $z \simeq 0 - 0.7$ depending on $\{ M_{\rm SMBH}, p_0, \mathcal{F}_{\rm BH} \}$, where the vast majority of GW capture BBHs merged in the performed MC experiments (Figure \ref{fig:ParamDistGWcBBHs_Volume_Limits}). The weak dependence of $\mathcal{F}_{\rm BH}$ on $z$ is a result of the fact that the fractional increase in the number of BHs is very similar by $z=1$ in the two models. Since DF has non-negligible impact on $N_{\rm BH}$, we compute its redshift dependence according to Equation \eqref{eq:NBH_z}, where we set the reference redshift to be $z = 0$ and the reference $N_{\rm BH}(0)$ according to Equation \eqref{eq:NtotBHappr}. Note that DF has a marginal effect on $r_{\rm min}$ and thereby on $r_{\rm min}^{A,B}$ as it weakly depends on both $N_{\rm BH}$ and $\mathcal{F}_{\rm BH}$ (Appendix \ref{subsec:GNs_BHPops}). Similarly, we find that DF has a negligible effect on MS, WD, and NS populations, both their adopted mass distribution and total number.
 
\begin{figure}
  \includegraphics[width=85mm]{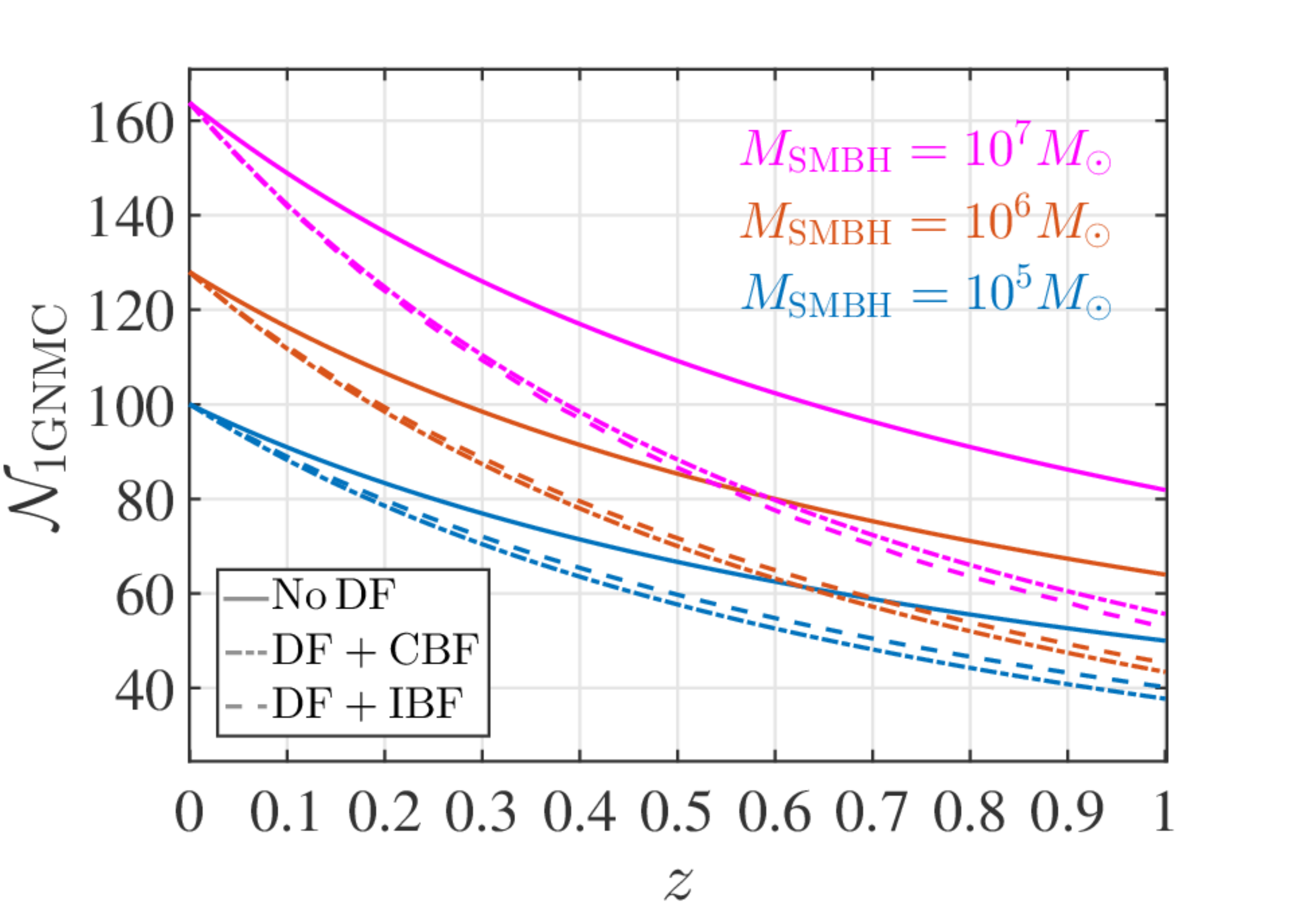}  
  \caption{Number of GW capture BBHs $(\mathcal{N}_{\rm 1GNMC})$ generated in single GNs in mock GN samples as a function of redshift $z$ out to $z = 1$ (Section \ref{subsec:DLdistribGNs}) for three different SMBH masses as labelled. We compute $\mathcal{N}_{\rm 1GNMC}$ according to Equation \eqref{eq:SampleSize_GNhost}. Three different models are adopted to calculate $\mathcal{N}_{\rm 1GNMC}$ across cosmic time as labeled: solid line does not account for the effect of dynamical friction (${\rm No \, DF}$), dash and dash-dotted lines account for the effect of DF by assuming "continuous" (${\rm DF + CBF}$) and "instantaneous" (${\rm DF + IBF}$) BH formation rates in the centers of galaxies. We plot results for the fiducial model ($\beta = 2$ and $p_0 = 0.5$), and find similar trends for other BH mass distributions and mass--segregation parameters (not shown). Results are very similar for the two models with DF because most of the BHs are delivered to the GN at higher redshift $z>1$ in both cases. \label{fig:NumbGenBBHsperGN} } 
\end{figure}

 Finally, let us examine the impact of DF on the number of MC samples in a single GN, $\mathcal{N}_{\rm 1GNMC}$, for different $\{ \mathcal{F}_{\rm BH}, p_0, \Msmbh, z \}$ by means of Equation (\ref{eq:SampleSize_GNhost}). To this end, we evaluate $\mathcal{N}_{\rm 1GNMC}$ for three different models: DF with ``continuous'' and ``instantaneous'' BH formation rate (Section \ref{subsec:MergRates}), and we consider the model in which DF is neglected \mbox{(i.e. $\zeta \equiv 1$)}. Numerical experiments show that $\mathcal{N}_{\rm 1GNMC}$ is nearly the same for the two DF models because most of the BHs are delivered to the GN at higher redshift $z>1$ in both cases. However, $\mathcal{N}_{\rm 1GNMC}$ systematically increases with $z$ and $\Msmbh$ for the model without DF compared to the models with DF, e.g. by $\sim 19 - 61\%$ ($\sim 11 - 27 \%$) at $z = 1$ ($z = 0.5$) depending on $\{ p_0, \mathcal{F}_{\rm BH}, \Msmbh \}$; see Figure \ref{fig:NumbGenBBHsperGN} for examples. This can be explained as follows. As for the $z$ dependence, the intrinsic merger rate $\Gamma_{\rm 1GN,DF}(\Msmbh, z)$ decreases with increasing $z$ for the two DF models, while $ \Gamma_{\rm 1GN,DF}(\Msmbh, z)$ is conserved in time for the model without DF (Section \ref{subsec:MergRates}). This leads to a steeper merger rate profile in $z$ for the models with DF. We find that the relative increment of $N_{\rm BH}$ with decreasing $z$ is higher for more massive SMBH masses for the two models with DF. In contrast, $N_{\rm BH}$ does not change in time when DF is neglected. Since $\Gamma_{\rm 1GN,DF}(\Msmbh, z) \propto N_{\rm BH}^2$ (Equation \ref{eq:rate_mAmB}), the $\Msmbh$ dependence of $N_{\rm BH}$ and thereby $\Gamma_{\rm 1GN,DF}(\Msmbh, z)$ leads to the obtained $\Msmbh$ dependence of $\mathcal{N}_{\rm 1GNMC}$. Since most of the BHs settle to the inner region due to DF at higher redshift in mock GN samples (Figure \ref{fig:PDF_Dlum}), DF could bias the results in a star formation dependent way at higher redshifts, but not at $z<1$. Thus, without loss of generality, we simply adopt the ``continuous'' BH formation model for concreteness.

%% file: Appendices/Appendix_RespGWdetector.tex
\section{Response of a Single Earth-based GW Detector}
\label{sec:CoordSyst_RespGWdet}
 
 We first introduce the adopted coordinate systems, then describe the interaction of an L-shaped Earth-based interferometer with GWs.
 
 We assume an L-shaped Earth-based interferometer whose arms define two-thirds of an orthonormal triple and adopt a Cartesian coordinate system with orthonormal basis vectors $\mathbf{x}$, $\mathbf{y}$, and $\mathbf{z}$ to the interferometer as follows. $\mathbf{x}$ and $\mathbf{y}$ define unit vectors along the arms of the interferometer that set $\mathbf{z} = \mathbf{x} \times \mathbf{y}$, and we orient the coordinate system such that $-\mathbf{z}$ points from the intersection of arms to the center of the Earth. We define a spherical coordinate system ($\theta, \phi$) fixed to the detector such that $\theta = 0$ is along $\mathbf{z}$, $(\theta, \phi) = (\pi/2, \, 0)$ is along $\mathbf{x}$, and $(\theta, \phi) = (\pi/2, \, \pi/2)$ is along $\mathbf{y}$. Next, we denote the unit vector pointing from the coordinate origin to a galaxy as $\mathbf{N}$, which also refers to the sky position unit vectors of binaries in the GN of a galaxy, and the angular momentum unit vector direction of a binary is given by $\mathbf{L}$,
\begin{align} 
 \mathbf{N} & = \sin \theta_N \cos \phi_N \, \mathbf{x} + \sin \theta_N 
  \sin \phi_N \, \mathbf{y} + \cos \theta_N \, \mathbf{z} \, ,
  \\
  \mathbf{L} & = \sin \theta_L \cos \phi_L \, \mathbf{x} + \sin \theta_L 
  \sin \phi_L  \, \mathbf{y} + \cos \theta_L \, \mathbf{z} \, ,
\end{align} 
 where the $(\theta_N, \, \phi_N)$ and $(\theta_L, \, \phi_L)$ coordinate pairs are the corresponding spherical coordinates. According to the adopted geometrical conventions $0 \leqslant \{ \theta_N, \theta_L \} \leqslant \pi$ and $0 \leqslant \{ \phi_N, \phi_L \} \leqslant 2 \pi$.
 
 We now describe how the GW waves interact with an advanced GW detector. The response of a detector to a GW can be given in time-domain by
\begin{equation} \label{eq:ht}
  h(t) = h_{+}(t) F_{+} + h_{\times}(t) F_{\times} \, , 
\end{equation} 
 where $h_+(t)$ and $h_{\times}(t)$ are the two polarization states of a GW waveform, and $F_+ $ and $F_{\times}$ are the antenna pattern functions for the two polarizations that are given for an L-shaped interferometer as
\begin{align} 
 \label{eq:Fplus}
 F_+ &= \frac{1}{2} \left( 1 + \cos^2 \theta_N \right) \cos{2 \phi_N} \cos{2 \psi} 
 - \cos{\theta_N} \sin{2 \phi_N} \sin{2 \psi} \, ,
  \\
 \label{eq:Fcross}
 F_{\times} &= \frac{1}{2} \left( 1 + \cos^2 \theta_N \right) \cos{2 \phi_N} \sin{2 \psi} 
 + \cos{\theta_N} \sin{2 \phi_N} \cos{2 \psi} 
\end{align} 
 \citep{Thorne1987}, where $\psi \in [0, \, 2 \pi]$ is the polarization angle. Note that $\psi$ is practically unknown because the GW's polarization itself is typically unknown.
 
 We use the Fourier transform of $h(t)$ to assign ${\rm S/N}$ values to binaries in the MC simulations. As we neglect spins, the angular momentum unit vector direction $(\theta_L, \phi_L)$ is conserved during the eccentric inspiral \citep{CutlerFlanagan1994}. Furthermore, \citet{KocsisLevin2012,Gondanetal2018a,GondanKocsis2019} showed that more than $99 \%$ of the ${\rm S/N}$ accumulates in only a few seconds to a few $\times 10$ of minutes in the GW detectors' sensitive frequency band for stellar-mass eccentric compact binaries with arbitrary $\rho_{\rm p0}$ and $e_0$, where lower timescales correspond to more massive binaries. Thus, $(\theta_N, \phi_N)$ are approximately conserved while the GW signal of such binaries is in the aLIGO/AdV/KAGRA band. Consequently, modulation effects on the ${\rm S/N}$ due to the Earth's rotation and orbit around the Sun and the evolution of $\{ \theta_N, \phi_N, \theta_L, \phi_L \}$ can be neglected. Thus, the Fourier transform of $h(t)$ is
\begin{equation}  \label{eq:hf}
  \tilde{h}(f) = F_{+} \, \tilde{h}_{+}(f) + F_{\times} \, \tilde{h}_{\times} (f) \, , 
\end{equation}
 where both $\tilde{h}_{+}(f)$ and $\tilde{h}_{\times}(f)$ are calculated for sources with fixed parameters as 
\begin{equation}  \label{eq:hf_pt}
  \tilde{h}_{+,\times}(f) = \int_{-\infty}^{+\infty} h_{+,\times}(t) \, e^{2 \pi i f} dt \, .
\end{equation}

%% file: Appendices/Appendix_DetDistance.tex
\section{Maximum Luminosity Distance of Detection with a single detector} 
\label{sec:HorizonDistance}

 We first calculate the aLIGO/AdV/KAGRA horizon distance from which the instrument can detect initially highly eccentric BBHs as a function of binary parameters, then determine the maximum luminosity distance of detection.
 
 The horizon distance is used to characterize the reach of a given search, which depends on the details of the search pipeline and the detector data quality. So far only unmodeled GW search methods have been developed to find GW signals of stellar-mass eccentric BBHs in data streams of LIGO/Virgo detectors \citep{Taietal2014,Coughlinetal2015,Tiwarietal2016,Abbottetal2019d}. This originates from the lack of reliable complete waveform models along with the support for both a large range of eccentricity and spin, which prevents the implementation of a matched-filtering search.\footnote{The impact of eccentricity on template-based GW searches used by Advanced LIGO and Advanced Virgo for quasi-circular compact binaries and on the unmodeled GW searches developed for eccentric compact binaries were thoroughly investigated in \citet{BrownZimmerman2010,HuertaBrown2013,RamosBuadesetal2020_2}.} A simplifying, while still satisfactorily accurate, detection criterion commonly used in the astrophysical literature (e.g., \citealt{Abadieetal2010,OShaughnessyetal2010,Belczynskietal2016b}) involves computing the ${\rm S/N}$ of events and assuming detection if ${\rm S/N} > {\rm S/N}_{\rm lim}=8$ for a single GW detector (e.g., \citealt{OLearyetal2006,Abadieetal2010,OShaughnessyetal2010,Dominiketal2015,Belczynskietal2016b,GerosaBerti2019b,Tagawaetal2020}). In this case, the \textit{horizon distance} is defined as the luminosity distance at which an optimally oriented ($\mathbf{z} \cdot \mathbf{L} \equiv \pm 1$ and $\psi \equiv 0$) and overhead ($\mathbf{z} \cdot \mathbf{N} \equiv \pm 1$) source can be detected with ${\rm S/N} = 8$. The ${\rm S/N}$ is computed as
\begin{equation}  \label{eq:SNR_onedet} 
  \mathrm{S/N} = \left( 4 \int_0 ^{+\infty } \frac{|\tilde{h}(f)|^2 }{S_h(f)} df \right)^{1/2} \, ,
\end{equation} 
 where $f$ is the GW frequency in the observer frame, $|\tilde{h}(f)|$ is the amplitude of adopted frequency-domain eccentric waveform model, and $S_h(f)$ is the one-sided power spectral density of an advanced GW detector. We assume the advanced GW detectors at design sensitivity \citep{Abbottetal2018} in this study.
 
 The recently developed inspiral-only and inspiral-merger-ringdown waveform models for eccentric BBHs span the eccentricity range up to $0.9$ \citep{Tanayetal2016,LoutrelYunes2017,Mooreetal2018,MooreYunes2019,Loutrel2020,Khaliletal2021} and $0.8$ \citep{HindererBabak2017}, respectively. However, as shown in Section \ref{subsec:ResMC_SingleGN}, GW capture BBHs in single GNs typically form above $e_0 \simeq 0.95$ and more than half of them enter the $10 \, \Hz$ frequency band with eccentricities above $0.9$ or form at higher eccentricities. Furthermore, the generation of these waveform models is computationally prohibitively expensive even for thousands of single systems. Currently, numerical relativity simulation-fitted ready--to--use full waveform models are unavailable for large eccentricity, arbitrary spin, and masses, but efforts for generating them are underway (e.g., \citealt{Tanayetal2016,CaoHanetal2017,HindererBabak2017,Huertaetal2017,LoutrelYunes2017,Hinderetal2018,Huertaetal2018,Mooreetal2018,Boetzeletal2019,HabibHuerta2019,MooreYunes2019,Tiwarietal2019,Chenetal2020,Chiaramelloetal2020,Liuetal2020,Loutrel2020,RamosBuadesetal2020,Khaliletal2021}). As a consequence, uncertainties in the ${\rm S/N}$ and thereby in $D_{\rm hor}$ arise even from the adopted waveform model itself because the accumulated ${\rm S/N}$ is sensitive to the order of PN corrections \citep{KocsisLevin2012} as well as templates for the merger/ringdown phases \citep{Eastetal2013}, even in the circular limit \citep{Dominiketal2015}. In this study, we intend to use the adopted eccentric waveform model only to estimate the ${\rm S/N}$ for GW capture BBHs as a proxy for binary detection with an aLIGO/AdV/KAGRA detector. Therefore, we adopt a simpler waveform model that can be used to efficiently compute the ${\rm S/N}$ for the large MC database of binaries with \mbox{$e_0 \gtrsim 0.95$}. To this end, we adopt the non-precessing version of the semi-analytical inspiral-only waveform model of \citet{MorenoGarridoetal1994,MorenoGarridoetal1995,Mikoczietal2012} in the frequency domain, which describes the quadrupole waveform in the inspiral phase emitted by a non-spinning binary on a Keplerian orbit and is applicable for arbitrary eccentricity. Note that this waveform model serves as a conservative estimate of the ${\rm S/N}$ and thereby $D_{\rm hor}$ because the ${\rm S/N}$ tends to be increased by pericenter precession effects \citep{Gondanetal2018a,GondanKocsis2019} particularly for relatively heavy BBHs.
 
 Using Equation \eqref{eq:SNR_onedet} and the adopted waveform model, a numerically efficient algorithm can be formulated to calculate the ${\rm S/N}$ for the inspiral phase of initially highly eccentric binaries; see Appendix D.1.1 in \citet{Gondanetal2018a}.\footnote{Validation tests for the applied formula were carried out in Appendix E.2.1 in \citet{Gondanetal2018a}.} We utilize this formula in further calculations after substituting $F_{+}$ and $F_{\times}$ in Equation (D4) in \citet{Gondanetal2018a} with that defined in Equations \eqref{eq:Fplus} and \eqref{eq:Fcross}. The adopted formula for the ${\rm S/N}$ depends on the parameters $\{ M_{\rm tot, z}, \eta, \rho_{\rm p0}, e_0, D_{\rm L}, \theta_N, \phi_N, \theta_L, \phi_L, \psi \}$\footnote{The adopted formula for the ${\rm S/N}$ depends on $M_{\rm tot, z}$ via the redshifted Keplerian mean orbital frequency, whereas $\eta$ appears by expressing the redshifted chirp mass in the amplitude of the adopted waveform model in terms of $M_{\rm tot, z}$ and $\eta$. The presence of other parameters is straightforward.}, therefore mock samples of GW capture BBHs generated in a given volume have to involve these parameters for each binary in order to account for the selection bias. Note that the scaling relation of the ${\rm S/N}$ can be expressed analytically as ${\rm S/N} \propto \sqrt{\eta}/D_{\rm L}$ in case of $\eta$ and $D_{\rm L}$. Moreover, we find that the ${\rm S/N}$ is marginally sensitive ($\lesssim 1 \%$) to $\psi$ even in the highly eccentric limit. Besides, increasing $e_0$ from $0.99$ ($0.95$) to higher values increase the ${\rm S/N}$ by less than $0.5 \%$ ($1 \%$). Thus, the ${\rm S/N}$ is also marginally sensitive to $e_0$ in the highly eccentric limit.
 
 $D_{\rm hor}$ can be obtained by first setting the applied ${\rm S/N}$ equal to ${\rm S/N}_{\rm lim}$, then assuming an optimally oriented ($\mathbf{z} \cdot \mathbf{L} \equiv \pm 1$ and $\psi \equiv 1$) and overhead ($\mathbf{z} \cdot \mathbf{N} \equiv \pm 1$) source, finally expressing $D_{\rm hor}$. Accordingly, $D_{\rm hor}$ depends on the parameters $\{ M_{\rm tot, z}, \eta, \rho_{\rm p0}, e_0 \}$. Since $D_{\rm hor}$ slightly increases above $e_0 = 0.99$ due to the marginal sensitivity of the ${\rm S/N}$ on $e_0\sim 1$ for any given $\{M_{\rm tot, z}, \eta, \rho_{\rm p0}\}$, an accurate estimate can be put on $D_{\rm hor}$ for initially highly eccentric systems by setting $e_0 = 0.99$. In Figure \ref{fig:HorizonDistGWcBBHs}, we display results for $D_{\rm hor}$ in case of aLIGO between $5 \leqslant \rho_{\rm p0} \leqslant 100$ and \mbox{$10 \, \Msun \leqslant M_{\rm tot, z} \leqslant 600 \, \Msun$} at fixed $\eta = 1/4$ as this choice leads to the largest $D_{\rm hor}$ values. Here, lower limits account for the lower cutoffs in $P(M_{\rm tot})$ and $P(\rho_{\rm p0})$ (Section \ref{subsec:ResMC_SingleGN}). We find that the maximum luminosity distance of detection is $\sim 6.71 \, \Gpc$, while binaries in the circular limit are expected to be detected out to $\sim 4.86 \, \Gpc$. Similarly, the maximum luminosity distance of detection is $\sim 4.12 \, \Gpc$ and $\sim 4.48 \, \Gpc$ for AdV and KAGRA, respectively. As seen analytically, $D_{\rm hor}$ scales in terms of ${\rm S/N}_{\rm lim}$ and $\eta$ as $D_{\rm hor} \propto \sqrt{\eta} / ({\rm S/N}_{\rm lim})$. Therefore, results for $D_{\rm hor}$ in the $\rho_{\rm p0} - M_{\rm tot,z}$ plane and thereby the maximum $D_{\rm hor}$ are reduced by $(4 \eta)^{1/2}$ or increased by $8 / ({\rm S/N}_{\rm lim})$ for other choices of $\eta$ and ${\rm S/N}_{\rm lim}$, respectively.
 
 Figure \ref{fig:HorizonDistGWcBBHs} also represents the $M_{\rm tot,z}$ and $\rho_{\rm p0}$ dependent characteristics of $D_{\rm hor}$. 
\begin{itemize}
     \item $D_{\rm hor}$ at fixed $M_{\rm tot,z}$ converges asymptotically to the circular limit for high $\rho_{\rm p0}$. Furthermore, it systematically increases with decreasing $\rho_{\rm p0}$ until $\rho_{\rm p0} = 5$, or increases with decreasing $\rho_{\rm p0}$ until a certain value between $5$ and $\approx 30$ then quickly drops off when approaching $\rho_{\rm p0} = 5$.
     
     \item $D_{\rm hor}$ at fixed $\rho_{\rm p0}$ peaks at a certain $M_{\rm tot,z}$ and systematically decreases below and above it.
\end{itemize} 
 Since $D_{\rm hor}$ is naturally connected to the ${\rm S/N}$ through  the relationship $D_{\rm hor} \propto {\rm S/N}$, we explain the $\rho_{\rm p0}$  and $M_{\rm tot,z}$ dependent characteristics of $D_{\rm hor}$  for arbitrary angular coordinates through the characteristics of the ${\rm S/N}$. This is possible because the intrinsic dynamics of the eccentric inspiral does not depend on sky position and inclination, the latter quantities introduce the same observational bias for any $\rho_{\rm p0},M_{\rm tot,z},\eta$. The $\rho_{\rm p0}$ dependence of the ${\rm S/N}$ at fixed $M_{\rm tot,z}$ is discussed in details in Section 6.1 in \citet{Gondanetal2018a}. To understand the $M_{\rm tot,z}$ dependence of the ${\rm S/N}$ at fixed $\rho_{\rm p0}$, we note the following arguments. On the one hand, the amplitude of emitted GWs systematically increases with $M_{\rm tot,z}$ leading to an increased ${\rm S/N}$. On the other hand, the GW spectrum systematically shifts toward lower frequencies with increasing $M_{\rm tot,z}$, thereby decreasing the amount of GW energy and hence the ${\rm S/N}$ in the detector's frequency band. The competition of these two effects leads to a peak in the ${\rm S/N}$ between $\sim 95 - 180 \, \Msun$ for aLIGO, where the peak is at $\sim 95 \, \Msun$ and $\sim 180 \, \Msun$ at $\rho_{\rm p0} \rightarrow \infty$ and $\rho_{\rm p0} = 6.5$, respectively (Figure \ref{fig:HorizonDistGWcBBHs}).
 
 We find that depending on $M_{\rm tot,z}$ binaries among those with $\rho_{\rm p0} \lesssim 20$ possess the largest ${\rm S/N}$ values for the advanced GW detectors. Since the $M_{\rm tot,z}$ and $\rho_{\rm p0}$ dependence of ${\rm S/N}$ is independent of $\{ \theta_N, \phi_N, \theta_L, \phi_L, \psi, \eta, {\rm S/N}_{\rm lim} \}$ and of $e_0$ in the highly eccentric limit, these types of binaries are preferentially selected by observational bias.

%% file: Appendices/Appendix_MCvolume.tex
\section{Monte Carlo sampling} 
\label{sec:MCsampling}
 
 We discuss the MC method with which we generate mock catalogs of GW capture BBHs detectable by a single GW detector in Appendix \ref{subsec:SetupMC}, and discuss the generation of merging populations in the local Universe in Appendix \ref{subsec:SetupMCLocUniv}.

\subsection{Setup of Monte Carlo simulations for binaries detectable by a single GW detector}
\label{subsec:SetupMC}

 Steps in the MC routine are as follows. 
\begin{enumerate}
    \item Generating a mock sample of GNs.
        \begin{itemize}
        \item We randomly draw $N_{\rm GN} = 5 \times 10^4$ SMBH masses from $P(\Msmbh)$ (Equation \ref{eq:P_Msmbh}) between $[10^5 \, \Msun, \, 10^7 \, \Msun]$ (Section \ref{subsec:SMBH_MassRange}) and luminosity distances from $P(D_{\rm L})$ between $[0 \, \Gpc, \, 6.71 \, \Gpc]$ (Equation \ref{eq:dens_DL}).
        
        \item We generate an isotropic random sample of their sky position angles $\theta_N$ and $\phi_N$ by drawing $\cos{\theta_N}$ and $\phi_N$ from a uniform distribution between $[-1, 1]$ and $[0, 2 \pi]$, respectively.
        
        \item We assign redshift to each GN by using the inverse luminosity distance--redshift relation (Equation \ref{eq:CovDLz}). Then, we compute the DF time for each GN utilizing the DF time--redshift relation given by Equation \eqref{eq:tDF_vs_z} to account for the impact of DF on the merger rates of GW capture BBHs and distributions of binary parameters.
        
        \item The sample size of a mock GW capture BBH population in each GN host is calculated using Equation \eqref{eq:SampleSize_GNhost}. 
        \end{itemize}
    
    \item Generating a mock sample of GW capture BBHs for each GN. 
        \begin{itemize}
            \item We use the methodology introduced in Section \ref{subsec:Setup_MC} to generate a MC sample of binaries with parameters $\{e_0, \rho_{\rm p0}, m_A, m_B \}$ (steps 2 - 6). To do so, we set the sample sizes $\{ N_{AB}, N_r, N_{w,b}, N_{\rm BBH} \}$ to be $\{ 1.5 \mathcal{N}_{\rm 1GNMC}, 1, 1, \mathcal{N}_{\rm 1GNMC} \}$, where we generate $1.5 \times $ more component mass pairs than the target sample size because some of them may be discarded in the simulation due to binary--single interaction (Section \ref{subsec:BBHevaporation}).
            
            \item Angular momentum unit vector direction angles of binaries $\{ \theta_L, \phi_L \}$ are drawn from an isotropic distribution similar to that given for $\theta_N$ and $\phi_N$, and the polarization angle, $\psi$, is sampled from a uniform distribution between $[0, \, 2 \pi]$.
            
            \item We assign the sky position, luminosity distance, and redshift of hosting GN to binaries in the generated sample, and compute $M_{\rm tot,z}$ and $\eta$ for each one of them.
        \end{itemize}
        
    \item Selecting the binaries detectable by a single GW detector.
        \begin{itemize}
            \item We discard binaries from the MC sample that are surely not detectable by a single GW detector. For this purpose, we use a selection criterion purely based on $D_{\rm L}$. First, we generate a $D_{\rm hor}$ map in the $\rho_{\rm p0} - M_{\rm tot,z}$ plane for equal-mass systems, shown specifically for aLIGO in Figure \ref{fig:HorizonDistGWcBBHs}. Next, we assign $D_{\rm hor}$ to each binary according to its coordinate pair $\{\rho_{\rm p0}, \, M_{\rm tot,z} \}$, then we reduce it by $(4 \eta)^{1/2}$ (Appendix \ref{sec:HorizonDistance}). Finally, we eliminate those binaries for which $ D_{\rm L}> D_{\rm hor}$.
            
            \item We follow the prescription introduced in Appendix \ref{sec:HorizonDistance} to obtain the ${\rm S/N}$ for each binary in the remaining sample, then keep only those that satisfy the detection criterion ${\rm S/N} > {\rm S/N}_{\rm lim}$.
        \end{itemize}
        
    \item Recording outputs.
        \begin{itemize}
            \item We record $e_{\rm 10 Hz} = e_0$ for each binary that forms with \mbox{$\rho_{\rm p0} \leqslant \rho_{\rm p0,10 Hz}$} (Equation \ref{eq:rhoaLIGO}). Binaries with $\rho_{\rm p0} > \rho_{\rm p0,10 Hz}$ are evolved from their initial orbit until $f_{\rm GW} = 10 \, \Hz$ using Equation \eqref{eq:drhopde_3p5PN}, then we record the obtained $e_{\rm 10 Hz}$ values.
            
            \item $e_{\rm 10 M} \equiv e_0$ is recorded for each binary with $\rho_{\rm p0} \leqslant 10$. For those with $\rho_{\rm p0} > 10$, we evolve them until $\rho_{\rm p} = 10$ using Equation \eqref{eq:drhopde_3p5PN} and record their eccentricity.
            
            \item The resulting sample contains the binary parameters $\{m_A, m_B, z, D_{\rm L}, e_0, \rho_{\rm p0}, e_{\rm 10 Hz}, e_{\rm 10 M} \}$.
        \end{itemize}
\end{enumerate}

 We tested the convergence of resulting distributions as a function of $N_{\rm GN}$ and $\mathcal{R}_{\rm GN,fid}$ by evaluating the Kolmogorov--Smirnov test with respect to the final distributions for various choices of $\mathcal{F}_{\rm BH}$ and $p_0$. We found that the settled sample sizes lead to convergent distributions, which contain \mbox{$\approx 1.5 \times 10^5 - 3.3 \times 10^5$} binaries per mock catalog depending on $\mathcal{F}_{\rm BH}$ and $p_0$.

\subsection{Setup of Monte Carlo simulations for binaries in the local Universe}
\label{subsec:SetupMCLocUniv}

 We simplify the MC method introduced in Appendix \ref{subsec:SetupMC} to generate the distributions of binary parameters for GW capture BBHs merging in the local Universe.
 
 The MC routine implements the following steps.
\begin{enumerate}
  \item We randomly draw $N_{\rm GN} = 5 \times 10^4$ SMBH masses from $P(\Msmbh)$ (Equation \ref{eq:P_Msmbh}) between $[10^5 \, \Msun, \, 10^7 \, \Msun]$ (Section \ref{subsec:SMBH_MassRange}), and neglect redshift for each GN by setting $z = 0$.
  
  \item We set $\{ \mathcal{N}_{\rm 1GNMC,fid}, z_{\rm fid}, M_{\rm SMBH,fid} \} = \{ 50, 0, 10^5 \, \Msun \}$, and determine the number of binaries to be generated for each GN host using Equation \eqref{eq:SampleSize_GNhost}.
  
  \item We use the methodology introduced in Section \ref{subsec:Setup_MC} to generate a MC sample of binaries with parameters $\{e_0, \rho_{\rm p0}, m_A, m_B \}$ for each GN host (steps 2 - 6). Sample sizes $\{ N_{AB}, N_r, N_{w,b}, N_{\rm BBH} \}$ are set as $\{ 1.5 \mathcal{N}_{\rm 1GNMC}, 1, 1, \mathcal{N}_{\rm 1GNMC} \}$ according to Appendix \ref{subsec:SetupMC}.
  
  \item We record $e_{\rm 10 Hz} \equiv e_0$ for each binary in the sample with \mbox{$\rho_{\rm p0} \leqslant \rho_{\rm p0,10 Hz}$} (Equation \ref{eq:rhoaLIGO}). Next, we evolve binaries with \mbox{$\rho_{\rm p0} > \rho_{\rm p0,10 Hz}$} from their initial orbit until $f_{\rm GW} = 10 \Hz$ using Equation (\ref{eq:drhopde_3p5PN}), then record their residual eccentricities.
  
  \item Similarly, we record $e_{\rm 10 M} \equiv e_0$ for each binary with $\rho_{\rm p0} \leqslant 10$. Finally, we evolve those with $10 < \rho_{\rm p0}$ until $\rho_{\rm p} = 10$ using Equation \eqref{eq:drhopde_3p5PN} and then record their eccentricity.
  
  \item The output parameters for the generated binaries are $\{m_A, m_B, e_0, \rho_{\rm p0}, e_{\rm 10 Hz}, e_{\rm 10 M} \}$.
\end{enumerate}
 The adopted sample sizes $N_{\rm GN}$ and $\mathcal{N}_{\rm 1GNMC,fid}$ lead to convergent distributions of binary parameters\footnote{We also evaluated the Kolmogorov--Smirnov test with respect to the final distributions similar to that in Appendix \ref{subsec:SetupMC}.}, and resulting catalogs contain $\approx 2.6 - 2.8 \times 10^5$ binaries depending on $p_0$ and $\mathcal{F}_{\rm BH}$.